\documentclass[10pt,aps,prd,twocolumn,groupedaddress,draft]{revtex4-1}
\usepackage{natbib}
\usepackage{dcolumn}
\usepackage{mathtools}
\usepackage{amsmath}
\usepackage{amssymb}
\usepackage{amsfonts}
\usepackage{mathrsfs}
\usepackage{graphicx}
\usepackage{bm}
\usepackage{xcolor} 
\usepackage{bigstrut}
\usepackage{tabularx}
\usepackage{upgreek}
\newcommand{\Lagr}{\mathcal{L}}
\newcommand\Lie{\pounds} 
  {\color{red}}%
  {}
  
\begin{document}

\title{Relativistic, finite temperature multifluid hydrodynamics in a neutron star from a variational principle}

\author{Peter B. Rau and Ira Wasserman}
\affiliation{Cornell Center for Astrophysics and Planetary Science and Department of Astronomy, Cornell University, Ithaca, New York 14853, USA}

\date{\today}

\begin{abstract}
We develop a relativistic multifluid dynamics appropriate for describing neutron star cores at finite temperatures based on Carter's convective variational procedure. The model includes seven fluids, accounting for both normal and superfluid/superconducting neutrons and protons, leptons (electrons and muons) and entropy. The formulation is compared to the non-variational relativistic multifluid hydrodynamics of Gusakov and collaborators and shown to be equivalent. Vortex lines and flux tubes, mutual friction, vortex pinning, heat conduction and viscosity are incorporated into the model in steps after the basic hydrodynamics is described. The multifluid system is then considered at the mesoscopic scale where the currents around individual vortex lines and flux tubes are important, and this mesoscopic theory is averaged to determine the detailed vortex line/flux tube contributions to the macroscopic ``effective'' theory. This matching procedure is partially successful, though obtaining full agreement between the averaged mesoscopic and macroscopic effective theory requires discarding subdominant terms. The matching procedure allow us to interpret the magnetic $H$-field inside a neutron star in a way that is consistent with condensed matter physics literature, and to clarify the difference between this interpretation and that in previous astrophysical works.
\end{abstract}

\maketitle

\section{Introduction}

Neutron stars are fundamentally relativistic objects, so it is necessary to have a relativistic hydrodynamic formalism to accurately model their internal dynamics. As it is believed that neutron star cores can consist of both superfluid neutrons and superconducting protons~\citep{Baym1969,Pines1985,Lombardo2001,Chamel2017b,Haskell2018,Sedrakian2019}, this formalism needs to incorporate multiple separately-moving fluids. An ideal formalism will also incorporate effects such as superfluid--normal fluid phase transitions, superfluid neutron vortex lines, type-II superconducting proton flux tubes and dissipation. Vortex lines/flux tubes can affect the fluid dynamics through mutual friction due to scattering between vortex lines and normal fluid particles, most importantly leptons, and pinning between neutron and proton vortex lines. These effects may be important in determining the oscillation modes of neutron stars that could be excited during binary inspiral~\citep{Yu2017,Char2018}, and in explaining pulsar glitches~\citep{Anderson1975,Sedrakian1999a,Sedrakian1999,Glampedakis2009}.

In this paper we develop a fully general relativistic formulation of finite temperature, multifluid hydrodynamics appropriate for neutron star cores. We consider a core consisting of four particle species: neutrons, protons, electrons and muons. The neutrons and protons exist in both superfluid/superconducting and normal phases, whose relative motions are dynamically connected through superfluid entrainment. Our approach has a few advantages compared to previous formulations of relativistic multifluid dynamics applied to neutron stars. We follow the convective variational approach originated by Taub~\cite{Taub1954} and later elaborated by Carter and collaborators~\citep{Carter1989,Carter1991,Carter1995,Carter1998}, making only limited assumptions about the dependence of the master function (Lagrangian) on Lorentz-invariant combinations of vectors and tensors. We thus retain the full symmetry of the variational procedure while being connected, through strategic rearrangement of terms, to relativistic formulations of the~\citet{Landau1941} (see also~\citep{Landau1987}) and~\citet{Khalatnikov2000} superfluid hydrodynamics based on Son's~\citep{Son2001} hybrid multifluid hydrodynamics, most notably that of Gusakov and collaborators~\citep{Gusakov2007,Gusakov2016,Gusakov2016a}. This formulation has been applied in numerous publications e.g.~\citep{Gusakov2013,Gualtieri2014}. We improve on the most similar existing works on the subject which use Carter-style variational procedures, by~\citet{Andersson2013} and~\citet{Glampedakis2011}, through the inclusion of finite temperature effects, relativity and quantized vortex lines/flux tubes, one or more of which is absent from each of the two references.  We carefully treat how flux tubes and magnetized vortex lines change the electromagnetic fields and the Maxwell equations and compare to previous versions~\citep{Carter1998,Carter2000,Prix2000,Glampedakis2011,Gusakov2016a} of relativistic and nonrelativistic superfluid-superconducting neutron star MHD. We also account for causal heat conduction, not assuming thermal excitations move with entropy as does~\cite{Andersson2013}. A final distinction between ours and previous versions of relativistic multifluid hydrodynamics is an explicit separation between the normal and superfluid degrees of freedom as separate current densities. We find this separation more physically intuitive than the Son hybrid multifluid hydrodynamics formulation.

In the first two sections, we describe the master function and the variational procedure used to determine the stress-energy tensor and equations of motion for the relativistic, finite temperature multifluid system, making as few assumptions about the dynamics as possible. We then connect our dynamics to those of~\citet{Gusakov2007} and collaborators, showing that the two formulations are similar, though ours is derivable from a variational principle. Forces between the fluids and vortex lines/flux tubes in the form of mutual friction and vortex pinning, viscosity, and conduction are then added, using the second law of thermodynamics to determine their form following~\citet{Carter1991}. We conclude by determining the form of the electromagnetic auxiliary field and vortex self-tension tensors, which are conjugate to the electromagnetic field tensor $F^{\mu\nu}$ and vorticity tensors respectively, by considering a simplified model of the multifluid hydrodynamics at the ``mesoscopic'' scale where currents around individual vortex lines and flux tubes are considered. The mesoscopic theory is then averaged to determine an effective macroscopic theory, with most of the details of this procedure relegated to an appendix. We are successful in averaging the mesoscopic theory, but only find an approximate match to the effective macroscopic theory. We conclude by using the results of the averaged mesoscopic-to-macroscopic matching procedure to resolve some disagreements about the interpretation of the magnetic $H$-field in a rotating superfluid--superconducting neutron star and also clarify the form of the Maxwell equations and Lorentz force acting on the charged fluids in neutron star MHD. An alternate form of the relativistic stress-energy tensor is included in an appendix. $c=G=1$ units and the $(-,+,+,+)$ metric convention are used throughout.

\section{Convective variational procedure}

Starting with a Lagrangian density describing the finite temperature multifluid in a neutron star core, we employ the convective variational procedure~\citep{Carter1989,Carter1991,Carter1995,Carter1998} to compute the relevant equations of motion. There has been recent interest in this formulation~\citep{Gavassino2020,Gavassino2020a} for application to problems involving neutron star asteroseismology, pulsar glitches and gravitational waves from binary neutron stars. Compared to other fluid variational methods~\citep{Taub1954,Schutz1970}, with the convective variational procedure we can transparently include additional forces between the fluids that are not obviously incorporated directly via a variational method. An additional advantage which we exploit is the ability to include viscosity using a convective variational-type method.

In the first subsection, we describe our Lagrangian density and define the dynamical variables, adding in steps the fluid number currents, electromagnetism and vorticity. In the second subsection we introduce the Lagrangian displacement fields employed in the convective variational procedure and derive the equations of motion.

\subsection{Lagrangian and its variation}
\label{sec:LagrangianAndVariation}

Consider a multifluid neutron star core consisting of neutrons ($n$), protons ($p$), electrons ($e$), muons ($m$) and entropy ($s$). The neutrons and protons will have both superfluid/superconducting and normal fluid excitation components, with the former being distinguished using an overline ($\overline{n},\overline{p}$). The notation $\overline{x}$ refers to either superfluid species. There exists a four-current $n^{\mu}_x$, $x\in\{n,p,e,m,\overline{n},\overline{p},s\}$, for each species/quantity. $x=s$ is the entropy four-current $s^{\mu}$, which will later be related to the four-currents of the entropy-carrying normal fluids. In principal each normal fluid could have its own corresponding entropy current, but as we will later restrict the normal fluids to move together, we introduce only a single entropy current here. The following Lorentz-invariant scalars can be constructed by contracting the four-currents:
\begin{align}
n_x^2={}&-n_x^{\mu}n^x_{\mu}, \quad \alpha_{xy}^2=-n_{x}^{\mu}n^{y}_{\mu}=\alpha_{yx}^2, 
\end{align}
where $y\in\{n,p,e,m,\overline{n},\overline{p},s\}\neq x$. $\alpha^2_{xy}$ is equivalent to the product of the Lorentz factor for the relative motion between fluids $x$ and $y$ and the two number densities $n_x$ and $n_y$ as measured in the respective fluids' rest frames, as will be clear from the definition of $n_x^{\mu}$ given in Eq.~(\ref{eq:SuperfluidCurrent}). $\alpha_{xy}$ with $y\neq s$ will be responsible for superfluid entrainment, while the $\alpha_{xs}$ are ``entropy entrainment'' terms representing heat convection by the particle currents. The $\alpha_{xs}$ will later allow for heat conduction independent of the particle currents. There will be 10 nonzero $\alpha_{xy}$ ($\alpha_{np}$, $\alpha_{n\overline{n}}$, $\alpha_{n\overline{p}}$, $\alpha_{p\overline{n}}$, $\alpha_{p\overline{p}}$, $\alpha_{\overline{n}\overline{p}}$, $\alpha_{sn}$, $\alpha_{sp}$, $\alpha_{se}$, $\alpha_{sm}$). The superfluids do not carry entropy, so $\alpha_{s\overline{n}}=\alpha_{s\overline{p}}=0$. The exclusion of entropy entrainment results in instabilities and causality violation~\citep{Olson1990,Lopez-Monsalvo2011}, and we discuss the effects of heat conduction on the entropy current in Section~\ref{sec:Conduction}.

The Lagrangian density will be a function of dynamical variables $n^{\mu}_x$, the electromagnetic field tensor $A_{\mu}$, and vorticity tensors $w^{\overline{x}}_{\mu\nu}$ associated with the vortex line/flux tube arrays for each superfluid species. We can split the Lagrangian density into a master function $\Lambda$, interaction terms and spacetime curvature terms. $\Lambda$ includes the thermodynamic internal energy density of the fluid, the electromagnetic field energy and the vortex line/flux tube energy, and is a function of Lorentz invariant scalars. To begin, we consider only the dependence of this master function on the number currents and the metric: 
\begin{equation}
\Lambda=\Lambda(n_x^2,\alpha^2_{xy}, g_{\mu\nu}),
\end{equation}
where all $x$ and distinct combinations of $x$ and $y$ are implicitly included. Varying this $\Lambda$ gives
\begin{equation}
\delta\Lambda = \frac{\partial\Lambda}{\partial n^2_x}\delta n^2_x+\frac{\partial\Lambda}{\partial \alpha^2_{xy}}\delta \alpha^2_{xy}+\frac{\partial \Lambda}{\partial g_{\mu\nu}}\delta g_{\mu\nu}.
\label{eq:DeltaLambda1}
\end{equation}
The variations with respect to the four-currents can be rewritten in terms of the number and entropy four-currents using
\begin{align}
\frac{\partial\Lambda}{\partial n^2_x}\delta n^2_x={}&\left(-2\frac{\partial\Lambda}{\partial n_x^2}n_{\mu}^x\right)\delta n_x^{\mu},
\\
\frac{\partial\Lambda}{\partial \alpha^2_{xy}}\delta \alpha^2_{xy}={}&\left(-\frac{\partial\Lambda}{\partial \alpha_{xy}^2}n_{\mu}^x\right)\delta n_y^{\mu}+\left(-\frac{\partial\Lambda}{\partial \alpha_{xy}^2}n_{\mu}^y\right)\delta n_x^{\mu},
\end{align}
where we adopt the convention of~\cite{Carter1989,Andersson2007,Andersson2013}, among others, in defining
\begin{equation}
\mathcal{B}^x\equiv-2\frac{\partial\Lambda}{\partial n_x^2}, \quad \mathcal{A}^{xy}=\mathcal{A}^{yx}\equiv-\frac{\partial\Lambda}{\partial \alpha_{xy}^2}.
\label{eq:ChemPotCoeffDefs} 
\end{equation}
There will be 7 $\mathcal{B}^x$, one for each particle current plus the entropy current, and 10 $\mathcal{A}^{xy}$ corresponding to each nonzero $\alpha_{xy}$. Using Eq.~(\ref{eq:ChemPotCoeffDefs}) and noting which $A^{xy}$ are zero, we can define the conjugate dynamical momenta or generalized chemical potential four-vectors
\begin{equation}
\mu_{\mu}^x = \mathcal{B}^xn^x_{\mu}+\sum_{y\neq x}\mathcal{A}^{xy}n^y_{\mu}+A^{sx}s_{\mu}, \label{eq:NFNeutronMuVec1}
\end{equation}
where $x,y\in\{n,p,e,m,\overline{n},\overline{p}\}$, and a conjugate ``thermal momentum''
\begin{equation}
\Theta_{\mu}=\mu^s_{\mu}=\mathcal{B}^ss_{\mu}+\sum_x\mathcal{A}^{sx}n^x_{\mu}, \label{eq:ThetaVec1}
\end{equation}
where $x\in\{n,p,e,m\}$. To determine $\partial\Lambda/\partial g_{\mu\nu}=\partial\Lambda/\partial g_{\nu\mu}$, following \citet{Carter1991}, the variations in Eq.~(\ref{eq:DeltaLambda1}) are specified by their Lie derivative $\Lie_{\xi}$ with respect to a single infinitesimal displacement field $\xi^{\rho}$ which acts on the background manifold. This displacement field is not the same as the displacement fields which specify the motion of the individual fluids and which are introduced in Section~\ref{sec:EquationsOfMotion}. For the purposes of determining $\partial\Lambda/\partial g_{\mu\nu}$, we use
\begin{subequations}
\begin{align}
\delta\Lambda ={}& \Lie_{\xi}\Lambda = \xi^{\rho}\nabla_{\rho}\Lambda,
\label{eq:DeltaLambda}
\\
\delta n_x^{\mu}={}& \Lie_{\xi}n_x^{\mu}=\xi^{\rho}\nabla_{\rho}n^{\mu}_x-n_x^{\rho}\nabla_{\rho}\xi^{\mu},
\label{eq:DeltaVector}
\\
\delta g_{\mu\nu}={}&\Lie_{\xi}g_{\mu\nu}=\nabla_{\mu}\xi_{\nu}+\nabla_{\nu}\xi_{\mu}=2\nabla_{(\mu}\xi_{\nu)},
\label{eq:DeltaG}
\end{align}
\end{subequations}
which, inserted into Eq.~(\ref{eq:DeltaLambda1}), give the following relation
\begin{align}
\Biggl(\sum_x \mu^x_{\nu}\nabla_{\mu}n_x^{\nu}-&\nabla_{\mu}\Lambda\Biggr)\xi^{\mu} \nonumber
\\
{}&=\left( \sum_x \mu_x^{\mu}n_x^{\nu}-2\frac{\partial\Lambda}{\partial g_{\mu\nu}}\right)\nabla_{\mu}\xi_{\nu}.
\label{eq:MasterFunctionMetricVariation}
\end{align}
Since this must be true for arbitrary $\xi^{\mu}$ and $\nabla_{\mu}\xi_{\nu}$, both sides of this must be zero independently, giving
\begin{align}
\nabla_{\mu}\Lambda=\sum_x\mu^x_{\nu}\nabla_{\mu}n_x^{\nu},
\label{eq:MasterFunctionGradient1}
\\
\frac{\partial\Lambda}{\partial g_{\mu\nu}}=\frac{1}{2}\sum_x\mu_x^{\mu}n_x^{\nu}.
\label{eq:MasterFunctionMetricDerivative1}
\end{align}
Inserting the second of these into Eq.~(\ref{eq:DeltaLambda1}) and using the definitions of the conjugate momenta, $\delta\Lambda$ becomes
\begin{equation}
\delta\Lambda = \sum_x\mu^{x}_{\mu}\delta n^{\mu}_x+\frac{1}{2}\sum_xn_x^{\mu}\mu_x^{\nu}\delta g_{\mu\nu}.
\end{equation}
As written, the extremization of the action with respect to each current density would require the conjugate momentum to be zero. This is of course too restrictive, and the correct variation of the current densities in terms of Lagrangian displacement fields is introduced in Section~\ref{sec:EquationsOfMotion}.

To include electromagnetism, we allow the master function to depend on the electromagnetic field tensor $F_{\mu\nu}=2\nabla_{[\mu}A_{\nu]}$ through a contraction with another antisymmetric rank-two tensor. The electromagnetic four-potential $A_{\mu}$ is minimally coupled to the total charge current
\begin{equation}
\Lagr_{\text{EM coup.}} = j^{\mu}_eA_{\mu},
\end{equation}
where the charge current is 
\begin{equation}
j^{\mu}_e=\sum_xq_xn^{\mu}_x,
\end{equation}
for $x$ including all species/quantities with $q_p=q_{\overline{p}}=e$, $q_e=q_m=-e$, $q_n=q_{\overline{n}}=q_s=0$. The variation of the action thus contains additional terms
\begin{equation}
\delta(\Lambda_{\text{EM}}+\Lagr_{\text{EM coup.}})=-\frac{1}{8\pi}\mathcal{K}^{\mu\nu}\delta F_{\mu\nu}+j^{\mu}_e\delta A_{\mu}+A_{\mu}\delta j_e^{\mu}
\end{equation}
where we have defined the (antisymmetric) \textit{electromagnetic auxiliary tensor}
\begin{equation}
\mathcal{K}^{\mu\nu}=-8\pi \left.\frac{\partial\Lambda}{\partial F_{\mu\nu}}\right|_{n^{\mu}_x,w^{\overline{x}}_{\mu\nu}}.
\label{eq:EMAuxiliaryTensor}
\end{equation}
This tensor has been defined as the \textit{electromagnetic displacement tensor} $\mathcal{H}^{\mu\nu}$ in previous works~\cite{Carter1998}, but for reasons explained in Section~\ref{sec:VLFTContribution}, we reserve this notation and nomenclature for a different quantity. We have explicitly denoted that all number currents $n^{\mu}_x$ and vorticity tensors $w^{\overline{x}}_{\mu\nu}$  are held fixed during this variation.

In a rotating superfluid-superconducting neutron star, there will be quantized neutron vortex lines. If the proton superconductivity is type-II in some or all regions of the core, there will also be quantized flux tubes in those regions. These are incorporated into the variational formalism by adding terms coupling the superfluid currents to the vorticity and allowing $\Lambda$ to depend on the vorticity tensors $w^{\overline{x}}_{\mu\nu}$, $\overline{x}\in\{\overline{n},\overline{p}\}$. This method was developed in~\citep{Carter1994,Carter1995,Carter1998}, though we take a somewhat different approach.

We first rewrite the vorticity tensor in terms of a lattice field $\mathcal{X}^{\overline{x}}_{\mu}$
\begin{equation}
w^{\overline{x}}_{\mu\nu}=2\nabla_{[\mu}\mathcal{X}^{\overline{x}}_{\nu]}.
\label{eq:VorticityTensor}
\end{equation}
$\mathcal{X}^{\overline{x}}_{\mu}$ will be dynamically identified with the canonical momentum four-vector $\pi^{\overline{x}}_{\mu}$. $w^{\overline{x}}_{\mu\nu}$ can also be expressed in terms of two lattice scalars $\chi^a_{\overline{x}}$, $a\in\{1,2\}$
\begin{equation}
w^{\overline{x}}_{\mu\nu}=2\nabla_{[\mu}\chi^1_{\overline{x}}\nabla_{\nu ]}\chi^2_{\overline{x}}.
\label{eq:VorticityTensor2}
\end{equation}
The gradients of $\chi^a_{\overline{x}}$ define a plane that is locally orthogonal to the vortex lines/flux tubes. We choose for a current-vorticity coupling
\begin{equation}
\Lagr_{\text{v coup.}}=-\sum_{\overline{x}}n^{\mu}_{\overline{x}}\mathcal{X}^{\overline{x}}_{\mu}.
\end{equation}
The variation of the action will thus contain the additional vorticity terms
\begin{align}
\delta(\Lambda_{\text{v}}\hspace{-0.1em}+\hspace{-0.1em}\Lagr_{\text{v coup.}})=&{}-\sum_{\overline{x}}\left[\frac{1}{2}\lambda^{\mu\nu}_{\overline{x}}\delta w^{\overline{x}}_{\mu\nu}\hspace{-0.1em}+\hspace{-0.1em}\mathcal{X}^{\overline{x}}_{\mu}\delta n^{\mu}_{\overline{x}}+n^{\mu}_{\overline{x}}\delta\mathcal{X}^{\overline{x}}_{\mu}\right]\hspace{-0.2em},
\end{align}
where we have defined the vortex line/flux tube self-tension~\cite{Carter1998} tensor
\begin{equation}
\lambda^{\mu\nu}_{\overline{x}}\equiv-2\left.\frac{\partial\Lambda}{\partial w^{\overline{x}}_{\mu\nu}}\right|_{{n^{\mu}_x,F_{\mu\nu}}}.
\label{eq:VortexTensionTensor}
\end{equation}

The generalization of Eq.~(\ref{eq:MasterFunctionMetricVariation}) to incorporate electromagnetism and vorticity modifies Eq.~(\ref{eq:MasterFunctionGradient1})--(\ref{eq:MasterFunctionMetricDerivative1}) into
\begin{align}
\nabla_{\mu}\Lambda={}&\sum_x\mu^x_{\nu}\nabla_{\mu}n_x^{\nu}-\frac{1}{8\pi}\mathcal{K}^{\rho\nu}\nabla_{\mu}F_{\rho\nu} \nonumber
\\
{}&-\frac{1}{2}\sum_{\overline{x}}\lambda^{\rho\nu}_{\overline{x}}\nabla_{\mu}w^{\overline{x}}_{\rho\nu},
\label{eq:MasterFunctionGradient2}
\\
\frac{\partial\Lambda}{\partial g_{\mu\nu}}={}&\frac{1}{2}\left(\sum_x\mu_x^{\mu}n_x^{\nu}+\frac{1}{4\pi}\mathcal{K}^{\mu\rho}F^{\nu}_{\ \rho}+\sum_{\overline{x}}\lambda_{\overline{x}}^{\mu\rho}w^{\nu}_{\overline{x}\rho}\right).
\label{eq:MasterFunctionMetricDerivative2}
\end{align}
This was found using as the variation for rank two tensors $\mathcal{Y}_{\mu\nu}=F_{\mu\nu},w^{\overline{x}}_{\mu\nu}$
\begin{equation}
\delta\mathcal{Y}_{\mu\nu}=\Lie_{\xi}\mathcal{Y}_{\mu\nu}=\xi^{\rho}\nabla_{\rho}\mathcal{Y}_{\mu\nu}+\mathcal{Y}_{\mu\rho}\nabla_{\nu}\xi^{\rho}+\mathcal{Y}_{\rho\nu}\nabla_{\mu}\xi^{\rho}.
\label{eq:DeltaRank2Tensor}
\end{equation}
The minimal coupling terms between the currents and both electromagnetic field and vorticity are not part of $\Lambda$ and hence do not contribute to Eq.~(\ref{eq:MasterFunctionGradient2}--\ref{eq:MasterFunctionMetricDerivative2}).

We incorporate general relativity by including the Einstein--Hilbert term in the action, which corresponds to adding the following term to the Lagrangian
\begin{equation}
\Lagr_{\text{EH}}=\frac{1}{16\pi}R,
\end{equation} 
for Ricci scalar $R$, which adds the expected additional terms to the variation of the action
\begin{equation}
\delta\Lagr_{\text{EH}}=-\frac{1}{16\pi}\left(R^{\mu\nu}-\frac{1}{2}Rg^{\mu\nu}\right)\delta g_{\mu\nu},
\end{equation} 
where $R^{\mu\nu}$ is the Ricci tensor. To account for the Jacobian in the action
\begin{equation}
S=\int \mathrm{d}^4x\sqrt{-g}\Lambda,
\end{equation}
for metric determinant $g$, we add a term $\frac{1}{2}\Lambda g^{\mu\nu}\delta g_{\mu\nu}$ to the variation of $\Lagr$. We thus end up with
\begin{equation}
\delta\Lagr=\delta\Lambda+\delta\Lagr_{\text{v coup.}}+\delta\Lagr_{\text{EM coup.}}+\delta\Lagr_{\text{EH}}+\frac{1}{2}\Lambda g^{\mu\nu}\delta g_{\mu\nu},
\label{eq:Lagrangian0}
\end{equation}
where $\Lambda$ includes $\Lambda_{\text{EM}}$ and $\Lambda_{\text{v}}$.

\subsection{Deriving the equations of motion}
\label{sec:EquationsOfMotion}

We review the convective variational procedure of~\citet{Carter1989}, which is further developed and expounded in later papers~\citep{Carter1991,Carter1998,Andersson2007}. The main result of interest is the variation of the number four-current $n_x^{\mu}$, given by
\begin{equation}
\delta n_x^{\mu}=\xi^{\sigma}_x\nabla_{\sigma}n^{\mu}_x-n_x^{\sigma}\nabla_{\sigma}\xi^{\mu}_x+n_x^{\mu}\nabla_{\sigma}\xi^{\sigma}_x-\frac{1}{2}n_x^{\mu}g^{\sigma\rho}\delta g_{\sigma\rho},
\label{eq:CurrentVariation}
\end{equation}
where $\xi^{\mu}_x$ is the Lagrangian infinitesimal displacement field specifying the variation of the four-current of species $x$. This expression differs from the Lie derivative of $n^{\mu}_x$ by the inclusion of the effects of gravitational perturbations. We use the sign convention of~\citet{Carter1998}, which differs by $\xi^{\mu}\rightarrow-\xi^{\mu}$ compared to the expected nonrelativistic limit and other references such as~\citet{Andersson2007}. This is derived by first starting with a dual to the number current (omitting species labels)
\begin{equation}
^*n_{\mu\nu\sigma}=\varepsilon_{\mu\nu\sigma\rho}n^{\rho}
\label{eq:CurrentDual1}
\end{equation}
where $\varepsilon_{\mu\nu\sigma\rho}$ is the Levi-Civita tensor. This three-form can be specified by the derivatives of three scalars $N^1$, $N^2$, $N^3$, which label the coordinates of a particular fluid element in ``matter space'' and which are the same for all time. These coordinates can be pushed forward to give the coordinates of the fluid element at any time slice. So $^*n_{\mu\nu\sigma}$ can be written as
\begin{equation}
^*n_{\mu\nu\sigma}=-f(N^1,N^2,N^3)_{123}\nabla_{\mu}N^1\nabla_{\nu}N^2\nabla_{\sigma}N^3,
\label{eq:CurrentDual2}
\end{equation}
where $f(N^1,N^2,N^3)_{123}$ is antisymmetric in the scalar indices $1$, $2$, $3$. The variations of the scalars can be expressed in terms of an infinitesimal displacement field
\begin{equation}
\delta N^a = \xi^{\rho}\nabla_{\rho}N^a,
\end{equation}
and so Eq.~(\ref{eq:CurrentVariation}) can be found by taking the variation of 
Eq.~(\ref{eq:CurrentDual2}) and using Eq.~(\ref{eq:CurrentDual1}) and
\begin{equation}
\delta\varepsilon_{\mu\nu\sigma\rho}=\frac{1}{2}\varepsilon_{\mu\nu\sigma\rho}g^{\lambda\eta}\delta g_{\lambda\eta}.
\end{equation}
Note that the form of $^*n_{\mu\nu\sigma}$ as given by Eq.~(\ref{eq:CurrentDual2}) is closed
\begin{equation}
\nabla_{[\lambda}(^*n_{\mu\nu\sigma]})=0,
\end{equation}
which automatically means that $n^{\mu}$ is conserved through Eq.~(\ref{eq:CurrentDual1}). We thus assume separate conservation of each current $\nabla_{\mu}n^{\mu}_{x}=0$ in the rest of this paper for those currents where the variation Eq.~(\ref{eq:CurrentVariation}) is used. Implicit in this is the assumption that the rate of interconversion between particle species is much slower than the dynamical timescales of interest, which is certainly true for weak interactions in cold neutron stars, but not necessarily true for the formation or breaking of Cooper pairs of neutrons or protons.

The vorticity tensor $w^{\overline{x}}_{\mu\nu}$ can be specified in a similar way to the dual number current $^*n_{\mu\nu\sigma}$, except now only with two lattice scalars $\chi^1_{\overline{x}}$ and $\chi^2_{\overline{x}}$. The variation for these scalars is simply their Lie derivative with respect to $\xi^{\mu}_{\overline{x}}$:
\begin{equation}
\delta \chi^a_{\overline{x}}=\xi^{\mu}_{\overline{x}}\nabla_{\mu}\chi^a_{\overline{x}}, \quad a\in\{1,2\},
\label{eq:DeltaChi}
\end{equation}
where $\xi^{\mu}_{\overline{x}}$ are Lagrangian displacement fields describing the spacetime motion of the vortex line/flux tube array associated with the superfluid of species $\overline{x}$. The vortex line/flux tube arrays do not in general move along with the relevant superfluid species. We have assumed here that the same infinitesimal displacement field describes the variations of both $\chi^1_{\overline{x}}$ and $\chi^2_{\overline{x}}$. Since it is these lattice scalars that are the freely-varying quantities relating to the vorticity~\cite{Carter1994}, we must write the variations of the vorticity tensor and lattice field in terms of $\delta \chi^a_{\overline{x}}$ and hence $\xi^{\mu}_{\overline{x}}$.  Eq.~(\ref{eq:DeltaChi}) gives the variation of the vorticity tensor to be
\begin{equation}
\delta w^{\overline{x}}_{\mu\nu}=-2\nabla_{[\mu}(w^{\overline{x}}_{\nu]\rho}\xi^{\rho}_{\overline{x}}).
\label{eq:DeltaW1}
\end{equation}
However, the perturbation of Eq.~(\ref{eq:VorticityTensor}) gives
\begin{equation}
\delta w^{\overline{x}}_{\mu\nu}=2\nabla_{[\mu}\delta\mathcal{X}^{\overline{x}}_{\nu]},
\label{eq:DeltaW2}
\end{equation}
so by comparison to Eq.~(\ref{eq:DeltaW1}) we have 
\begin{equation}
\delta\mathcal{X}^{\overline{x}}_{\mu}=-w_{\mu\nu}^{\overline{x}}\xi^{\nu}_{\overline{x}}+\nabla_{\mu}\delta\phi_{\overline{x}}
\label{eq:Deltabeta}
\end{equation}
for a scalar $\delta\phi_{\overline{x}}$ that can be thought of as a gauge field. If we postulate the form $\mathcal{X}^{\overline{x}}_{\mu}=\chi_{\overline{x}}^1\nabla_{\mu}\chi_{\overline{x}}^2$ based on Eq.~(\ref{eq:VorticityTensor}--\ref{eq:VorticityTensor2}), then we find $\delta\phi_{\overline{x}}=\xi^{\nu}_{\overline{x}}\mathcal{X}^{\overline{x}}_{\nu}$. Note that $\mathcal{X}^{\overline{x}}_{\mu}$ itself is not determined uniquely, but only up to a physical unimportant gradient of a scalar which we set to zero here.

Combining these Lagrangian variations plus $\delta F_{\mu\nu}=2\nabla_{[\mu}\delta A_{\nu]}$ and inserting into Eq.~(\ref{eq:Lagrangian0}), we obtain
\begin{align}
\delta\Lagr ={}& \sum_{x\neq \overline{x}}\pi^{x}_{\mu}\delta n^{\mu}_x +\left(j^{\mu}_e-\frac{1}{4\pi}\nabla_{\nu}\mathcal{K}^{\mu\nu}\right)\delta A_{\mu} \nonumber
\\
{}&-\sum_{\overline{x}}\left[\frac{1}{2}\lambda^{\mu\nu}_{\overline{x}}\delta w^{\overline{x}}_{\mu\nu}+(\mathcal{X}^{\overline{x}}_{\mu}-\pi^{\overline{x}}_{\mu})\delta n^{\mu}_{\overline{x}}+n^{\mu}_{\overline{x}}\delta\mathcal{X}^{\overline{x}}_{\mu}\right]\nonumber
\\
{}&+ \frac{1}{2}\Biggl[\sum_xn_x^{\mu}\mu^x_{\sigma}g^{\sigma\nu}+\frac{1}{4\pi}\mathcal{K}^{\mu\rho}F^{\nu}_{\ \rho}+\sum_{\overline{x}}\lambda^{\mu\rho}_{\overline{x}} w^{\nu}_{{\overline{x}}\rho} \nonumber
\\
{}& \quad\qquad -\frac{1}{8\pi}R^{\mu\nu}+\left(\Lambda+\frac{1}{16\pi}R\right)g^{\mu\nu}\Biggr]\delta g_{\mu\nu}
\label{eq:Lagrangian1}
\end{align}
where we have defined the gauge-dependent canonical momentum covectors
\begin{align}
\pi^{x}_{\mu}\equiv{}&\mu^x_{\mu}+q_xA_{\mu}
\label{eq:CanonicalMomentumCovector}
\end{align}
For the normal fluids, we use Eq.~(\ref{eq:CurrentVariation}) to constrain the current variations, implying that the first term in Eq.~(\ref{eq:Lagrangian1}) becomes
\begin{align}
&\pi^x_{\mu}\xi^{\sigma}_x\nabla_{\sigma}n^{\mu}_x-\pi^x_{\mu}n_x^{\sigma}\nabla_{\sigma}\xi^{\mu}_x+\pi^x_{\mu}n_x^{\mu}\nabla_{\sigma}\xi^{\sigma}_x-\frac{1}{2}\pi^x_{\mu}n_x^{\mu}g^{\sigma\rho}\delta g_{\sigma\rho}\nonumber
\\
{}&=2\xi^{\mu}_xn_x^{\sigma}\nabla_{[\sigma}\pi^x_{\mu]}+\xi^{\mu}_x\pi_{\mu}^x\nabla_{\sigma}n^{\sigma}_x-\frac{1}{2}\pi^x_{\mu}n_x^{\mu}g^{\sigma\rho}\delta g_{\sigma\rho}
\label{eq:NormalFluidExpansion}
\end{align}
where we integrated by parts and dropped total derivative terms. Use of Eq.~(\ref{eq:CurrentVariation}) is unnecessary for the superfluid components, since the variation of the superfluid number currents is already constrained. However, this means we must enforce conservation of the separate superfluid current densities in a different manner. A simple way to do this is by adding a Schutz-type~\citep{Schutz1970} term for each superfluid to the Lagrangian:
\begin{equation}
\Lagr_{\text{S}}=-\sum_{\overline{x}}n^{\mu}_{\overline{x}}\nabla_{\mu}\varphi_{\overline{x}},
\end{equation}
where  $\varphi_{\overline{x}}$ is a scalar phase. Taking the variation of this and setting the coefficient of $\delta\varphi_{\overline{x}}$ equal to zero gives, after an integration by parts, $\nabla_{\mu}n^{\mu}_{\overline{x}}=0$. The variation with respect to $n^{\mu}_{\overline{x}}$ adds the additional term 
\begin{equation}
-\sum_{\overline{x}}\delta n^{\mu}_{\overline{x}}\nabla_{\mu}\varphi_{\overline{x}}
\end{equation}
to Eq.~(\ref{eq:Lagrangian1}). Setting the coefficient of $\delta n^{\mu}_{\overline{x}}$ equal to zero now gives
\begin{equation}
\pi^{\overline{x}}_{\mu}=\nabla_{\mu}\varphi_{\overline{x}}+\mathcal{X}^{\overline{x}}_{\mu}
\label{eq:LatticeFieldIdentity}
\end{equation}
which correctly gives us as the vorticity tensor the covariant curl of the canonical momentum covector. $\nabla_{\mu}\varphi_{\overline{x}}$ will not contribute to the vorticity, and can thus safely be set to zero. Microscopically, $\pi^{\overline{x}}_{\mu}$ is the gradient of a potential for superfluid neutrons and the gradient of a potential plus $eA_{\mu}$ for superconducting protons. This equation represents a macroscopic average. Using Eq.~(\ref{eq:DeltaW1}) and~(\ref{eq:Deltabeta}), the third and fifth terms in Eq.~(\ref{eq:Lagrangian1}) become
\begin{align}
-\frac{1}{2}\lambda_{\overline{x}}^{\mu\nu}\delta w^{\overline{x}}_{\mu\nu}{}&-n^{\mu}_{\overline{x}}\delta\mathcal{X}^{\overline{x}}_{\mu}
\nonumber
\\
={}&\lambda^{\mu\nu}_{\overline{x}}\nabla_{[\mu}(w_{\nu]\rho}\xi^{\rho}_{\overline{x}})+n^{\mu}_{\overline{x}}w_{\mu\nu}^{\overline{x}}\xi^{\nu}_{\overline{x}}-n^{\mu}_{\overline{x}}\nabla_{\mu}(\xi^{\nu}_{\overline{x}}\mathcal{X}^{\overline{x}}_{\mu})
\nonumber
\\
={}&\xi^{\mu}_{\overline{x}}\left[w_{\rho\mu}^{\overline{x}}(n^{\rho}_{\overline{x}}+\nabla_{\nu}\lambda_{\overline{x}}^{\rho\nu})+\pi^{\overline{x}}_{\mu}\nabla_{\nu}n^{\nu}_{\overline{x}}\right],
\label{eq:VorticityTermSimplify}
\end{align}
where we integrated by parts, dropping total derivative terms, and used Eq.~(\ref{eq:LatticeFieldIdentity}) in the last line.

Returning to Eq.~(\ref{eq:Lagrangian1}), $\delta\Lagr/\delta A_{\mu}=0$ gives the sourced Maxwell equations in a continuous medium
\begin{equation}
\nabla_{\nu}\mathcal{K}^{\mu\nu}=4\pi j^{\mu}_e,
\label{eq:SourcedMaxwellEquations}
\end{equation}
which also guarantees charge conservation $\nabla_{\mu}j^{\mu}_e=0$ due to the asymmetry of $\mathcal{K}^{\mu\nu}$. Both $F_{\mu\nu}$ and $w^{\overline{x}}_{\mu\nu}$ satisfy the closure condition
\begin{equation}
\nabla_{[\lambda}F_{\mu\nu]}=0,\qquad \nabla_{[\lambda}w^{\overline{x}}_{\mu\nu]}=0.
\label{eq:Closure}
\end{equation}
For $F_{\mu\nu}$, this is just the source-free Maxwell equations. The remainder of Eq.~(\ref{eq:Lagrangian1}) becomes, using Eqs.~(\ref{eq:NormalFluidExpansion}--\ref{eq:VorticityTermSimplify})
\begin{align}
\delta\Lagr ={}& \sum_{x\neq\overline{x}}\xi^{\mu}_x\left(2n_x^{\nu}\nabla_{[\nu}\pi^x_{\mu]}+\pi_{\mu}^x\nabla_{\nu}n^{\nu}_x\right)\nonumber
\\
&{}+\sum_{\overline{x}}\xi^{\mu}_{\overline{x}}\left[w^{\overline{x}}_{\rho\mu}\left(n^{\rho}_{\overline{x}}+\nabla_{\nu}\lambda^{\rho\nu}_{\overline{x}}\right)+\pi_{\mu}^{\overline{x}}\nabla_{\nu}n^{\nu}_{\overline{x}}\right]\nonumber 
\\
&{}+\frac{1}{2}\Biggl[\sum_xn_x^{\mu}\mu^x_{\rho}g^{\rho\nu}+\frac{1}{4\pi}\mathcal{K}^{\mu\rho}F^{\nu}_{\ \rho}+\sum_{\overline{x}}\lambda^{\mu\rho}_{\overline{x}} w^{\nu}_{{\overline{x}}\rho} \nonumber
\\
{}&\quad\quad\quad+\Psi g^{\mu\nu}-\frac{1}{8\pi}\left(R^{\mu\nu}-\frac{1}{2}Rg^{\mu\nu}\right)\Biggr]\delta g_{\mu\nu}
\nonumber
\\
={}&\sum_x\xi^{\mu}_xf^x_{\mu}+\frac{1}{2}\left(T^{\mu\nu}-\frac{1}{8\pi}\left(R^{\mu\nu}-\frac{1}{2}Rg^{\mu\nu}\right)\right)\delta g_{\mu\nu},
\label{eq:Lagrangian2}
\end{align}
where $f^{x}_{\mu}$ is the generalized force (density) acting on fluid $x$, the generalized pressure $\Psi$ is defined as
\begin{equation}
\Psi=\Lambda-\sum_x\mu^x_{\rho}n_x^{\rho},
\label{eq:GeneralizedPressure}
\end{equation}
and the stress-energy tensor is
\begin{equation}
T^{\mu\nu}=\sum_xn_x^{\mu}\mu^x_{\nu}+\frac{1}{4\pi}\mathcal{K}^{\mu\rho}F^{\nu}_{\ \rho}+\sum_{\overline{x}}\lambda^{\mu\rho}_{\overline{x}} w^{\nu}_{{\overline{x}}\rho}+\Psi g^{\mu\nu}.
\label{eq:StressEnergyTensor}
\end{equation}
$T^{\mu\nu}$, of course, satisfies the Einstein field equations
\begin{equation}
R_{\mu\nu}-\frac{1}{2}R g_{\mu\nu}=8\pi T_{\mu\nu}.
\end{equation}
This stress-energy tensor does not appear to be explicitly symmetric. We will discuss in Section~\ref{sec:VLFTContribution} why it is symmetric regardless by a comparison between this ``macroscopic'' stress-energy tensor and an average ``mesoscopic'' stress-energy tensor which accounts for small-scale motion around vortex lines and flux tubes. Using the forms of the four currents and conjugate four-momenta introduced in Section~\ref{sec:RelationtoPhysParams}, $T^{\mu\nu}$ is expanded in Appendix~\ref{app:SETensorExpansion} in a more explicitly symmetric form which is compared to the stress-energy tensor for a single perfect fluid.

The identification of the generalized forces in Eq.~(\ref{eq:Lagrangian2}) gives the equations of motion for the normal fluids and vortex line/flux tube arrays associated to each superfluid
\begin{align}
f_{\mu}^x={}&2n_x^{\nu}\nabla_{[\nu}\pi^x_{\mu]}+\pi^{x}_{\mu}\nabla_{\nu}n^{\nu}_{x},
\label{eq:ForceNormalFluid}
\\
f^{\overline{x}}_{\mu}={}&w^{\overline{x}}_{\rho\mu}\left(n^{\rho}_{\overline{x}}+\nabla_{\nu}\lambda^{\rho\nu}_{\overline{x}}\right)+\pi^{\overline{x}}_{\mu}\nabla_{\nu}n^{\nu}_{\overline{x}}.
\label{eq:ForceSuperfluid}
\end{align}
Note that, because the variational procedure assumes conserved currents (or imposes it via Lagrange multipliers for the superfluids), the final terms on the right-hand side of both of these equations is zero. We will, however, allow for non-conservation of the entropy current $x=s$, which allows us to discuss entropy generation in Section~\ref{sec:Dissipation}. In Section~\ref{sec:MFandVP}, we relate the equations of motion of the vortex line/flux tube arrays to those of the associated superfluids.

Finally, summing Eqs.~(\ref{eq:ForceNormalFluid}--\ref{eq:ForceSuperfluid}) for each fluid and using Eq.~(\ref{eq:MasterFunctionGradient2},\ref{eq:Closure},\ref{eq:GeneralizedPressure},\ref{eq:StressEnergyTensor}), we can show that the stress-energy tensor is conserved up to external forces acting on the fluids:
\begin{equation}
\nabla_{\nu}T^{\nu}_{\ \mu}=\sum_x f^x_{\mu}.
\label{eq:StressEnergyConservation}
\end{equation}
The right-hand side of this equation should equal zero if energy and momentum are conserved in this system, so in that case the $f^x_{\mu}$ must sum to zero. This can be accomplished if they are all zero individually, or if they cancel each other, which corresponds to forces which act between the fluid constituents. We can also add forces to the right-hand side here as long as they act on multiple fluid constituents and hence mutually cancel. This will allow us to insert forces that we are unable to derive from a variational principle.

\section{Relation to physical parameters}
\label{sec:RelationtoPhysParams}

Our discussion so far has focused on a somewhat abstract variational principle and the resulting equations of motion and stress-energy tensor. To proceed, we need to relate the variables in the previous section to physical quantities. First we introduce the four-velocities of the fluids. Because of short collisional coupling times~\cite{Shternin2007,Shternin2008,Bertoni2015}, it is expected that all four normal fluid components $x=n,p,e,m$ will comove and have common four-velocity $u^{\mu}$, normalized in the standard manner $u^{\mu}u_{\mu}=-1$. Their currents are defined as
\begin{equation}
n^{\mu}_x=n_xu^{\mu}, \quad x\in\{n,p,e,m\},
\label{eq:NormalFluidCurrent}
\end{equation}
where $n_x$ is the number density of species $x$ defined in the normal fluid rest frame. We thus work in the Eckart frame~\citep{Eckart1940}.

The superfluids do not have to comove with the normal fluid, and we specify their four-currents by
\begin{align}
n^{\mu}_{\overline{x}}={}&n_{\overline{x}}u^{\mu}_{\overline{x}}=n_{\overline{x}}\gamma(v^2_{\overline{x}})(u^{\mu}+v_{\overline{x}}^{\mu}), \quad \overline{x}\in\{\overline{n},\overline{p}\}, 
\label{eq:SuperfluidCurrent}
\end{align}
where $u^{\mu}_{\overline{x}}$ is the four-velocity of the superfluid species $\overline{x}$ and $v_{\overline{x}}^{\mu}$ is a spacelike relative four-velocity between the normal fluid and the superfluid $\overline{x}$. $n_{\overline{x}}$ is the number density of species $\overline{x}$ in its own rest frame, equal to twice the density of Cooper pairs. The $u^{\mu}_{\overline{x}}$ are defined in this way so that they are normalized in the same way as the normal fluid four-velocity. We will use a subscript $^*$ to indicate a quantity measured in the normal fluid rest frame, so the superfluid density in this frame is
\begin{equation}
n^*_{\overline{x}}\equiv n_{\overline{x}}\gamma(v^2_{\overline{x}})=-u^{\mu}n^{\overline{x}}_{\mu}, \quad \gamma(v^2_{\overline{x}})=(1-v^2_{\overline{x}})^{-1/2},
\end{equation}
where $v^2_{\overline{x}}=v^{\mu}_{\overline{x}}v_{\mu}^{\overline{x}}$ and $v^{\overline{x}}_{\mu}u^{\mu}=0$. Strong electrostatic coupling between the normal fluid leptons and the superconducting protons means that the latter will also likely move collisionlessly with the normal fluid, but for now we permit the superconducting protons to move independently of the normal fluid.

Like the superfluids, the entropy current can move independently of the normal fluids, and is specified by
\begin{equation}
s^{\mu}=s\gamma(w^2)(u^{\mu}+w^{\mu})=s^*(u^{\mu}+w^{\mu}),
\label{eq:EntropyCurrent}
\end{equation}
where $u^{\mu}w_{\mu}=0$, $\gamma(w^2)=(1-w^2)^{-1/2}$ and $w^2=w^{\mu}w_{\mu}$.  The heat flux four-vector $q^{\mu}$ is related to $w^{\mu}$ by
\begin{equation}
q^{\mu}=s^*T^*w^{\mu},
\label{eq:HeatFluxVector}
\end{equation}
Here $s^*$ and $T^*$ are the entropy density and temperature measured in the normal fluid rest frame, while $s$ is the entropy density in the comoving frame.

Using Eqs.~(\ref{eq:NormalFluidCurrent},\ref{eq:SuperfluidCurrent},\ref{eq:EntropyCurrent},\ref{eq:HeatFluxVector}) in Eqs.~(\ref{eq:NFNeutronMuVec1}--\ref{eq:ThetaVec1}), the conjugate momentum covectors can be rewritten as
\begin{subequations}
\begin{align}
\mu_{\mu}^n {}&= \mu_nu_{\mu} + \mathcal{A}^{n\overline{n}}n_{\overline{n}}^*v^{\overline{n}}_{\mu}+ \mathcal{A}^{n\overline{p}}n_{\overline{p}}^*v^{\overline{p}}_{\mu}+\frac{\mathcal{A}^{sn}}{T^*}q_{\mu}, \label{eq:NFNeutronMuVec2}
\\ 
\mu_{\mu}^p {}&= \mu_pu_{\mu}+ \mathcal{A}^{p\overline{n}}n_{\overline{n}}^*v^{\overline{n}}_{\mu}+ \mathcal{A}^{p\overline{p}}n_{\overline{p}}^*v^{\overline{p}}_{\mu}+\frac{\mathcal{A}^{sp}}{T^*}q_{\mu},
\\
\mu_{\mu}^{\overline{n}} {}&= \mu_{\overline{n}}^*u_{\mu}+\mathcal{B}^{\overline{n}}n^*_{\overline{n}}v^{\overline{n}}_{\mu}+\mathcal{A}^{\overline{n}\overline{p}}n^*_{\overline{p}}v^{\overline{p}}_{\mu},
\label{eq:SFNeutronMuVec2}
\\
\mu_{\mu}^{\overline{p}} {}&= \mu^*_{\overline{p}}u_{\mu}+\mathcal{B}^{\overline{p}}n^*_{\overline{p}}v^{\overline{p}}_{\mu}+\mathcal{A}^{\overline{n}\overline{p}}n^*_{\overline{n}}v^{\overline{n}}_{\mu},
\label{eq:SFProtonMuVec2}
\\
\mu^e_{\mu} {}&= \mu_eu_{\mu}+\frac{\mathcal{A}^{se}}{T^*}q_{\mu},
\\
\mu^m_{\mu} {}&= \mu_mu_{\mu}+\frac{\mathcal{A}^{sm}}{T^*}q_{\mu},
\\
\Theta_{\mu} {}&= T^*u_{\mu}+\frac{\mathcal{B}^s}{T^*}q_{\mu},
\label{eq:ThetaVec2}
\end{align}
\end{subequations}
where we have defined the following chemical potentials/temperature measured in the normal fluid rest frame
\begin{subequations}
\begin{align}
\mu_n \equiv{}& \mathcal{B}^nn_n+\mathcal{A}^{np}n_p+\mathcal{A}^{n\overline{n}}n_{\overline{n}}^*+\mathcal{A}^{n\overline{p}}n_{\overline{p}}^*+\mathcal{A}^{sn}s^*,
\label{eq:ChemPotDefNFNeutron}
\\
\mu_p \equiv{}& \mathcal{B}^pn_p+\mathcal{A}^{np}n_n+\mathcal{A}^{p\overline{n}}n^*_{\overline{n}}+\mathcal{A}^{n\overline{p}}n^*_{\overline{p}} + \mathcal{A}^{sp}s^*,
\\
\mu_{\overline{n}}^* \equiv{}& \mathcal{B}^{\overline{n}}n_{\overline{n}}^*+\mathcal{A}^{n\overline{n}}n_n+\mathcal{A}^{p\overline{n}}n_p+\mathcal{A}^{\overline{n}\overline{p}}n^*_{\overline{p}},
\label{eq:SFNChemPotNFFrame}
\\
\mu_{\overline{p}}^* \equiv{}& \mathcal{B}^{\overline{p}}n_{\overline{p}}^*+\mathcal{A}^{n\overline{p}}n_n+\mathcal{A}^{p\overline{p}}n_p+\mathcal{A}^{\overline{n}\overline{p}}n^*_{\overline{n}},
\label{eq:SFPChemPotNFFrame}
\\
\mu_e \equiv{}& \mathcal{B}^en_e + \mathcal{A}^{se}s^*,
\\
\mu_m \equiv{}& \mathcal{B}^mn_m + \mathcal{A}^{sm}s^*,
\\
T^*\equiv{}& \mathcal{B}^ss^* + \mathcal{A}^{sn}n_n + \mathcal{A}^{sp}n_p + \mathcal{A}^{se}n_e + \mathcal{A}^{sm}n_m.
\label{eq:ChemPotDefT}
\end{align}
\end{subequations}
The superfluid chemical potentials in the rest frames of the respective superfluids, and the temperature in the rest frame of the entropy, are 
\begin{subequations}
\begin{align}
\mu_{\overline{n}}\equiv{}&-u_{\overline{n}}^{\mu}\mu^{\overline{n}}_{\mu}=\gamma(v^2_{\overline{n}})(\mathcal{A}^{n\overline{n}}n_n+\mathcal{A}^{p\overline{n}}n_p)+\mathcal{B}^{\overline{n}}n_{\overline{n}} \nonumber
\\
{}& \qquad\qquad\quad +\gamma(v^2_{\overline{n}})(1-v^{\mu}_{\overline{n}}v_{\mu}^{\overline{p}})\mathcal{A}^{\overline{n}\overline{p}}n^*_{\overline{p}} \nonumber
\\
={}&\gamma(v^2_{\overline{n}})[\mu^*_{\overline{n}}-v_{\overline{n}}^2\mathcal{B}^{\overline{n}}n^*_{\overline{n}}-v^{\mu}_{\overline{n}}v_{\mu}^{\overline{p}}\mathcal{A}^{\overline{n}\overline{p}}n^*_{\overline{p}}],
\label{eq:SFNChemPotSFFrame}
\\
\mu_{\overline{p}}\equiv{}&-u_{\overline{p}}^{\mu}\mu^{\overline{p}}_{\mu}=\gamma(v^2_{\overline{p}})(\mathcal{A}^{p\overline{p}}n_p+\mathcal{A}^{n\overline{p}}n_n)+\mathcal{B}^{\overline{p}}n_{\overline{p}} \nonumber 
\\
{}& \qquad\qquad\quad +\gamma(v^2_{\overline{p}})(1-v^{\mu}_{\overline{p}}v_{\mu}^{\overline{n}})\mathcal{A}^{\overline{n}\overline{p}}n^*_{\overline{n}} \nonumber
\\
={}&\gamma(v^2_{\overline{p}})[\mu_{\overline{p}}^*-v_{\overline{p}}^2\mathcal{B}^{\overline{p}}n^*_{\overline{p}}-v^{\mu}_{\overline{p}}v_{\mu}^{\overline{n}}\mathcal{A}^{\overline{n}\overline{p}}n^*_{\overline{n}}],
\label{eq:SFPChemPotSFFrame}
\\
T\equiv{}&-\gamma(w^2)(u^{\mu}+w^{\mu})\Theta_{\mu} \nonumber
\\
={}&\mathcal{B}^ss+\gamma(w^2)(\mathcal{A}^{sn}n_n+\mathcal{A}^{sp}n_p+\mathcal{A}^{se}n_e+\mathcal{A}^{sm}n_m) \nonumber
\\
={}& \gamma(w^2)[T^*-w^2\mathcal{B}^ss^*].
\end{align}
\end{subequations}

If all of the normal fluids comove, the same Lagrangian displacement field $\xi^{\mu}_n=\xi^{\mu}_p=\xi^{\mu}_e=\xi^{\mu}_m\equiv\xi^{\mu}_r$ must be used to describe their variations, and a single generalized force $f^r_{\mu}$ acts on this combined normal fluid. This force is
\begin{align}
f^r_{\mu}={}&2u^{\sigma}\nabla_{[\sigma}\Pi_{\mu]}+\Pi_{\mu}\nabla_{\sigma}u^{\sigma}+u^{\sigma}\mu_{\sigma}^n\nabla_{\mu}n_n \nonumber
\\
{}&+u^{\sigma}\pi_{\sigma}^p\nabla_{\mu}n_p+u^{\sigma}\pi_{\sigma}^e\nabla_{\mu}n_e+u^{\sigma}\pi_{\sigma}^m\nabla_{\mu}n_m,
\label{eq:NormalFluidForceWEntropy}
\end{align}
where $\Pi_{\mu}\equiv sTu_{\mu}+n_n\mu^n_{\mu}+n_p\pi^p_{\mu}+n_e\pi^e_{\mu}+n_m\pi^m_{\mu}$ is the effective momentum for the normal fluid. In the absence of dissipation, the entropy will move with the same four-velocity $u^{\mu}$ as the normal fluids since the superfluids carry no entropy. In that case, $s=s^*$, $T=T^*$, $\xi^{\mu}_s=\xi^{\mu}_r$ and there is an entropy contribution to $f^r_{\mu}$~\citep{Carter1998}.

The coefficients $\mathcal{B}^x$, $\mathcal{A}^{xy}$ need to be calculated using microphysics. Previously, relativistic entrainment coefficients have been computed using Landau Fermi liquid theory~\citep{Gusakov2009,Gusakov2009a}, though these references employ a different formulation of the hydrodynamics and their relativistic entrainment coefficients thus differ from the $\mathcal{B}^x$, $\mathcal{A}^{xy}$ used here. We invert our definitions of the conjugate four-momenta and determine how these previously calculated entrainment coefficients could be used in the more symmetric hydrodynamics of this paper. 

We assume $q^{\mu}=0$, which is implied in~\citet{Gusakov2009,Gusakov2009a}. Inverting $\mu_{\mu}^{\overline{n}}$ and $\mu_{\mu}^{\overline{p}}$ to obtain equations for the superfluid number currents and then adding to these the equations for the normal fluid current of each species gives
\begin{align}
(n^{\mu}_{\overline{n}})_{\text{total}}={}&\frac{\mathcal{B}^{\overline{p}}}{\det(\mathbb{A})}\mu^{\overline{n}}_{\mu}-\frac{\mathcal{A}^{\overline{n}\overline{p}}}{\det(\mathbb{A})}\mu^{\overline{p}}_{\mu}+n_nu^{\mu} \nonumber
\\
{}&-\frac{\mathcal{B}^{\overline{p}}(\mathcal{A}^{n\overline{n}}n_n+\mathcal{A}^{p\overline{n}}n_p)}{\det(\mathbb{A})}u^{\mu} \nonumber
\\
{}&+\frac{\mathcal{A}^{\overline{n}\overline{p}}(\mathcal{A}^{n\overline{p}}n_n+\mathcal{A}^{p\overline{p}}n_p)}{\det(\mathbb{A})}u^{\mu}
\label{eq:RWNCurrent}
\\
(n^{\mu}_{\overline{p}})_{\text{total}}={}&\frac{\mathcal{B}^{\overline{n}}}{\det(\mathbb{A})}\mu^{\overline{p}}_{\mu}-\frac{\mathcal{A}^{\overline{n}\overline{p}}}{\det(\mathbb{A})}\mu^{\overline{n}}_{\mu}+n_pu^{\mu} \nonumber
\\
{}&-\frac{\mathcal{B}^{\overline{n}}(\mathcal{A}^{n\overline{p}}n_n+\mathcal{A}^{p\overline{p}}n_p)}{\det(\mathbb{A})}u^{\mu} \nonumber
\\
{}&+\frac{\mathcal{A}^{\overline{n}\overline{p}}(\mathcal{A}^{n\overline{n}}n_n+\mathcal{A}^{p\overline{n}}n_p)}{\det(\mathbb{A})}u^{\mu}
\label{eq:RWPCurrent}
\end{align}
where we have explicitly shown dependence on the $\mathcal{B}^x$, $\mathcal{A}^{xy}$, and where
\begin{equation}
\mathbb{A}\equiv\left(\begin{array}{cc} \mathcal{B}^{\overline{n}} & \mathcal{A}^{\overline{n}\overline{p}} \\ \mathcal{A}^{\overline{n}\overline{p}} & \mathcal{B}^{\overline{p}}\end{array}\right).
\end{equation}

\citet{Gusakov2009,Gusakov2009a} use as the total (normal plus superfluid) baryon number currents
\begin{align}
(n_n^{\mu})_{\text{total}}={}&[(n_n)_{\text{total}}-\mu_nY_{np}-\mu_pY_{np}]u^{\mu} \nonumber
\\
{}&+Y_{nn}Q_{\overline{n}}^{\mu}+Y_{np}Q_{\overline{p}}^{\mu}
\label{eq:GKHNCurrent}
\\
(n_p^{\mu})_{\text{total}}={}&[(n_p)_{\text{total}}-\mu_pY_{pp}-\mu_nY_{np}]u^{\mu} \nonumber
\\
{}&+Y_{pp}Q_{\overline{p}}^{\mu}+Y_{np}Q_{\overline{n}}^{\mu}
\label{eq:GKHPCurrent}
\end{align}
where the (symmetric, relativistic) entrainment matrix is
\begin{equation}
\mathbb{Y}=\left(\begin{array}{cc} Y_{nn} & Y_{np} \\ Y_{np} & Y_{pp} \end{array}\right)
\end{equation}
and the number densities and chemical potentials are measured in the rest frame of the normal fluid. The $Q_{\overline{x}}^{\mu}$ of the references are written in terms of superfluid ``velocities'' $V^{\mu}_{x}$ (actually the conjugate four-momentum divided by the chemical potential)
\begin{equation}
Q^{\mu}_{\overline{x}}=\mu_{\overline{x}}V^{\mu}_{\overline{x}}=\mu^{\overline{x}}_{\nu}g^{\nu\mu},
\end{equation}
where recall that $\mu_{\overline{x}}$ is measured in the rest frame of the superfluid of species $\overline{x}$. Comparing Eqs.~(\ref{eq:RWNCurrent})--(\ref{eq:RWPCurrent}) and~(\ref{eq:GKHNCurrent})--(\ref{eq:GKHPCurrent}), it is obvious that 
\begin{equation}
Y_{nn}=\frac{\mathcal{B}^{\overline{p}}}{\det(\mathbb{A})}, \quad Y_{pp}=\frac{\mathcal{B}^{\overline{n}}}{\det(\mathbb{A})}, \quad Y_{np}=Y_{pn}=\frac{-\mathcal{A}^{\overline{n}\overline{p}}}{\det(\mathbb{A})},
\label{eq:GusakovEntrainmentParameters}
\end{equation}
which can be inverted to give
\begin{equation}
\mathcal{B}^{\overline{n}}=\frac{Y_{pp}}{\det(\mathbb{Y})}, \quad \mathcal{B}^{\overline{p}}=\frac{Y_{nn}}{\det(\mathbb{Y})}, \quad \mathcal{A}^{\overline{n}\overline{p}}=\frac{-Y_{np}}{\det(\mathbb{Y})}.
\label{eq:AnderssonEntrainmentParameters}
\end{equation}
The total baryon currents in our notation are thus
\begin{align}
(n_n^{\mu})_{\text{total}}={}&=Y_{nn}\mu_{\overline{n}}^{\mu}+Y_{np}\mu_{\overline{p}}^{\mu}\nonumber
\\
{}&+\Bigl[n_n-Y_{nn}(\mathcal{A}^{n\overline{n}}n_n+\mathcal{A}^{p\overline{n}}n_p) \nonumber
\\
{}&\qquad-Y_{np}(\mathcal{A}^{n\overline{p}}n_n+\mathcal{A}^{p\overline{p}}n_p)\Bigr]u^{\mu},
\label{eq:TotalNCurrent}
\\
(n_p^{\mu})_{\text{total}}={}&Y_{pp}\mu_{\overline{p}}^{\mu}+Y_{np}\mu_{\overline{n}}^{\mu} \nonumber
\\
{}&+\Bigl[n_p-Y_{pp}(\mathcal{A}^{n\overline{p}}n_n+\mathcal{A}^{p\overline{p}}n_p) \nonumber
\\
{}&\qquad-Y_{np}(\mathcal{A}^{n\overline{n}}n_n+\mathcal{A}^{p\overline{n}}n_p)\Bigr]u^{\mu}.
\label{eq:TotalPCurrent}
\end{align}
Using Eqs.~(\ref{eq:SFNChemPotNFFrame},\ref{eq:SFPChemPotNFFrame},\ref{eq:GusakovEntrainmentParameters}), we can rewrite Eqs.~(\ref{eq:TotalNCurrent}) and~(\ref{eq:TotalPCurrent}) as 
\begin{align}
(n_n^{\mu})_{\text{total}}={}&\left[n_n+n_{\overline{n}}^*-Y_{nn}\mu^*_{\overline{n}}-Y_{np}\mu^*_{\overline{p}}\right]u^{\mu} \nonumber
\\
{}&+Y_{nn}\mu_{\overline{n}}^{\mu}+Y_{np}\mu_{\overline{p}}^{\mu},
\\
(n_p^{\mu})_{\text{total}}={}&\left[n_p+n_{\overline{p}}^*-Y_{pp}\mu^*_{\overline{p}}-Y_{np}\mu^*_{\overline{n}}\right]u^{\mu} \nonumber
\\
{}&+Y_{pp}\mu_{\overline{p}}^{\mu}+Y_{np}\mu_{\overline{n}}^{\mu},
\end{align}
which are nearly identical to Eq.~(\ref{eq:GKHNCurrent}) and~(\ref{eq:GKHPCurrent}) except for including additional relativistic corrections due to the relative motion between the normal and superfluid components of each baryon species. The difference in the species labels between normal and superfluid baryons on the chemical potentials is not a concern since $\mu_x=\mu_{\overline{x}}^*$ should be true in chemical equilibrium i.e. in equilibrium, the protons and neutrons should have no preference between the paired (superfluid) and unpaired (normal fluid) phases.

While we have so far been as general as possible with regards to the coefficients $\mathcal{A}^{xy}$, simple physical arguments allow us to reduce their number. If we assume that the entrainment coefficients parameterize a coupling between the \textit{total} (normal and superfluid) neutron current and the total proton current, we will have $\mathcal{A}^{\overline{np}}=\mathcal{A}^{n\overline{p}}=\mathcal{A}^{p\overline{n}}=\mathcal{A}^{np}$. If the total current of each baryon species is coupled to itself, then we might expect $\mathcal{B}^n=\mathcal{B}^{\overline{n}}=\mathcal{A}^{n\overline{n}}$ and likewise for the protons. However, we cannot have $\mathcal{B}^x=\mathcal{B}^{\overline{x}}$ unless $\mathcal{A}^{sx}s^*=0$, since this would prevent $\mu_x=\mu_{\overline{x}}^*$ in equilibrium. The Lagrangian could also include $n^{\mu}_xn_{\mu}^x$ and $n^{\mu}_{\overline{x}}n_{\mu}^{\overline{x}}$ terms which would allow $\mathcal{B}^x\neq\mathcal{B}^{\overline{x}}$ for the baryons.

\section{Dissipation}
\label{sec:Dissipation}

\subsection{Heat conduction}
\label{sec:Conduction}

We begin our discussion of dissipation by determining the allowed form of the heat flux $q^{\mu}$ introduced in Eq.~(\ref{eq:HeatFluxVector}). Its form is found by enforcing the positive definiteness of the entropy generation $\Gamma_s=\nabla_{\mu}s^{\mu}$ using a standard procedure in relativistic dissipative hydrodynamics (see e.g.~\citep{Weinberg1972,Israel1976,Stewart1977,Olson1990,Carter1991,Gusakov2007,Lopez-Monsalvo2011}). Like~\citet{Olson1990},~\citet{Priou1991} and~\citet{Lopez-Monsalvo2011}, we are careful to note that the ``regular'' Carter formulation of relativistic finite temperature fluid dynamics, corresponding to setting the parameters $\mathcal{A}^{sx}=0$, is acausal, which is why we have included entropy entrainment.

The most general way to obtain the form of the heat flux is to start with the equation of motion for the entropy current
\begin{equation}
2s^{\sigma}\nabla_{[\sigma}\Theta_{\mu]}+\Theta_{\mu}\nabla_{\sigma}s^{\sigma}=f_{\mu}^s.
\end{equation}
Contraction with $u^{\mu}$ and rearranging gives
\begin{align}
T^*\nabla_{\sigma}s^{\sigma}=-\frac{q^{\sigma}}{T^*}\Biggl[{}&\nabla_{\sigma}T^*+T^*\dot{u}_{\sigma}+\frac{2\mathcal{B}^s}{T^*}u^{\mu}\nabla_{[\mu}q_{\sigma]} \nonumber 
\\
{}&+\left(\dot{\mathcal{B}^s}-\frac{\mathcal{B}^s\dot{T^*}}{T^*}\right)\frac{q_{\sigma}}{T^*}\Biggr]-u^{\mu}f^s_{\mu},
\label{eq:EntropyGenerationSOnly}
\end{align}
where $\dot{a}=u^{\mu}\nabla_{\mu}a$. The easiest way to enforce that the entropy generation from heat conduction is positive definite is to make
\begin{align}
q^{\mu}=-\kappa\perp^{\mu\nu}\Biggl[{}&\nabla_{\nu}T^*+T^*\dot{u}_{\nu}+\frac{2\mathcal{B}^s}{T^*}u^{\sigma}\nabla_{[\sigma}q_{\nu]} \nonumber
\\
{}&+\left(\dot{\mathcal{B}^s}-\frac{\mathcal{B}^s\dot{T^*}}{T^*}\right)\frac{q_{\nu}}{T^*}\Biggr],
\label{eq:HeatFlowVector}
\end{align}
where $\perp^{\mu\nu}=g^{\mu\nu}+u^{\mu}u^{\nu}$. This matches~\citet{Lopez-Monsalvo2011} and gives the same entropy generation term due to heat conduction as~\citet{Weinberg1972} up to the additional terms which are higher-order in $q^{\mu}$. These terms are necessary for causal heat conduction, since rearranging Eq.~(\ref{eq:HeatFlowVector}) following~\citep{Lopez-Monsalvo2011} gives a relativistic version of the Cattaneo--Vernotte equation
\begin{equation}
t_h(\dot{q}^{\mu}+q^{\nu}\nabla_{\nu}u^{\mu})+q^{\mu}=-\tilde{\kappa}\perp^{\mu\nu}\left(\nabla_{\nu}T^*+T^*u_{\nu}\right),
\end{equation}
where $t_h$ is a heat conduction timescale and $\tilde{\kappa}$ is a modified heat conductivity, which are given by
\begin{align}
t_h={}&\frac{\mathcal{B}^s/T^*}{1+\kappa\dot{\left(\frac{\mathcal{B}^s}{T^*}\right)}}\approx\frac{\mathcal{B}^s}{T^*},
\\
\tilde{\kappa}={}&\frac{\kappa}{1+\kappa\dot{\left(\frac{\mathcal{B}^s}{T^*}\right)}}\approx\kappa,
\end{align}
where the approximate forms are valid if we drop higher-order terms in an expansion in the mean free collision time. The entropy entrainment parameters which appear in the definition of $T^*$ thus clearly affect $t_h$. Causal heat conductivity is absent from the treatment of dissipation in previous papers on relativistic multifluid neutron stars~\citep{Gusakov2007,Gusakov2016a}, which use the treatment of dissipation in~\citet{Weinberg1972}.

The remaining term on the right-hand side of Eq.~(\ref{eq:EntropyGenerationSOnly}) is due to the generalized force on the entropy current $f^s_{\mu}$. Using conservation of energy-momentum, we can rewrite $f^s_{\mu}$ in terms of the generalized forces on the other fluids. The viscous contributions to entropy generation will be included in this manner by modifying the stress-energy tensor and hence the generalized forces. We next discuss the inclusion of mutual friction and vortex pinning forces which act between the fluids and vortex line/flux tube arrays, and then incorporate viscosity.

\subsection{Mutual friction and vortex pinning}
\label{sec:MFandVP}

Mutual friction is a dissipative drag force acting on vortex lines/flux tubes, and hence on their associated superfluids, due to scattering off of the normal fluid. Vortex pinning is an attractive force between neutron vortex lines and proton flux tubes that, in different limits based on the relative velocity between the two arrays, either make them move together or acts as an additional drag force. Both of these interactions are incorporated into the hydrodynamics by adding additional forces between the fluid constituents to the generalized forces $f^x_{\mu}$, $f^{\overline{x}}_{\mu}$ appearing on the right-hand side of Eq.~(\ref{eq:ForceNormalFluid}--\ref{eq:ForceSuperfluid}). We follow a relativistic version of the Hall--Vinen procedure~\cite{Hall1956a} to write the new equations of motion with these forces included. Our procedure is similar to~\citet{Andersson2016a}, but differs in the definitions of the fluid velocities so as to be consistent with Section~\ref{sec:RelationtoPhysParams}, and also in our inclusion of the vortex line self-tension, and later vortex pinning.

Consider a properly-normalized four-velocity for the vortex lines/flux tubes within superfluid $\overline{x}$, using subscript $L$ to denote vortex lines/flux tubes
\begin{equation}
u^{\mu}_{L,\overline{x}}=\gamma(\beta^2_{\overline{x}})(u^{\mu}_{\overline{x}}+\beta^{\mu}_{\overline{x}}), \quad \gamma(\beta^2_{\overline{x}})=(1-\beta^2_{\overline{x}})^{-1/2}, \qquad 
\label{eq:LineVelocity}
\end{equation}
where $\beta^2_{\overline{x}}=\beta^{\mu}_{\overline{x}}\beta_{\mu}^{\overline{x}}$, $\beta^{\mu}_{\overline{x}}u^{\overline{x}}_{\mu}=0$. $\beta^{\mu}_{\overline{x}}$ is the relative (spacelike) four-velocity of the vortex lines of species $\overline{x}$ with respect to the corresponding superfluid. Since the vorticity moves with the vortex lines, 
\begin{equation}
u^{\mu}_{L,\overline{x}}w_{\mu\nu}^{\overline{x}}=0,
\end{equation}
and~\citep{Langlois1998}
\begin{equation}
n^{\mu}_{\overline{x}}w_{\mu\nu}^{\overline{x}}=-n_{\overline{x}}\beta^{\mu}_{\overline{x}}w_{\mu\nu}^{\overline{x}}=f_{\nu}^{M,\overline{x}},
\label{eq:MagnusForce}
\end{equation}
where $f_{\nu}^{M,\overline{x}}$ is the Magnus force acting on the superfluid $\overline{x}$ due to the relative motion between it and the vortex lines/flux tubes within it. Note that for the superconducting proton fluid, the Lorentz force is included in this definition of the Magnus force. In the absence of additional forces and assuming current conservation, Eq.~(\ref{eq:ForceSuperfluid}) and~(\ref{eq:MagnusForce}) say that the Magnus force on the superfluid is balanced by a tension force due to the vortex lines/flux tubes, which is represented by the second term on the right-hand side of the equation. Due to this interpretation, we can use Eq.~(\ref{eq:ForceSuperfluid}) as a force balance equation for the superfluid of species $\overline{x}$ instead of its associated vortex line/flux tube array. The force balance equation for the array only differs from that for the superfluid by an irrelevant overall minus sign.

The vortex lines/flux tubes would move along with their associated superfluid if not for their scattering off of the normal fluid (mutual friction) or due to pinning to the vortex lines/flux tubes associated with the other superfluid (vortex pinning). We consider the mutual friction first, and represent it in Eq.~(\ref{eq:StressEnergyConservation}) and Eq.~(\ref{eq:ForceNormalFluid}--\ref{eq:ForceSuperfluid}) through equal but opposite contributions to $f^{\overline{x}}_{\mu}$ and $f^r_{\mu}$, the generalized force on the combined normal fluid. To lowest order, this force should depend only on the relative velocity between the normal fluid and the vortex lines/flux tubes of species $q^{\mu}_{\overline{x}}$, which we define analogously to~\citet{Andersson2016a}
\begin{equation}
u^{\mu}=\gamma(q^2_{\overline{x}})(u^{\mu}_{L,\overline{x}}+q_{\overline{x}}^{\mu}), \qquad \gamma(q^2_{\overline{x}})=(1-q_{\overline{x}}^2)^{-1/2},
\label{eq:NormalFluidVelocityULQ}
\end{equation}
where $u^{\mu}_{L,\overline{x}}q_{\mu}^{\overline{x}}=0$, $q^2_{\overline{x}}=q^{\mu}_{\overline{x}}q_{\mu}^{\overline{x}}$. So we modify the generalized force on superfluid $\overline{x}$ and the combined normal fluid by setting
\begin{equation}
f^{\overline{x}}_{\mu}=f^{\text{mf},\overline{x}}_{\mu}\equiv\mathcal{R}^{\text{mf},\overline{x}}_{\mu\nu}q^{\nu}_{\overline{x}}, \quad f^r_{\mu}= f^{\text{mf},r}_{\mu}\equiv-\sum_{\overline{x}}\mathcal{R}^{\text{mf},\overline{x}}_{\mu\nu}q^{\nu}_{\overline{x}},
\end{equation}
where the $\mathcal{R}^{\text{mf},\overline{x}}_{\mu\nu}$ projects out components of $q^{\nu}_{\overline{x}}$ either along the direction tangent to the corresponding vortex line/flux tube array or along the respective vortex line/flux tube array velocity 
\begin{equation}
\mathcal{R}^{\text{mf},\overline{x}}_{\mu\nu}\equiv\mathcal{R}^{\text{mf},\overline{x}}\left(g_{\mu\nu}+u^{L,\overline{x}}_{\mu}u^{L,\overline{x}}_{\nu}-\hat{t}^{\overline{x}}_{\mu}\hat{t}^{\overline{x}}_{\nu}\right),
\end{equation}
where $\hat{t}^{\overline{x}}_{\mu}$ is the average spacelike tangent vector to the vortex lines/flux tube array. $\mathcal{R}^{\text{mf},\overline{x}}$ are dissipative coefficients parameterizing the mutual friction. Since these additional forces cancel out in the right-hand side of Eq.~(\ref{eq:StressEnergyConservation}), the total stress-energy tensor is still conserved. In this case the equation of motion for a superfluid becomes
\begin{equation}
\mathcal{R}^{\text{mf},\overline{x}}_{\mu\nu}q^{\nu}_{\overline{x}}=n_{\overline{x}}^{\rho}w^{\overline{x}}_{\rho\mu}+w^{\overline{x}}_{\rho\mu}\nabla_{\nu}\lambda_{\overline{x}}^{\rho\nu}.
\label{eq:ForceSuperFluid2}
\end{equation}
We would like to remove references to the vortex line velocity and $q^{\mu}_{\overline{x}}$ from Eq.~(\ref{eq:ForceSuperFluid2}) and rewrite it in the form of Eq.~(\ref{eq:MagnusForce}). Equating the two forms of $u^{\mu}_{L,\overline{x}}$ using Eq.~(\ref{eq:LineVelocity},\ref{eq:NormalFluidVelocityULQ}) gives 
\begin{equation}
q^{\mu}_{\overline{x}}=-\frac{1}{\hat{\gamma}_{\overline{x}}}v^{\mu}_{\overline{x}}-\tilde{\gamma}_{\overline{x}}\beta^{\mu}_{\overline{x}}+\left(\frac{1}{\gamma_{\overline{x}}\hat{\gamma}_{\overline{x}}}-\tilde{\gamma}_{\overline{x}}\right)u^{\mu}_{\overline{x}},
\label{eq:QVector}
\end{equation}
where $\tilde{\gamma}_{\overline{x}}\equiv\gamma(\beta^2_{\overline{x}})$, $\hat{\gamma}_{\overline{x}}\equiv\gamma(q^2_{\overline{x}})$, $\gamma_{\overline{x}}\equiv\gamma(v^2_{\overline{x}})$. To perform the necessary manipulations, it will be convenient to rewrite the vorticity tensor in terms of the corresponding ``electric'' and ``magnetic'' four-fields in the frame comoving with the vortex lines,
\begin{equation}
W_{\mu}^{E,\overline{x}}=u^{\nu}_{L,\overline{x}}w_{\mu\nu}^{\overline{x}}=0, \quad  W^{\mu}_{B,\overline{x}}=\frac{1}{2}\varepsilon^{\mu\nu\sigma\rho}u_{\nu}^{L,\overline{x}}w_{\sigma\rho}^{\overline{x}}\equiv W^{\mu}_{\overline{x}},
\label{eq:WVector}
\end{equation}
in terms of which we can write $\hat{t}^{\overline{x}}_{\mu}$ as
\begin{equation}
\hat{t}^{\overline{x}}_{\mu}=W^{\overline{x}}_{\mu}/\sqrt{W^{\overline{x}}_{\sigma}W_{\overline{x}}^{\sigma}}.
\label{eq:ArrayTangentVector}
\end{equation}
We can of course invert $W^{\mu}_{\overline{x}}$ to find
\begin{equation}
w_{\mu\nu}^{\overline{x}}=-\varepsilon_{\mu\nu\sigma\rho}u^{\sigma}_{L,\overline{x}}W^{\rho}_{\overline{x}}.
\label{eq:WVectorInversion}
\end{equation}
Using Eqs.~(\ref{eq:ForceSuperFluid2}--\ref{eq:QVector},\ref{eq:WVectorInversion}), we solve for $v_{\mu}^{\overline{x}}$,
\begin{align}
v_{\mu}^{\overline{x}}={}&\frac{1}{\eta_{\overline{x}}}\varepsilon_{\mu\nu\rho}\beta^{\nu}_{\overline{x}}W^{\rho}_{\overline{x}}-\frac{\hat{\gamma}_{\overline{x}}}{\mathcal{R}^{\text{mf},\overline{x}}}w_{\rho\mu}^{\overline{x}}\nabla_{\nu}\lambda_{\overline{x}}^{\rho\nu} \nonumber
\\
{}&+\left(\frac{1}{\gamma_{\overline{x}}}-\tau_{\overline{x}}\right)u_{\mu}^{\overline{x}}-\tau_{\overline{x}}\beta_{\mu}^{\overline{x}}-(\hat{t}^{\overline{x}}_{\nu}u^{\nu})\hat{t}^{\overline{x}}_{\mu},
\label{eq:VVectorRearrangement}
\end{align}
where for simplicity we have defined, 
\begin{equation}
\varepsilon_{\mu\nu\rho}\equiv \varepsilon_{\sigma\mu\nu\rho}u^{\sigma}_{\overline{x}}, \quad \eta_{\overline{x}}\equiv \frac{\mathcal{R}_{\text{mf},\overline{x}}}{\tilde{\gamma}_{\overline{x}}\hat{\gamma}_{\overline{x}}n_{\overline{x}}},\qquad \tau_{\overline{x}}\equiv\tilde{\gamma}_{\overline{x}}\hat{\gamma}_{\overline{x}}.
\end{equation}
Contracting Eq.~(\ref{eq:VVectorRearrangement}) with $\varepsilon^{\lambda\eta\mu}W^{\overline{x}}_{\eta}$ and then using the same equation to replace $\varepsilon_{\mu\nu\rho}\beta^{\nu}_{\overline{x}}W^{\rho}_{\overline{x}}$ gives
\begin{align}
\varepsilon^{\lambda\eta\mu}W^{\overline{x}}_{\eta}v^{\overline{x}}_{\mu}={}&-\eta_{\overline{x}}\tau_{\overline{x}}v^{\lambda}_{\overline{x}}+\left(\frac{\eta_{\overline{x}}\tau_{\overline{x}}}{W^2_{\overline{x}}}(W^{\nu}_{\overline{x}}u_{\nu})-\frac{W_{\overline{x},\parallel}}{\eta_{\overline{x}}}\right)W^{\lambda}_{\overline{x}}\nonumber
\\
{}&+\left(\tau_{\overline{x}}-\frac{1}{\gamma_{\overline{x}}}+\frac{W_{\overline{x},\parallel}^2}{\eta_{\overline{x}}}\right)u^{\lambda}_{\overline{x}} \nonumber
\\
{}&+\frac{1}{\eta_{\overline{x}}}\left(W^2_{\overline{x},\perp}+\eta^2_{\overline{x}}\tau^2_{\overline{x}}\right)\beta^{\lambda}_{\overline{x}} \nonumber
\\
{}&+\frac{\hat{\gamma}_{\overline{x}}}{\mathcal{R}^{\text{mf},\overline{x}}}\left(\eta_{\overline{x}}\tau_{\overline{x}} w^{\lambda}_{\overline{x}\rho}+\varepsilon^{\lambda\eta\mu}W_{\eta}^{\overline{x}}w^{\overline{x}}_{\mu\rho}\right)\nabla_{\nu}\lambda_{\overline{x}}^{\rho\nu},
\end{align}
where $W^2_{\overline{x}}=w^{\overline{x}}_{\mu\nu}w^{\nu\mu}_{\overline{x}}/2$, and where we have used a split of $W^{\mu}_{\overline{x}}$ into components parallel and perpendicular to $u^{\mu}_{\overline{x}}$
\begin{equation}
W^{\mu}_{\overline{x}}=W_{\overline{x},\parallel}u^{\mu}_{\overline{x}}+W^{\mu}_{\overline{x},\perp}, \quad W^{\mu}_{\overline{x},\perp}u_{\mu}^{\overline{x}}=0.
\end{equation}
Contracting with $\varepsilon_{\lambda\sigma\alpha}W_{\overline{x}}^{\sigma}$ again yields an equation from which the Magnus force can be isolated:
\begin{widetext}
\begin{align}
f_{M,\overline{x}}^{\mu}=(\tilde{g}_{\mu\nu})^{-1}\frac{\tilde{\gamma}_{\overline{x}} n_{\overline{x}}\eta_{\overline{x}}}{W_{\overline{x},\perp}^2+\eta_{\overline{x}}^2\tau_{\overline{x}}^2}\Bigg[{}&(v^{\overline{x}}_{\sigma}W^{\sigma}_{\overline{x}})W_{\nu}^{\overline{x},\perp}+\eta_{\overline{x}}\tau_{\overline{x}}\varepsilon_{\nu\sigma\rho}W^{\sigma}_{\overline{x},\perp}v^{\rho}_{\overline{x}}
-W_{\overline{x},\perp}^2v^{\overline{x}}_{\nu}-\gamma_{\overline{x}}v_{\overline{x}}^2(W_{\overline{x},\perp}^2u^{\overline{x}}_{\nu}+W_{\overline{x},\perp}W_{\nu}^{\overline{x},\perp}) \nonumber 
\\
{}& -\frac{\hat{\gamma}_{\overline{x}}}{\mathcal{R}^{\text{mf},\overline{x}}}\left(\eta_{\overline{x}}\tau_{\overline{x}}\varepsilon_{\nu\sigma\rho}W^{\sigma}w^{\rho}_{\overline{x}\alpha}+W^2_{\overline{x},\perp}w^{\overline{x}}_{\nu\alpha}\right)\nabla_{\eta}\lambda_{\overline{x}}^{\alpha\eta}\Bigg].
\label{eq:MagnusForceSelfTension}
\end{align}
\end{widetext}
where we have replaced $\tilde{\gamma}_{\overline{x}}n_{\overline{x}}\varepsilon_{\mu\nu\rho}W^{\nu}_{\overline{x}}\beta^{\rho}_{\overline{x}}$ with $f^{\text{M},\overline{x}}_{\mu}$ and where we have defined
\begin{equation}
\tilde{g}_{\mu\nu}\equiv g_{\mu\nu}-\frac{\eta_{\overline{x}}\tau_{\overline{x}}}{W^2_{\overline{x},\perp}+\eta^2_{\overline{x}}\tau^2_{\overline{x}}}\frac{W^2_{\overline{x},\perp}u^{\overline{x}}_{\mu}+W_{\overline{x},\parallel}W^{\overline{x},\perp}_{\mu}}{\mathcal{R}^{\text{mf},\overline{x}}}\nabla_{\rho}\lambda^{\overline{x}\rho}_{\nu}.
\end{equation}
Note that Eq.~(\ref{eq:MagnusForceSelfTension}) still depends on gamma factors that are functions of $\beta_{\overline{x}}^2$ and $q_{\overline{x}}^2$. If the relative velocities are assumed to be small, these gamma factors can simply be approximated to be unity. In general, it is theoretically possible to solve for these gamma factors in terms of only the normal fluid velocity $u^{\mu}$ and the superfluid relative velocities $v^{\mu}_{\overline{x}}$, but we do not attempt such a calculation here.

Vortex pinning can be incorporated by adding a force which acts between the vortex line and flux tube arrays in the neutron superfluid and proton superconductor. This force should behave like a drag force for intermediate relative velocities between the two arrays and should force the two arrays to move together for small relative velocities. The pinning force $f^{\text{pin}}_{\mu}$ acting on the neutron vortex lines due to the proton vortex lines is incorporated into the force balance equations on the two arrays as
\begin{equation}
f^{\overline{n}}_{\mu}= f^{\text{mf},\overline{n}}_{\mu}+f^{\text{pin}}_{\mu}, \qquad f^{\overline{p}}_{\mu}=f^{\text{mf},\overline{p}}_{\mu}-f^{\text{pin}}_{\mu}.
\label{eq:MFplusVPForces}
\end{equation}
Since $f^{\overline{x}}_{\mu}$ are force densities, the force per unit length on a vortex line/flux tube equals $f^{\overline{x}}_{\mu}/\mathcal{N}_{\overline{x}}$, 
where $\mathcal{N}_{\overline{x}}$ is the areal number density of vortex lines/flux tubes of species $\overline{x}$ measured perpendicular to them (we give a relativistic definition of $\mathcal{N}_{\overline{x}}$ in Section~\ref{sec:EffectiveTheoryFT}). It is reasonable to expect that the vortex pinning force should be proportional to the product of $\mathcal{N}_{\overline{p}}$ and $\mathcal{N}_{\overline{n}}$, so the vortex pinning force per unit length acting on a proton flux tube will be proportional to $\mathcal{N}_{\overline{n}}\sim 2\Omega/\kappa_{\overline{n}}\sim 10^4(\Omega/10 \text{ s}^{-1})$ cm$^{-2}$ where $\Omega$ is the angular rotational frequency of the neutron star and $\kappa_{\overline{n}}$ is the circulation quantum. This is much smaller than the number density of proton flux tubes $\mathcal{N}_{\overline{p}}\sim B/\Phi_{\overline{p}}\sim 5\times10^{18}(B/10^{12}\text{ G})$ cm$^{-2}$ where $B$ is the magnetic field strength and $\Phi_{\overline{p}}$ is the flux quantum. $\kappa_{\overline{n}}$ and $\Phi_{\overline{p}}$ are also defined in Section~\ref{sec:EffectiveTheoryFT}. For this reason, the vortex pinning force acting on a single proton flux tube is negligible and often ignored. However, as we are interested in force densities, we will retain the pinning force acting on the proton flux tubes.

To lowest order, the vortex pinning force depends only on the (average) relative velocity between the two vortex line arrays contracted into an as yet undetermined rank two tensor:
\begin{equation}
f^{\text{pin}}_{\mu}\equiv\mathcal{R}^{\text{pin}}_{\mu\nu}b^{\nu},
\end{equation}
where $b^{\nu}$ is the (spacelike, average) relative velocity of the proton vortex lines in the (average) neutron vortex line rest frame defined such that 
\begin{equation}
u^{\mu}_{L,\overline{p}}=\gamma(b^2)(u^{\mu}_{L,\overline{n}}+b^{\mu}), \quad \gamma(b^2)=(1-b^2)^{-1/2},
\end{equation}
where $b^{\mu}u^{L,\overline{n}}_{\mu}=0$ and $b^2=b^{\mu}b_{\mu}$. A reasonable nonrelativistic version of vortex pinning drag force would point in the direction defined by the cross product of the tangent vectors to both arrays, and only the component of the relative velocity between the two arrays that is in this direction will contribute to a drag force. One possible relativistic generalization of this is
\begin{equation}
\mathcal{R}^{\text{pin}}_{\mu\nu}=-\mathcal{R}^{\text{pin}}\varepsilon_{\alpha\sigma\rho(\mu}\varepsilon_{\nu)\beta\eta\lambda}u^{\alpha}_{L,\overline{n}}\hat{t}^{\sigma}_{\overline{n}}\hat{t}^{\rho}_{\overline{p}}u^{\beta}_{L,\overline{p}}\hat{t}^{\eta}_{\overline{n}}\hat{t}^{\lambda}_{\overline{p}}.
\end{equation}
The coefficient $\mathcal{R}_{\text{pin}}$ should be a function of $b=\sqrt{b^{\mu}b_{\mu}}$, the relative orientation between the vortex line/flux tube arrays or $\hat{t}^{\mu}_{\overline{n}}\hat{t}^{\overline{p}}_{\mu}$, and should scale linearly with both $\mathcal{N}_{\overline{n}}$ and $\mathcal{N}_{\overline{p}}$ as discussed previously. The dependence on $b$ should be $b^{-1/2}$~\citep{Epstein1992,Jones1992,Gugercinoglu2020} when the linear $b$-dependence of the pinning energy is considered, as this will give the correct behaviour for the pinning force: at large $b^{\mu}$, the vortex pinning drag becomes insignificant compared to the mutual friction drag, while for small $b^{\mu}$, the vortex lines become pinned to the flux tubes~\citep{Link2009}. The principal dissipation mechanism in the drag regime of vortex pinning is the excitation of kelvons, and in calculations like those in~\citep{Epstein1992,Jones1992}, the interactions exciting the kelvons were with individual nuclei. However, in the core the pinning interaction is of course between lines of macroscopic extent, so a modification of $\mathcal{R}_{\text{pin}}$ may be required when the finite length of the lines is considered~\citep{Seveso2016,Graber2018}.

It should be noted that the pinning drag force would be relevant only to a precessing neutron star with sufficiently large precession amplitude. Even in that case, the drag force estimated by~\citet{Link2003} is large enough for pinning to happen on rather short timescales of days to weeks. Simple relative motion with energy stored in the Baym--Chandler kinetic energy~\citep{Baym1983} would damp away almost instantly.

The Magnus force acting on superfluid $\overline{x}$ can thus be written as
\begin{align}
f^{\text{M},\overline{x}}_{\mu}=n^{\rho}_{\overline{x}}w_{\rho\mu}^{\overline{x}}=w_{\rho\mu}^{\overline{x}}\nabla_{\nu}\lambda^{\nu\rho}_{\overline{x}}+\mathcal{R}^{\text{mf},\overline{x}}_{\mu\nu}q^{\nu}_{\overline{x}}\pm\mathcal{R}^{\text{pin}}_{\mu\nu}b^{\nu}.
\end{align}
with $\pm$ corresponding to $\overline{x}=\overline{n}$ and $\overline{p}$ respectively. It should be possible in principle to rewrite this equation in terms of only the vorticity tensor or vector, the normal fluid velocity and the superfluid relative velocities $v^{\mu}_{\overline{x}}$ in a manner similar to what was done in Eq.~(\ref{eq:VVectorRearrangement}--\ref{eq:MagnusForceSelfTension}). We do not attempt this calculation here because of the unessential complication it would add to this paper.

\subsection{Bulk and shear viscosity}
\label{sec:Viscosity}

To incorporate viscosity into this variational formalism, we follow~\citet{Carter1991}, the review of his work in~\citet{Andersson2007} and the nonrelativistic generalization by~\citet{Andersson2006}, though we specify to the fluids expected in a superfluid--superconducting neutron star core. We also neglect chemical reactions that convert between fluid species as we have implicitly assumed current conservation for the separate species. 

Introducing the (assumed symmetric) viscosity tensor $\tau_{\Sigma}^{\mu\nu}$, where the label $\Sigma$ is used to specify the different fluid constituents contributing to the viscosity. The variation of the master function to include viscosity takes the form (summing over $\Sigma$)
\begin{equation}
\delta\Lambda_{\text{vis}}=\frac{1}{2}\kappa_{\mu\nu}^{\Sigma}\delta\tau_{\Sigma}^{\mu\nu},\qquad \kappa_{\mu\nu}^{\Sigma}\equiv 2\frac{\partial\Lambda}{\partial\tau^{\mu\nu}_{\Sigma}},
\end{equation}
where $\kappa_{\mu\nu}^{\Sigma}$ is a strain tensor. The new form of Eqs.~(\ref{eq:MasterFunctionMetricDerivative1}) and~(\ref{eq:MasterFunctionGradient2}), giving the new form of $\partial\Lambda/\partial g_{\mu\nu}$, is
\begin{align}
\nabla_{\mu}\Lambda={}&\sum_x\mu^x_{\nu}\nabla_{\mu}n_x^{\nu}-\frac{1}{8\pi}\mathcal{K}^{\rho\nu}\nabla_{\mu}F_{\rho\nu}-\frac{1}{2}\lambda^{\rho\nu}_{\overline{x}}\nabla_{\mu}w^{\overline{x}}_{\rho\nu}\nonumber
\\
{}&+\frac{1}{2}\kappa^{\Sigma}_{\nu\rho}\nabla_{\mu}\tau^{\nu\rho}_{\Sigma},
\label{eq:MasterFunctionGradient3}
\\
\frac{\partial\Lambda}{\partial g_{\mu\nu}}={}&\frac{1}{2}\Bigg(\sum_x\mu_x^{\mu}n_x^{\nu}+\frac{1}{4\pi}\mathcal{K}^{\mu\rho}F^{\nu}_{\ \rho}+\lambda_{\overline{x}}^{\mu\rho}w^{\nu}_{\overline{x}\rho} \nonumber
\\
{}&\qquad+\kappa^{\mu}_{\Sigma\rho}\tau_{\Sigma}^{\rho\nu}\Bigg),
\label{eq:MasterFunctionMetricDerivative3}
\end{align}
where we used 
\begin{equation}
\delta\tau_{\Sigma}^{\mu\nu}=\Lie_{\xi}\tau_{\Sigma}^{\mu\nu}=\xi^{\rho}\nabla_{\rho}\tau_{\Sigma}^{\mu\nu}-2\tau_{\Sigma}^{\rho(\mu}\nabla_{\rho}\xi^{\nu)}.
\end{equation}
The full variation of $\tau_{\Sigma}^{\mu\nu}$ is, from~\citet{Carter1991}
\begin{equation}
\delta\tau_{\Sigma}^{\mu\nu}=\xi^{\sigma}_{\Sigma}\nabla_{\sigma}\tau_{\Sigma}^{\mu\nu}-2\tau_{\Sigma}^{\sigma(\mu}\nabla_{\sigma}\xi^{\nu)}_{\Sigma}+\tau_{\Sigma}^{\mu\nu}\nabla_{\sigma}\xi^{\sigma}_{\Sigma}-\frac{1}{2}\tau_{\Sigma}^{\mu\nu}g^{\sigma\rho}\delta g_{\sigma\rho},
\end{equation}
so $\delta\Lagr$ becomes
\begin{align}
\delta\Lagr ={}&\xi^{\mu}f^r_{\mu}+\xi^{\mu}_sf^s_{\mu}+\sum_{\overline{x}}\xi^{\mu}_{\overline{x}}f^{\overline{x}}_{\mu}+\sum_{\Sigma}\xi^{\mu}_{\Sigma}f^{\Sigma}_{\mu} \nonumber
\\
{}&+\frac{1}{2}\left(T^{\mu\nu}-\frac{1}{8\pi}\left(R^{\mu\nu}-\frac{1}{2}Rg^{\mu\nu}\right)\right)\delta g_{\mu\nu},
\label{eq:Lagrangian5}
\end{align}
where $T^{\mu\nu}$, $\Psi$ and $f^{\Sigma}_{\mu}$ are given by
\begin{align}
T^{\mu\nu}={}&\sum_x n_x^{\mu}\mu^x_{\rho}g^{\rho\nu}+\frac{1}{4\pi}\mathcal{K}^{\mu\rho}F^{\nu}_{\ \rho}+\sum_{\overline{x}}\lambda^{\mu\rho}_{\overline{x}} w^{\nu}_{{\overline{x}}\rho} \nonumber
\\
{}&+\sum_{\Sigma}\kappa^{\mu}_{\Sigma\rho}\tau^{\rho\nu}_{\Sigma}+\Psi g^{\mu\nu},
\label{eq:StressEnergyTensorwithViscosity}
\\
\Psi ={}&\Lambda-\sum_x\mu^x_{\rho}n_x^{\rho}-\frac{1}{2}\sum_{\Sigma}\tau^{\rho\sigma}_{\Sigma}\kappa^{\Sigma}_{\rho\sigma},
\label{eq:GeneralizedPressurewithViscosity}
\\
f^{\Sigma}_{\mu}={}&\kappa^{\Sigma}_{\mu\nu}\nabla_{\rho}\tau^{\rho\nu}_{\Sigma}+\tau^{\nu\rho}_{\Sigma}\left(\nabla_{\nu}\kappa^{\Sigma}_{\mu\rho}-\frac{1}{2}\nabla_{\mu}\kappa^{\Sigma}_{\nu\rho}\right),
\end{align}
$\xi^{\mu}$ is the common displacement field for the normal fluid and $f^r_{\mu}=f^{n}_{\mu}+f^{p}_{\mu}+f^{e}_{\mu}+f^{m}_{\mu}$. 

We now look at the $u^{\mu}f^s_{\mu}$ term in  Eq.~(\ref{eq:EntropyGenerationSOnly}). Conservation of energy-momentum implies
\begin{equation}
f^s_{\mu}=-f^r_{\mu}-\sum_{\overline{x}}f^{\overline{x}}_{\mu}-\sum_{\Sigma}f^{\Sigma}_{\mu},
\label{eq:ForceSum}
\end{equation}
so contracting with $u^{\mu}$ and then using
\begin{align}
u^{\mu}f^{r}_{\mu} ={}&\sum_x\gamma(q^2_{\overline{x}})\mathcal{R}^{\text{mf},\overline{x}}_{\mu\nu}q^{\mu}_{\overline{x}}q^{\nu}_{\overline{x}},
\\
u_{\overline{x}}^{\mu}f^{\overline{x}}_{\mu}={}&u^{\mu}_{\overline{x}}w^{\overline{x}}_{\sigma\mu}\nabla_{\nu}\lambda^{\sigma\nu}_{\overline{x}} \nonumber
\\
\rightarrow u^{\mu}f^{\overline{x}}_{\mu}={}&-v^{\mu}_{\overline{x}}f^{\overline{x}}_{\mu}-\frac{1}{n^*_{\overline{x}}}f^{\text{M},\overline{x}}_{\sigma}\nabla_{\nu}\lambda_{\overline{x}}^{\sigma\nu},
\end{align}
Eq.~(\ref{eq:EntropyGenerationSOnly}) becomes
\begin{align}
T^*\nabla_{\sigma}s^{\sigma}={}&\frac{1}{\kappa T^*}\perp_{\mu\nu}q^{\mu}q^{\nu}+\sum_{\overline{x}}\bigg[ \gamma(q^2_{\overline{x}})\mathcal{R}^{\text{mf},\overline{x}}_{\mu\nu}q^{\mu}_{\overline{x}}q^{\nu}_{\overline{x}}
\nonumber
\\
{}&+\frac{1}{n^*_{\overline{x}}}f^{\text{M},\overline{x}}_{\mu}\nabla_{\nu}\lambda^{\nu\mu}_{\overline{x}}-v^{\mu}_{\overline{x}}f_{\mu}^{\overline{x}}\bigg]+\sum_{\Sigma}u^{\mu}f_{\mu}^{\Sigma}.
\label{eq:EntropyGenerationSOnly2}
\end{align}
The second law of thermodynamics requires that $\nabla_{\sigma}s^{\sigma}\geq0$, which is most easily satisfied if each term on the right hand side of Eq.~(\ref{eq:EntropyGenerationSOnly2}) is individually greater than or equal to zero. 

Define four-vectors
\begin{equation}
u^{\mu}_{\Sigma}=\gamma(v^2_{\Sigma})(u^{\mu}+v^{\mu}_{\Sigma}), \quad \gamma(v^2_{\Sigma})=(1-v^2_{\Sigma})^{-1/2},
\end{equation}
where $u^{\mu}v_{\mu}^{\Sigma}=0$ and $v^2_{\Sigma}=v^{\mu}_{\Sigma}v_{\mu}^{\Sigma}$, such that
\begin{equation}
\tau_{\Sigma}^{\mu\nu}u_{\nu}^{\Sigma}=0,\qquad \kappa^{\Sigma}_{\mu\nu}u^{\nu}_{\Sigma}=0.
\end{equation}
That is, the viscosity tensor and the strain tensor are both purely spacelike in the frame moving with $u^{\mu}_{\Sigma}$, and the viscous and strain tensors have been constrained to have only six independent components. The entropy generation equation can be rewritten as
\begin{align}
T^*\nabla_{\sigma}s^{\sigma}={}&\frac{1}{\kappa T^*}\perp_{\mu\nu}q^{\mu}q^{\nu}+\sum_{\overline{x}}\bigg[ \gamma(q^2_{\overline{x}})\mathcal{R}^{\text{mf},\overline{x}}_{\mu\nu}q^{\mu}_{\overline{x}}q^{\nu}_{\overline{x}}
\nonumber
\\
{}&+\frac{1}{n^*_{\overline{x}}}f^{\text{M},\overline{x}}_{\mu}\nabla_{\nu}\lambda^{\nu\mu}_{\overline{x}}-v^{\mu}_{\overline{x}}f_{\mu}^{\overline{x}}\bigg] \nonumber
\\
{}&-\sum_{\Sigma}\left[v^{\mu}_{\Sigma}f_{\mu}^{\Sigma}+\frac{1}{2\gamma(v_{\Sigma}^2)}\tau^{\mu\nu}_{\Sigma}\Lie_{u_{\Sigma}}\kappa^{\Sigma}_{\mu\nu}\right].
\label{eq:EntropyGeneration2}
\end{align}
where $v^{\mu}_s=w^{\mu}$ and 
\begin{align}
\tau^{\mu\nu}_{\Sigma}\Lie_{u_{\Sigma}}\kappa^{\Sigma}_{\mu\nu}{}&=-2u^{\mu}_{\Sigma}f^{\Sigma}_{\mu} \nonumber
\\
{}&=\tau^{\mu\nu}_{\Sigma}\left(u_{\Sigma}^{\rho}\nabla_{\rho}\kappa^{\Sigma}_{\mu\nu}+2\kappa^{\Sigma}_{\rho(\mu}\nabla_{\nu)}u^{\rho}_{\Sigma}\right).
\end{align}

Analogously to~\citet{Carter1991}, introduce linear combinations of the $v^{\mu}_{\overline{x}}$ and $v^{\mu}_s=w^{\mu}$ such that
\begin{equation}
\sum_{a=\overline{x},s}\varsigma_{\Sigma}^av^{\mu}_a=v^{\mu}_{\Sigma},\qquad \sum_{a}\varsigma^a_{\Sigma}=1,
\label{eq:SumVarSigma}
\end{equation}
so the terms depending on the forces can be combined using
\begin{equation}
f^a_{\mu}+\sum_{\Sigma}\varsigma_{\Sigma}^af^{\Sigma}_{\mu}\equiv-\sum_{b=s,\overline{x}}\mathcal{R}^{ab}_{\mu\nu}v^{\nu}_{b},
\label{eq:GeneralizedResistivityTensor}
\end{equation}
where $\mathcal{R}^{ab}_{\mu\nu}$ is a positive-definitive symmetric generalized resistivity tensor. This tensor must be symmetric by the Onsager reciprocal relations. This procedure assumes that there are no other dynamical velocities in the problem than $v^{\mu}_{\overline{p}}$, $v^{\mu}_{\overline{p}}$ and $w^{\mu}$. There will also be a contribution to the viscosity from the normal fluid 
\begin{equation}
\sum_{a=\overline{x},s} \varsigma^av_a^{\mu}=0,
\end{equation}
corresponding to $u_{\Sigma}^{\mu}\equiv u^{\mu}_{r}=u^{\mu}$. To make the viscosity term look more like a standard entropy generation equation, we use
\begin{subequations}
\begin{align}
\kappa_{\mu\nu}^{\Sigma}={}&\perp_{\mu\nu}^{\Sigma} = g_{\mu\nu}+u^{\Sigma}_{\mu}u^{\Sigma}_{\nu},
\label{eq:KappaDefinition}
\\ 
f^{\Sigma}_{\mu}={}&\nabla_{\rho}\tau^{\rho}_{\Sigma\mu},
\\
\tau^{\mu\nu}_{\Sigma}={}&-\eta^{\mu\nu\rho\sigma}_{\Sigma}\Lie_{u_{\Sigma}}\kappa^{\Sigma}_{\rho\sigma},
\\
\eta^{\mu\nu\rho\sigma}_{\Sigma}={}&\eta_{\Sigma}\perp^{\mu(\rho}_{\Sigma}\perp^{\sigma)\nu}_{\Sigma}+\left(\frac{\zeta_{\Sigma}}{2}-\frac{\eta_{\Sigma}}{3}\right)\perp^{\mu\nu}_{\Sigma}\perp^{\rho\sigma}_{\Sigma},
\end{align}
\end{subequations}
where $\eta_{\Sigma}$ and $\zeta_{\Sigma}$ are (dynamic) shear and bulk viscosity coefficients, respectively. This form ensures that the entropy generation is positive definite. We do not include the higher-order corrections to $\tau^{\mu\nu}_{\Sigma}$ discussed in~\citet{Carter1991} and hence assume that the we only have viscosity linear in the fluid velocities. As written, the viscous forces are causal for small perturbations from thermal equilibrium~\citep{Priou1991}. The viscous tensor can also be rewritten as
\begin{align}
\tau_{\Sigma}^{\mu\nu}={}&-2\eta_{\Sigma}\left(\nabla^{(\mu}u^{\nu)}_{\Sigma}+u^{(\mu}_{\Sigma}\dot{u}^{\nu)}_{\Sigma}-\frac{1}{3}\perp^{\mu\nu}_{\Sigma}\nabla_{\sigma}u^{\sigma}_{\Sigma}\right) \nonumber \\
{}&-\zeta_{\Sigma}\perp^{\mu\nu}_{\Sigma}\nabla_{\sigma}u^{\sigma}_{\Sigma}\nonumber
\\
={}&-\eta_{\Sigma}\perp^{\mu\alpha}_{\Sigma}\perp^{\nu\beta}_{\Sigma}W^{\Sigma}_{\alpha\beta}-\zeta_{\Sigma}\perp^{\mu\nu}_{\Sigma}\nabla_{\sigma}u^{\sigma}_{\Sigma},
\label{eq:ViscosityTensor}
\end{align} 
where $\dot{u}^{\mu}_{\Sigma}=u^{\rho}_{\Sigma}\nabla_{\rho}u^{\mu}_{\Sigma}$ and $W^{\Sigma}_{\mu\nu}=\nabla_{\mu}u_{\nu}^{\Sigma}+\nabla_{\nu}u_{\mu}^{\Sigma}-2/3g_{\mu\nu}\nabla_{\sigma}u^{\sigma}_{\Sigma}$ is the shear tensor. Eq.~(\ref{eq:EntropyGeneration2}) becomes
\begin{widetext}
\begin{align}
T^*\nabla_{\sigma}s^{\sigma}={}&\frac{1}{\kappa T^*}\perp_{\mu\nu}q^{\mu}q^{\nu}+\sum_{\overline{x}}\bigg[ \gamma(q^2_{\overline{x}})\mathcal{R}^{\text{mf},\overline{x}}_{\mu\nu}q^{\mu}_{\overline{x}}q^{\nu}_{\overline{x}}
+\frac{1}{n^*_{\overline{x}}}f^{\text{M},\overline{x}}_{\mu}\nabla_{\nu}\lambda^{\nu\mu}_{\overline{x}}\bigg] +\sum_{a,b=s,\overline{x}}\mathcal{R}^{ab}_{\mu\nu}v^{\mu}_av^{\nu}_b\nonumber
\\
{}&+\sum_{\Sigma}\frac{1}{\gamma(v_{\Sigma}^2)}\left[\frac{\eta_{\Sigma}}{2}\left((\nabla_{\mu}u_{\nu}^{\Sigma}+\nabla_{\nu}u_{\mu}^{\Sigma})(\nabla^{\mu}u^{\nu}_{\Sigma}+\nabla^{\nu}u^{\mu}_{\Sigma})-\frac{4}{3}(\nabla_{\mu}u^{\mu}_{\Sigma})^2\right)+\eta_{\Sigma} \dot{u}_{\mu}^{\Sigma}\dot{u}^{\mu}_{\Sigma}+\zeta_{\Sigma}(\nabla_{\mu}u^{\mu}_{\Sigma})^2\right].
\label{eq:EntropyGeneration3}
\end{align}
\end{widetext}
In the case of the shear and bulk viscosity of the normal fluid, $u^{\mu}_{\Sigma}=u^{\mu}$, $\gamma(v^2_{\Sigma})=1$. We also expect a bulk viscosity term from the superfluids~\citep{Khalatnikov2000,Landau1987,Gusakov2007}. The most general form of the viscosity contribution to the entropy generation (the second line on the right-hand side of Eq.~(\ref{eq:EntropyGeneration3})) should thus be of the form
\begin{widetext}
\begin{align}
\left[T^*\nabla_{\sigma}s^{\sigma}\right]_{\text{visc.}}={}&\frac{\eta_r}{2}\left((\nabla_{\mu}u_{\nu}+\nabla_{\nu}u_{\mu})(\nabla^{\mu}u^{\nu}+\nabla^{\nu}u^{\mu})-\frac{4}{3}(\nabla_{\sigma}u^{\sigma})^2\right)+\eta_r \dot{u}_{\mu}\dot{u}^{\mu}+\zeta_r(\nabla_{\sigma}u^{\sigma})^2+\sum_{\Sigma\neq r}\frac{1}{\gamma(v_{\Sigma}^2)}\zeta_{\Sigma}(\nabla_{\sigma}u^{\sigma}_{\Sigma})^2,
\label{eq:TotalViscousEntropyGen}
\end{align}
\end{widetext}
where the subscript $r$ is used to specify the viscosity coefficients for the normal fluid. Only the normal fluid $\Sigma=r$ contributes to the shear viscosity, and it also gives a contribution to the bulk viscosity from species-converting reactions between the normal fluid constituents. The terms with $\Sigma\neq r$ represent the bulk viscosity contributions from species-converting reactions involving the superfluids. These reactions are: (1) conversion between the normal and superfluid neutrons; (2) conversion between the normal and superconducting protons; (3) between the neutron superfluid and superconducting protons; (4) between the (non-neutron) normal fluid and neutron superfluid; and (5) between the (non-proton) normal fluid and superconducting protons. We thus require five distinct $\Sigma\neq r$ such that the five $\zeta_{\Sigma}\neq\zeta_{r}$ can represent these five bulk viscosity contributions. The corresponding $v^{\mu}_{\Sigma}$ will be linear combinations only of the relative superfluid velocities $v^{\mu}_{\overline{n}}$ and $v^{\mu}_{\overline{p}}$, but the exact specification of the $v^{\mu}_{\Sigma}$ is somewhat arbitrary as long as Eq.~(\ref{eq:SumVarSigma}) is satisfied. However, the bulk viscosity coefficients will be completely determined by the microphysics.

Comparing our formulation of the viscosity to the relativistic version of the Landau--Khalatnikov superfluid viscosity~\citep{Landau1987,Khalatnikov2000,Gusakov2007}, both formulations have six bulk viscosity coefficients. In a realistic neutron star core, with the superconducting protons comoving with the normal fluid due to electrostatic attraction, $v^{\mu}_{\overline{p}}=0$ and there will be only three distinct bulk viscosity coefficients parameterizing the reactions (1) between normal fluid constituents; (2) conversion between normal and superfluid neutrons; and (3) between the neutron superfluid and (non-neutron) normal fluid constituents. However, as shown by~\citet{Gusakov2007}, only two of these bulk viscosity coefficients will be independent of each other.

The different viscosity coefficients are, in principle, possible to calculate from microphysics. The shear viscosity will have contributions from lepton-lepton, lepton-proton, nucleon-nucleon~\citep{Shternin2008,Shternin2013,Kolomeitsev2015,Schmitt2018,Shternin2020} and proton-mediated lepton-neutron scattering~\citep{Bertoni2015}. The bulk viscosity in both the normal fluids and superfluids is due to modified and direct Urca processes~\citep{Haensel2000,Haensel2001,Gusakov2007}. Superfluidity generally increases the shear viscosity of the normal fluid and lowers the bulk viscosity.

\subsection{Electrical conductivity}

The generalized resistivity tensor introduced in Eq.~(\ref{eq:GeneralizedResistivityTensor}) cannot fully account for electrical conductivity because the only relative velocities in this equation are the $v^{\mu}_{\overline{x}}$ and $w^{\mu}$. To properly incorporate electrical conductivity we must relax our assumption that the normal fluid components are comoving. Reserving $u^{\mu}$ to denote the rest frame of the normal fluid neutrons, the dominant normal fluid component in a neutron star core, Eq.~(\ref{eq:NormalFluidCurrent}) is replaced by
\begin{equation}
n_x^{\mu}=n_x\gamma(v_x^2)(u^{\mu}+v^{\mu}_x), \qquad x\in\{p,e,m\},
\end{equation}
where the relative velocities $v^{\mu}_x$ are all fractionally small compared to $u^{\mu}$ and satisfy $v_x^{\mu}u_{\mu}=0$. In this case, the sum on the right side of Eq.~(\ref{eq:GeneralizedResistivityTensor}) runs over the normal fluid species in addition to $s,\overline{x}$.  The generalized resistivity forces can then be included in the equations of motion by solving Eq.~(\ref{eq:GeneralizedResistivityTensor}) for the generalized force $f^{a}_{\mu}$ and inserting into Eq.~(\ref{eq:ForceNormalFluid}--\ref{eq:ForceSuperfluid}), noting that forces such as mutual friction can in principle be included within the generalized resistivity forces, though this may require rewriting velocities such as the vortex line/flux tube velocities in terms of the velocities of the different fluids.

The generalized Ohm's law can thus be derived by appropriately combining the equations of motion for the charged fluids, but this is beyond the scope of this paper. The generalized Ohm's law is discussed in more detail for nonrelativistic non-superfluid neutron stars in~\citep{Easson1979,Goldreich1992},for superfluid neutron stars in~\citep{Glampedakis2011a}, and for relativistic multifluids in~\citep{Andersson2012,Dommes2020}.

\section{Vortex line/flux tube contribution and the magnetic field problem}
\label{sec:VLFTContribution}

A remaining question is how to interpret and compute the tensors $\mathcal{K}^{\mu\nu}$ and $\lambda^{\mu\nu}_{\overline{x}}$, and to determine if they can be written in terms of $F^{\mu\nu}$ and $w^{\mu\nu}_{\overline{x}}$. Since it is impossible to account for the dynamics of individual vortex lines and flux tubes in a macroscopic fluid dynamics, the vorticity tensors $w^{\overline{x}}_{\mu\nu}$ and the electromagnetic field tensor $F^{\mu\nu}$ should be considered as macroscopic averaged quantities. The electromagnetic field tensor $F^{\mu\nu}$ should somehow depend on the $w^{\overline{x}}_{\mu\nu}$, since assuming type--II proton superconductivity, the magnetic field inside the star is largely confined to proton flux tubes, plus neutron vortex lines that are magnetized through superfluid entrainment. 

We first consider this problem at the mesoscopic scale of individual or small numbers of vortex lines and flux tubes. By averaging over a large number of flux tubes in the mesoscopic theory, we find an averaged mesoscopic stress-energy tensor, which is then matched term-by-term to the completely macroscopic stress-energy tensor derived in Section~\ref{sec:EquationsOfMotion}. This allow us to find an macroscopic ``effective'' theory in the form of the electromagnetic and vorticity-dependent contribution to the master function $\Lambda_{\text{EM+V}}$, which fixes the forms of $\mathcal{K}^{\mu\nu}$ and $\lambda^{\mu\nu}_{\overline{x}}$. We match to the stress-energy tensor as opposed to simply the master function because the former also contains information about the partial derivatives of the latter. A summary of this calculation is presented in the main text, reserving the full calculation for Appendix~\ref{app:MesoscopicAveraging}.

This section is concluded by discussing the relation of these quantities to the electromagnetic displacement tensor $\mathcal{H}^{\mu\nu}$, which we show is distinct from $\mathcal{K}^{\mu\nu}$. We compare the resulting electromagnetism to previous studies of superconducting neutron star cores with flux tubes and magnetized vortex lines. Finally, we discuss how to compute the magnetic field in a superconducting neutron star core given an electric current density, and the form of the Lorentz force in the total equation of motion for the charged fluids.

\subsection{Mesoscopic stress-energy tensor, averaging procedure and effective theory}
\label{sec:EffectiveTheoryFT}

We postulate the following Lorentz-invariant splitting of the (macroscopic) master function $\Lambda$ as a function of the contractions of $F^{\mu\nu}$ and $w^{\mu\mu}_{\overline{x}}$:
\begin{equation}
\Lambda=\Lambda_0+\Lambda_{\text{EM+V}}(X_F,X_{\overline{n}},X_{\overline{p}},Y_{\overline{n}},Y_{\overline{p}},Z)
\label{eq:MacroscopicMasterFunctionSplit}
\end{equation}
where the scalars of which $\Lambda_{\text{EM+V}}$ is a function are defined as
\begin{align}
X_F={}&\frac{1}{4}F^{\mu\nu}F_{\mu\nu}, \quad X_{\overline{n}}=\frac{1}{4}w^{\mu\nu}_{\overline{n}}w_{\mu\nu}^{\overline{n}},
\nonumber
\\
X_{\overline{p}}={}&\frac{1}{4}w^{\mu\nu}_{\overline{p}}w^{\overline{p}}_{\mu\nu}, \quad Y_{\overline{n}}=\frac{1}{2}F^{\mu\nu}w_{\mu\nu}^{\overline{n}},
\nonumber
\\
Y_{\overline{p}}={}&\frac{1}{2}F^{\mu\nu}w^{\overline{p}}_{\mu\nu}, \quad Z=\frac{1}{2}w^{\mu\nu}_{\overline{n}}w_{\mu\nu}^{\overline{p}}.
\nonumber
\end{align}
$\Lambda_0$ is the contribution to the master function from the four-currents alone, while $\Lambda_{\text{EM+V}}$ contains all contributions from flux tubes/vortex lines and electromagnetic fields. $\Lambda_{\text{EM+V}}$ will also contain functional dependence on contractions of the superfluid/superconducting four-currents, since the flux tube/vortex line energies will depend on number densities through dependence on the London length $\Lambda_*$ and coherence lengths $\xi_x$, but we have assumed that there are no terms involving contractions between the number currents and the tensors $F^{\mu\nu}$ and $w^{\mu\mu}_{\overline{x}}$. According to Eq.~(\ref{eq:EMAuxiliaryTensor},\ref{eq:VortexTensionTensor}) $\mathcal{K}^{\mu\nu}$ and $\lambda^{\mu\nu}_{\overline{x}}$ will then take the forms
\begin{align}
\mathcal{K}^{\mu\nu}\equiv{}&-4\pi\left(\frac{\partial\Lambda_{\text{EM+V}}}{\partial X_F}F^{\mu\nu}+\frac{\partial\Lambda_{\text{EM+V}}}{\partial Y_{\overline{p}}}w^{\mu\nu}_{\overline{p}}\right.\nonumber
\\
{}&\quad\qquad\left.+\frac{\partial\Lambda_{\text{EM+V}}}{\partial Y_{\overline{n}}}w^{\mu\nu}_{\overline{n}}\right)
\label{eq:KDefinition}
\\
\lambda^{\mu\nu}_{\overline{p}}\equiv{}&-\frac{\partial\Lambda_{\text{EM+V}}}{\partial X_{\overline{p}}}w_{\overline{p}}^{\mu\nu}-\frac{\partial\Lambda_{\text{EM+V}}}{\partial Z}w_{\overline{n}}^{\mu\nu}-\frac{\partial\Lambda_{\text{EM+V}}}{\partial Y_{\overline{p}}}F^{\mu\nu},
\label{eq:LambdaPDefinition}
\\
\lambda^{\mu\nu}_{\overline{n}}\equiv{}&-\frac{\partial\Lambda_{\text{EM+V}}}{\partial X_{\overline{n}}}w_{\overline{n}}^{\mu\nu}-\frac{\partial\Lambda_{\text{EM+V}}}{\partial Z}w_{\overline{p}}^{\mu\nu}-\frac{\partial\Lambda_{\text{EM+V}}}{\partial Y_{\overline{n}}}F^{\mu\nu}.
\label{eq:LambdaNDefinition}
\end{align}
The goal of the mesoscopic averaging procedure is to determine what $\Lambda_{\text{EM+V}}$ and its partial derivatives are.

We define the mesoscopic scale $\ell$ such that there are many vortex lines and flux tubes within an area $\ell^2$. $\ell$ obeys the following hierarchy of length scales:
\begin{equation}
\ell_g \gg \ell \gg d_n \gg d_p > \Lambda_* > \xi_p,\xi_n.
\end{equation}
$\ell_g$ is some characteristic length scale of the spacetime curvature, $d_n$ and $d_p$ are the spacings between neutron vortex lines/proton flux tubes, $\Lambda_*\equiv(4\pi e^2Y_{pp})^{-1/2}$ is the London length, and $\xi_n$/$\xi_p$ are the neutron vortex line/proton flux tube coherence lengths. We assume that physical properties like $n_x$, $\mu_x$, $Y_{xy}$, etc. are uniform over mesoscopic scales.

The system we consider is a simple configuration of two vortex line/flux tubes arrays, one for each superfluid/superconducting species. The vortex line array results from the rotation of the star, while the flux tube array is a result of a combination of a remnant magnetic field and field generation mechanisms early in the neutron star's life~\citep{Spruit2008}. We consider only the strong type-II limit of the superconducting protons i.e. $H^{\overline{p}}_{c1}\lesssim H\ll H^{\overline{p}}_{c2}$, where $H$ is the macroscopic average magnetic field and where $H^{\overline{p}}_{c1}$ and $H^{\overline{p}}_{c2}$ are the proton type--II superconductivity critical fields. Strong vortex pinning due to the significant outnumbering of neutron vortex lines by proton flux tubes could modify this simple model by distorting the vortex line lattice, but we ignore vortex pinning here. We also ignore mutual friction, heat conduction (including ``entropy entrainment'' $\mathcal{A}^{sx}$) and viscosity as a first approximation.

Denoting with a tilde a mesoscopic quantity, we take as the mesoscopic master function $\tilde{\Lambda}=\tilde{\Lambda}(\tilde{n}_x^2,\tilde{\alpha}^2_{xy})-\tilde{F}^{\rho\sigma}\tilde{F}_{\rho\sigma}/16\pi$, using the microscopic electromagnetic field Lagrangian in place of arbitrary dependence of $\tilde{\Lambda}$ on $\tilde{F}_{\mu\nu}$.  Following the same procedure used to derive the stress-energy tensor in Section~\ref{sec:EquationsOfMotion}, the mesoscopic stress-energy tensor is
\begin{align}
\tilde{T}^{\mu\nu}={}&\sum_x\tilde{n}_x^{\mu}\tilde{\mu}_x^{\nu}+\frac{1}{4\pi}\tilde{F}^{\mu\rho}\tilde{F}^{\nu}_{\ \rho}\nonumber
\\
{}&+\left(\tilde{\Lambda}(\tilde{n}_x^2,\tilde{\alpha}^2_{xy})-\frac{1}{16\pi}\tilde{F}^{\rho\sigma}\tilde{F}_{\rho\sigma}-\sum_x \tilde{n}^{\rho}_x\tilde{\mu}_{\rho}^x\right)g^{\mu\nu}.
\label{eq:MesoscopicSETensor}
\end{align}
This result combines the stress-energy tensors of a perfect multifluid plus that of vacuum electromagnetism. We also have the equations of motion
\begin{align}
2\tilde{n}^{\mu}_x\nabla_{[\mu}\tilde{\pi}^x_{\nu]}={}&0,
\\
\nabla_{\nu}\tilde{F}^{\mu\nu}={}&4\pi \tilde{j}_e^{\mu}=4\pi e(\tilde{n}_p^{\mu}+\tilde{n}_{\overline{p}}^{\mu}-\tilde{n}_e^{\mu}-\tilde{n}_m^{\mu}),
\label{eq:MesoscopicMaxwellEquation}
\end{align}
plus the Bianchi identity for $\tilde{F}^{\mu\nu}$. $\tilde{\pi}^x_{\nu}$ is defined as in Eq.~(\ref{eq:CanonicalMomentumCovector}). In this case, we used Eq.~(\ref{eq:CurrentVariation}) for all of the currents in deriving their equations of motion, not just for the normal fluids.

The averaging procedure first splits the number currents into large-scale and small-scale contributions. The latter represent currents around vortex lines and flux tubes and hence source the magnetic field associated with flux tubes and magnetized vortex lines. The mesoscopic stress-energy tensor is then averaged over an area $\sim\ell^2$ perpendicular to the vortex line/flux tube array in a procedure similar to~\cite{Prix2000} and Appendix E of~\cite{Gusakov2016}, allowing us to replace the sum over vortex lines/flux tubes with multiplication by the relevant areal number density $\mathcal{N}_{\overline{x}}$. This averaged mesoscopic stress-energy tensor is then compared to the macroscopic stress-energy tensor to determine the macroscopic effective master function $\Lambda_{\text{EM+V}}$ and its partial derivatives.

We relegate most of the details of the calculation to Appendix~\ref{app:MesoscopicAveraging}, but discuss the averaging procedure for the electromagnetic field and vorticity here. The canonical four momenta for the superfluid neutrons and superconducting protons, and hence the vorticity tensors, are quantized
\begin{align}
\oint\pi^{\overline{x}}_{\mu}dx^{\mu}=\int w^{\overline{x}}_{\mu\nu}dS^{\mu\nu}=hN^{\overline{x}},
\label{eq:QuantizationofW}
\end{align}
where $N^{\overline{x}}\in\mathbb{Z}$, $h=2\pi\hbar$ and the generalized Stokes' theorem~\cite{Poisson2004} was used. Recall that in Section~\ref{sec:MFandVP} we defined a vorticity vector
\begin{equation}
W^{\alpha}_{\overline{x}}\equiv\frac{1}{2}\varepsilon^{\alpha\beta\mu\nu}u_{\beta}^{L,\overline{x}}w^{\overline{x}}_{\mu\nu},
\label{eq:VorticityVector1}
\end{equation}
where $u_{\beta}^{L,\overline{x}}$ is the average four-velocity of the vortex lines/flux tubes of species $\overline{x}$. Since we are ignoring mutual friction and vortex pinning in the averaging calculation, to lowest order the vortex lines/flux tubes comove with their corresponding fluids i.e. $u_{\mu}^{L,\overline{p}}=u^{\overline{p}}_{\mu}$ and $u_{\mu}^{L,\overline{n}}=u_{\mu}^{\overline{n}}$. This assumption is equivalent to assuming that the vortex lines are straight and uniformly distributed, since we are ignoring the vortex line self-tension force in Eq.~(\ref{eq:ForceSuperfluid}). Since we will find a general expression for the $\lambda^{\mu\nu}_{\overline{x}}$ that does not necessarily correspond to zero vortex line self-tension force, this force can be considered a first-order correction to the equations of motion. Over length scales $\ell$ much larger than the separation of the vortex lines/flux tubes, the quantization condition allows us to write $W^{\alpha}_{\overline{x}}$ as
\begin{equation}
W^{\alpha}_{\overline{x}}=\mathcal{N}_{\overline{x}}\kappa_{\overline{x}}\mu_{\overline{x}}^*\hat{t}^{\alpha}_{\overline{x}}=\mathcal{N}_{\overline{x}}\Phi_{\overline{p}}e\hat{t}^{\alpha}_{\overline{x}}.
\label{eq:VorticityVector2}
\end{equation}
$\kappa_{\overline{x}}=h/(2\mu^*_{\overline{x}})$ is the relativistic generalization of the quantum of circulation with a factor of 2 because the superfluid neutrons/superconducting protons will form Cooper pairs, $\Phi_{\overline{p}}=h/(2e)$ is the flux quantum associated with a proton flux tube, $\hat{t}^{\alpha}_{\overline{x}}$ is the average spatial tangent vector to the vortex lines/flux tubes defined in Eq.~(\ref{eq:ArrayTangentVector}), and $\mathcal{N}_{\overline{x}}$ is the areal number density of vortex lines/flux tubes in the spatial plane perpendicular to $\hat{t}^{\alpha}_{\overline{x}}$. $\mathcal{N}_{\overline{x}}$ is Lorentz-invariant and defined by
\begin{equation}
\mathcal{N}_{\overline{x}}\equiv\frac{1}{\kappa_{\overline{x}}\mu^*_{\overline{x}}}\sqrt{W^{\mu}_{\overline{x}}W^{\overline{x}}_{\mu}}=\frac{1}{\kappa_{\overline{x}}\mu^*_{\overline{x}}}\sqrt{\frac{1}{2}w^{\mu\nu}_{\overline{x}}w^{\overline{x}}_{\mu\nu}}.
\label{eq:ArealFTDensityDefinition}
\end{equation}
We expect contributions to $\Lambda_{\text{EM+V}}$ that will be proportional to the $\mathcal{N}_{\overline{x}}$ times an energy per unit length. The electromagnetic field contributions due to flux tubes/magnetized vortex lines should also be linearly proportional to $\mathcal{N}_{\overline{x}}$. The separations between proton flux tubes/neutron vortex lines $d_p$/$d_n$ are defined by $\mathcal{N}_{\overline{p}}$/$\mathcal{N}_{\overline{n}}$ through
\begin{equation}
\mathcal{N}_{\overline{x}}=\frac{2}{\sqrt{3}d_x^2},
\end{equation}
assuming equilateral triangular lattices.

Because of entrainment, the neutron vortex lines will become magnetized. This is made apparent by combining the vorticity tensors $w_{\mu\nu}^{\overline{p}}$ and $w_{\mu\nu}^{\overline{n}}$ in a way to eliminate the superfluid neutron current, which itself does not source a magnetic field. In our formulation, this corresponds to eliminating $v^{\mu}_{\overline{n}}$. We thus add the two tensors in such a way as to give
\begin{equation}
F_{\mu\nu}=F^{\text{L}}_{\mu\nu}+\frac{1}{e}w_{\mu\nu}^{\overline{p}}+\frac{Y_{np}}{eY_{pp}}w_{\mu\nu}^{\overline{n}},
\label{eq:FmunuMesoscopicAverage}
\end{equation}
where $F^{\text{L}}_{\mu\nu}$ is the \textit{London electromagnetic field tensor}. If we can ignore derivatives of $\mu_x$, $n_x$ and the coefficients $\mathcal{B}^x$, $\mathcal{A}^{xy}$, it takes the form
\begin{equation}
F^{\text{L}}_{\mu\nu}\approx-\frac{2}{e}\left(\mu_{\overline{p}}^*+\frac{Y_{np}}{Y_{pp}}\mu^*_{\overline{n}}\right)\partial_{[\mu}u_{\nu]}-\frac{2n^*_{\overline{p}}}{eY_{pp}}\partial_{[\mu}v^{\overline{p}}_{\nu]},
\label{eq:LondonFmunu}
\end{equation}
where $v^{\mu}_{\overline{n}}$ has canceled out as expected. In the general case where the gradients of  $\mu_x$, $n_x$, $\mathcal{B}^x$, $\mathcal{A}^{xy}$ cannot be ignored, this will not be true.

Based on Eq.~(\ref{eq:FmunuMesoscopicAverage}), we split the mesoscopic electromagnetic field tensor $\tilde{F}^{\mu\nu}$ into
\begin{equation}
\tilde{F}^{\mu\nu}=\tilde{F}^{\mu\nu}_{\text{L}}+\tilde{F}^{\mu\nu}_{\overline{p}}+\tilde{F}^{\mu\nu}_{\overline{n}},
\label{eq:FmunuMesoscopicDivision}
\end{equation}
with the right-hand side terms corresponding to the London field, proton flux tube field and magnetized neutron vortex line field, respectively. $F^{\mu\nu}_{\text{L}}$ is a large-scale quantity and is the same when averaged i.e. $\langle\tilde{F}^{\mu\nu}_{\text{L}}\rangle=F^{\mu\nu}_{\text{L}}$. The average of the second and third terms on the right-hand side of Eq.~(\ref{eq:FmunuMesoscopicDivision}) can be identified with the second and third terms on the right-hand side of Eq.~(\ref{eq:FmunuMesoscopicAverage}) i.e. $\langle \tilde{F}^{\mu\nu}_{\overline{p}}\rangle=w^{\mu\nu}_{\overline{p}}/e$, $\langle \tilde{F}^{\mu\nu}_{\overline{n}}\rangle=Y_{np}w^{\mu\nu}_{\overline{n}}/(Y_{pp}e)$. The invertability of Eq.~(\ref{eq:VorticityVector1}) thus says that $\langle \tilde{F}^{\mu\nu}_{\overline{x}}\rangle\propto\mathcal{N}_{\overline{x}}\Phi_{\overline{x}}$ as expected. Note that $\Phi_{\overline{n}}=Y_{np}\Phi_{\overline{p}}/Y_{pp}$, so $\langle \tilde{F}^{\mu\nu}_{\overline{n}}\rangle\rightarrow 0$ when the entrainment is zero ($Y_{np}=0$) as required. Finally, since $\langle\tilde{F}^{\mu\nu}_{\overline{x}}\rangle\propto w^{\mu\nu}_{\overline{x}}$, we enforce
\begin{equation}
u^{L,\overline{x}}_{\mu}\tilde{F}^{\mu\nu}_{\overline{x}}=0,
\label{eq:NoElectricField}
\end{equation}
that is, there is no electric field due to the flux tubes/magnetized vortex lines in their respective rest frames.

After performing the averaging procedure on the mesoscopic stress-energy tensor, we obtain the following averaged mesoscopic stress-energy tensor
\begin{align}
\langle\tilde{T}^{\mu\nu}\rangle={}&\sum_x n_x^{\mu}\mu_x^{\nu}+\left(\tilde{\Lambda}_0+\sum_xn_x\mu_x\right)g^{\mu\nu}
\nonumber
\\
{}&+\sum_{\overline{x}}\left\langle\frac{1}{4\pi}\tilde{F}^{(\mu}_{\overline{x}\rho}\tilde{F}^{\nu)\rho}_{\text{L}}\right\rangle
\nonumber 
\\
{}&+\frac{1}{4\pi}\left(F^{\mu\rho}_{\text{L}}F^{\nu}_{\text{L}\rho}-\frac{1}{4}F^{\sigma\rho}_{\text{L}}F^{\text{L}}_{\sigma\rho}g^{\mu\nu}\right)
\nonumber
\\
{}&+\sum_{\overline{x}}\frac{\mathcal{E}_{v,\overline{x}}}{\mathcal{N}_{\overline{x}}(\Phi_{\overline{p}}e)^2}\left(w^{\mu\rho}_{\overline{x}}w^{\nu}_{\overline{x}\rho}-\frac{1}{2}w^{\sigma\rho}_{\overline{x}}w^{\overline{x}}_{\sigma\rho}g^{\mu\nu}\right)
 \nonumber
\\
{}&+\sum_{\overline{x}}\frac{1}{32\pi^2\mathcal{N}_{\overline{x}}\Lambda^2_*e^2}w^{\mu\rho}_{\overline{x}}w^{\nu}_{\overline{x}\rho},
\label{eq:MesoscopicSETensor4Main}
\end{align}
where $n_x^{\mu}$ and $\mu_x^{\nu}$ are the macroscopic number currents and conjugate momenta defined in terms of four-velocities in Section~\ref{sec:RelationtoPhysParams} and $\tilde{\Lambda}_0$ only includes dependence on those macroscopic number currents. We have defined the energy per unit length per flux tube/vortex line
\begin{align}
\mathcal{E}_{v,\overline{p}}\equiv{}&\frac{\Phi^2_{\overline{p}}}{16\pi^2\Lambda^2_*}\ln\left(\frac{1.12\Lambda_*}{\xi_p}\right)
\label{eq:VLEnergyP}
\\
\mathcal{E}_{v,\overline{n}}\equiv{}&\frac{\pi\hbar^2}{8\mathcal{B}^{\overline{n}}}\ln\left(\frac{0.0712}{\mathcal{N}_{\overline{n}}\xi^2_n}\right)+\frac{\Phi^2_{\overline{n}}}{16\pi^2\Lambda^2_*}\ln\left(\frac{1.12\Lambda_*}{\xi_n}\right).
\label{eq:VLEnergyN}
\end{align}
We ignore condensation energy in the $\mathcal{E}_{v,\overline{x}}$, which is much smaller than the other contributions. Since $w^{\mu\nu}_{\overline{x}}\propto\mathcal{N}_{\overline{x}}$, the final two terms in Eq.~(\ref{eq:MesoscopicSETensor4Main}) are proportional to the areal density of vortex lines/flux tubes as expected.

Eq.~(\ref{eq:MesoscopicSETensor4Main}) is then matched to Eq.~(\ref{eq:StressEnergyTensor}). The averaged mesoscopic-macroscopic stress energy tensor matching procedure is described in full detail in Appendix~\ref{app:MatchingSETensors}.
The resulting $\Lambda_{\text{EM+V}}$ is
\begin{equation}
\Lambda_{\text{EM+V}}=-\frac{F_{\text{L}}^{\mu\nu}F^{\text{L}}_{\mu\nu}}{16\pi}-\sum_{\overline{x}}\mathcal{N}_{\overline{x}}\mathcal{E}_{v,\overline{x}},
\label{eq:LambdaEMV1}
\end{equation}
or in terms of the scalars $X_F$, $X_{\overline{x}}$, $Y_{\overline{x}}$ and $Z$ and using
\begin{align}
F^{\mu\rho}_{\text{L}}F^{\text{L}}_{\mu\rho}={}&4X_F+\frac{4}{e^2}X_{\overline{p}}+\frac{4Y_{np}^2}{e^2Y_{pp}^2}X_{\overline{n}}-\frac{4}{e}Y_{\overline{p}}-\frac{4Y_{np}}{eY_{pp}}Y_{\overline{n}}
\nonumber
\\
{}&+\frac{4Y_{np}}{e^2Y_{pp}}Z,
\label{eq:FmunuLSquared}
\end{align}
plus Eq.~(\ref{eq:ArealFTDensityDefinition},\ref{eq:FmunuMesoscopicAverage}), we can write
\begin{align}
\Lambda_{\text{EM+V}}={}&-\frac{1}{4\pi}X_F-\frac{Y_{np}}{4\pi e^2Y_{pp}}Z
\nonumber
\\
{}&-\sum_{\overline{x}}\left(\frac{\sqrt{2X_{\overline{x}}}}{\Phi_{\overline{p}}e}\mathcal{E}_{v,\overline{x}}+\hspace{-0.4mm}\frac{\Phi_{\overline{x}}}{4\pi e\Phi_{\overline{p}}}\left(\frac{\Phi_{\overline{x}}}{\Phi_{\overline{p}}}X_{\overline{x}}-Y_{\overline{x}}\right)\hspace{-0.5mm}\right).
\label{eq:LambdaEMV2}
\end{align}

The final term on the fourth line of Eq.~(\ref{eq:MesoscopicSETensor4Main}) does not have a corresponding term in the macroscopic effective theory for reasons which we discuss in Appendix~\ref{app:MatchingSETensors}, and is thus not included in the averaged mesoscopic-macroscopic stress energy tensor matching procedure. We also exclude the final term in Eq.~(\ref{eq:MesoscopicSETensor4Main}) from the matching procedure, which is certainly legitimate in the strong type-II limit where the kinetic energy associated with flux tubes $\approx\mathcal{E}_{v,\overline{x}}$ is much larger than the flux tube/vortex line magnetic field energy per unit length $\Phi_{\overline{x}}^2/(32\pi^2\Lambda^2_*)$. If this term is not removed, there would be an inconsistency between (a) the $\Lambda_{\text{EM+V}}$ found by comparing the terms proportional to the metric in Eq.~(\ref{eq:MesoscopicSETensor4Main}) to those in Eq.~(\ref{eq:StressEnergyTensor}), and (b) the partial derivatives of $\Lambda_{\text{EM+V}}$ found by comparing the rest of the terms in Eq.~(\ref{eq:MesoscopicSETensor4Main}) and Eq.~(\ref{eq:StressEnergyTensor}). For consistency we also must ignore the derivative of $\mathcal{E}_{v,\overline{n}}$ with respect to $X_{\overline{n}}\propto\mathcal{N}^2_{\overline{n}}$, which is justified since
\begin{equation}
\left|\mathcal{N}_{\overline{n}}\frac{\partial\mathcal{E}_{v,\overline{n}}}{\mathcal{N}_{\overline{n}}}\right|=\frac{\pi\hbar^2}{8\mathcal{B}^{\overline{n}}}\ll\mathcal{E}_{v,\overline{n}}\sim \frac{\pi\hbar^2}{4\mathcal{B}^{\overline{n}}}\ln\left(\frac{d_n}{\xi_n}\right)
\label{eq:EvnDerivativeArgument}
\end{equation}
since $d_n\gg\xi_n$. That some terms in either the averaged mesoscopic stress-energy tensor or the partial derivatives of $\Lambda_{\text{EM+V}}$ must be ignored to obtain a consistent $\Lambda_{\text{EM+V}}$ is not unexpected, as there was no guarantee that an exact macroscopic effective action could be found to reproduce the averaged mesoscopic action and stress-energy tensor. That this procedure works so well suggests that we could simply use the averaging method as a motivation for an effective theory, for which we would use Eq.~(\ref{eq:LambdaEMV1}) as the macroscopic master function, and then use this to derive the macroscopic stress-energy tensor.

Eq.~(\ref{eq:LambdaEMV1}) agrees with the vortex line-flux tube-electromagnetic energy density obtained in~\citep{Carter2000,Prix2000,Glampedakis2011}, including in the lack of terms coupling the London field to the flux tube/magnetized vortex line fields. Such terms were eliminated in the references by the \textit{rotation energy cancellation lemma}. In our paper this lemma is used to eliminate the London field--flux tube/magnetized vortex line field cross terms that are proportional to the metric as discussed Appendix~\ref{app:AveragingSETensor}. However, this lemma requires making additional approximations to those used in the rest of the averaging procedure. First, the velocity differences between the normal fluid and superconducting protons $v^{\mu}_{\overline{p}}$ is negligible. Second, that $v^{\mu}_{\overline{n}}$ is negligible \textit{or} there is zero entrainment and hence no electromagnetic field associated with neutron vortex lines. These conditions are true in the fully pinned, lines comoving with superfluid case $u^{\mu}_{L,\overline{n}}=u^{\mu}_{L,\overline{p}}=u^{\mu}_{\overline{p}}$ as long as Eq.~(\ref{eq:NoElectricField}) is true, but is not necessarily true otherwise.

As the final result of this section, we calculate $\mathcal{K}^{\mu\nu}$ and the $\lambda^{\mu\nu}_{\overline{x}}$. Using Eq.~(\ref{eq:LambdaEMV2}) and the argument in Eq.~(\ref{eq:EvnDerivativeArgument}), Eq.~(\ref{eq:KDefinition}--\ref{eq:LambdaNDefinition}) gives
\begin{align}
\mathcal{K}^{\mu\nu}={}&F_{\text{L}}^{\mu\nu},
\label{eq:KDefinition2}
\\
\lambda^{\mu\nu}_{\overline{p}}={}&\frac{\mathcal{E}_{v,\overline{p}}}{\mathcal{N}_{\overline{p}}\Phi_{\overline{p}}^2e^2}w^{\mu\nu}_{\overline{p}}-\frac{1}{4\pi e^2}F^{\mu\nu}_{\text{L}},
\label{eq:LambdaPDefinition2}
\\
\lambda^{\mu\nu}_{\overline{n}}={}&\frac{\mathcal{E}_{v,\overline{n}}}{\mathcal{N}_{\overline{n}}\Phi_{\overline{p}}^2e^2}w^{\mu\nu}_{\overline{n}}-\frac{Y_{np}}{4\pi e^2Y_{pp}}F^{\mu\nu}_{\text{L}}.
\label{eq:LambdaNDefinition2}
\end{align}

\subsection{Magnetic H-field and Maxwell equations in a neutron star}

There is disagreement in the literature about what the electromagnetic displacement field tensor $\mathcal{H}^{\mu\nu}$, or equivalently the magnetic $\mathbf{H}$-field  (and electric displacement field $\mathbf{D}$ if we were concerned about electric fields) is inside a superconducting neutron star core. One of the early studies of neutron star MHD by~\citet{Mendell1991b} found $\mathbf{H}=\mathbf{B}$. This result was contradicted by later studies, the first of which appears to be~\citet{Carter1998}, who argued that $\mathbf{H}=\mathbf{B}_{\text{L}}$, where $\mathbf{B}_{\text{L}}$ is the London field which has approximate nonrelativistic, zero entrainment form $\mathbf{B}_{\text{L}}\approx-2m_p\boldsymbol{\Omega}/e$ for proton mass $m_p$ and uniform stellar rotation rate $\boldsymbol{\Omega}$. This result has been the standard since then~\citep{Carter2000,Glampedakis2011,Gusakov2016a}. However, $\mathbf{H}=\mathbf{B}_{\text{L}}$ disagrees with the accepted value for a type--II superconductor in the condensed matter literature: in the nonrotating case it suggests $\mathbf{H}=0$, while in the low flux tube density limit the standard electronic superconductivity result is~\citep{Tinkham1996} $H=H^{\overline{p}}_{c1}$ where $H^{\overline{p}}_{c1}$ is the first critical field for proton superconductivity. We clarify this disagreement below, and further discuss its implications for the Maxwell equations inside a neutron star.

According to~\cite{Landau1960}, the thermodynamic definition of the magnetic $\mathbf{H}$-field is
\begin{equation}
\mathbf{H}_T=4\pi\left.\frac{\partial u}{\partial \mathbf{B}}\right|_{s,n_i},
\label{eq:HfieldThermodynamicDef}
\end{equation}
for internal energy density $u$, average magnetic field $B$, entropy density $s$, and number density $n_i$. The subscript $T$ is used to denote the thermodynamic definition. In our formulation, the analog to the internal energy density is the master function $\Lambda$, and the analog to the entropy and number densities are the currents $n_x^{\mu}$, including the entropy current $s^{\mu}$. This means that the electromagnetic displacement tensor $\mathcal{H}^{\mu\nu}$, whose components in the fluid rest frame are the electric displacement field $\mathbf{D}$ and magnetic $\mathbf{H}$-field, is not equal to the electromagnetic auxiliary tensor defined in Eq.~(\ref{eq:EMAuxiliaryTensor}), but is instead defined through the variation
\begin{align}
\mathcal{H}^{\mu\nu}={}&-8\pi\left.\frac{\partial\Lambda}{\partial F_{\mu\nu}}\right|_{n^{\mu}_x} \nonumber
\\
={}&-8\pi\left[\left.\frac{\partial\Lambda}{\partial F_{\mu\nu}}\right|_{n^{\mu}_x,w^{\overline{x}}_{\mu\nu}}\hspace{-1em}+\sum_{\overline{x}}\left.\frac{\partial\Lambda}{\partial w^{\overline{x}}_{\mu\nu}}\right|_{n^{\mu}_x,F_{\mu\nu}}\left.\frac{\partial w^{\overline{x}}_{\mu\nu}}{\partial F_{\mu\nu}}\right|_{n^{\mu}_x}\right] \nonumber
\\
={}&\mathcal{K}^{\mu\nu}+4\pi e\lambda^{\mu\nu}_{\overline{p}},
\label{eq:EMDisplacementTensor}
\end{align}
where we use Eq.~(\ref{eq:VorticityTensor},\ref{eq:CanonicalMomentumCovector},\ref{eq:LatticeFieldIdentity}) in computing $\partial w^{\overline{x}}_{\mu\nu}/\partial F_{\mu\nu}|_{n^{\mu}_x}$. $\mathcal{H}^{\mu\nu}$ can then be related to $F^{\mu\nu}$ by defining a magnetization-polarization tensor $\mathcal{M}^{\mu\nu}$ and writing
\begin{equation}
F^{\mu\nu}=\mathcal{H}^{\mu\nu}+4\pi\mathcal{M}^{\mu\nu}.
\end{equation}
This subtle distinction between $\mathcal{K}^{\mu\nu}$ (which has often been called $\mathcal{H}^{\mu\nu}$) and $\mathcal{H}^{\mu\nu}$ as defined in Eq.~(\ref{eq:EMDisplacementTensor}), to be the source of disagreement between neutron star MHD and condensed matter superconductivity literature regarding the magnetic $\mathbf{H}$-field in a type--II proton superconducting neutron star. Based on Eq.~(\ref{eq:KDefinition2}--\ref{eq:LambdaPDefinition2}), $\mathcal{H}^{\mu\nu}$ is
\begin{equation}
\mathcal{H}^{\mu\nu}=H^{\overline{p}}_{c1}\hat{w}^{\mu\nu}_{\overline{p}}.
\label{eq:EMDisplacementTensor2}
\end{equation}
where the first critical field for proton superconductivity is
\begin{equation}
H^{\overline{p}}_{c1}\equiv\frac{4\pi\mathcal{E}_{v,\overline{p}}}{\Phi_{\overline{p}}},
\end{equation}
and $\hat{w}^{\mu\nu}_{\overline{p}}\equiv w^{\mu\nu}_{\overline{p}}/(\mathcal{N}_{\overline{p}}\Phi_{\overline{p}}e)$. Eq.~(\ref{eq:EMDisplacementTensor2}) agrees with the standard condensed matter result in the strong type-II limit.

The distinction between $\mathcal{K}^{\mu\nu}$ and $\mathcal{H}^{\mu\nu}$ as we define them here has implications on the interpretation of the Maxwell equations. The variation of the Lagrangian with respect to $A_{\mu}$ in Section~\ref{sec:EquationsOfMotion} gives Eq.~(\ref{eq:SourcedMaxwellEquations}), which by Eq.~(\ref{eq:KDefinition2}) gives as the sourced Maxwell equations
\begin{equation}
\nabla_{\nu}F_{\text{L}}^{\mu\nu}=4\pi j_e^{\mu}.
\label{eq:LondonFieldMaxwellEquation}
\end{equation}
If the $\mathbf{H}$-field is interpreted as the field whose curl is proportional to the current density, this suggests that $\mathbf{H}=\mathbf{B}_{\text{L}}$ and agrees with the Maxwell equations in~\citep{Carter1998,Glampedakis2011,Gusakov2016a}. Using Eq.~(\ref{eq:LondonFmunu}) and working in the zero temperature approximation such that $n_{\overline{p}}^*=Y_{pp}\mu^*_{\overline{p}}+Y_{np}\mu^*_{\overline{n}}$~\cite{Gusakov2009}, Eq.~(\ref{eq:LondonFieldMaxwellEquation}) then implies
\begin{align}
n_e^{\mu}+n_m^{\mu}-n_p^{\mu}\approx{}& n^{\mu}_{\overline{p}}+\Lambda_*^2\partial_{\nu}\left(\partial^{\mu}n^{\nu}_{\overline{p}}-\partial^{\mu}n^{\nu}_{\overline{p}}\right)
\nonumber
\\
\approx{}& n^{\mu}_{\overline{p}}(1+L^{-2}\Lambda_*^2),
\end{align}
where $L$ is the hydrodynamic length scale $\sim10^5$ cm. Since $\Lambda_*\sim 10^{-12}$ cm, the right-hand side of Eq.~(\ref{eq:LondonFieldMaxwellEquation}) is very close to $n^{\mu}_{\overline{p}}$. So to a very good approximation, $j_e^{\mu}=0$ inside a neutron star core, a conclusion drawn by~\citet{Jones1991} and which is a consequence of the proton superconductivity. We can thus interpret Eq.~(\ref{eq:LondonFieldMaxwellEquation}) as telling us how to compute $j_e^{\mu}$ given $F^{\mu\nu}_{\text{L}}$. We note that the source term on the right-hand side of Eq.~(\ref{eq:LondonFieldMaxwellEquation}) may need to be augmented by surface currents as in the original derivation of the London field by~\citet{London1950}; as in London's original derivation, the surface currents are actually currents in a boundary layer whose thickness is of order $\Lambda_*$.

We conclude our discussion on electromagnetism in the presence of vortex lines and flux tubes by finding the Lorentz force acting on the charged fluids. Using Eq.~(\ref{eq:SourcedMaxwellEquations},\ref{eq:ForceNormalFluid},\ref{eq:ForceSuperfluid},\ref{eq:EMDisplacementTensor}), the combined force acting on the charged fluids $x\in\{p,\overline{p},e,m\}$ is
\begin{align}
\sum_{x=p,\overline{p},e,m}\hspace{-2mm}f^x_{\mu}={}&\hspace{-2mm}\sum_{x=p,\overline{p},e,m}\hspace{-2mm}2n^{\nu}_x\nabla_{[\nu}\mu^x_{\mu]}+2\nabla_{[\nu}\mu^{\overline{p}}_{\mu]}\nabla_{\rho}\lambda^{\nu\rho}_{\overline{p}}
\nonumber
\\
{}&+\frac{1}{4\pi}F_{\nu\mu}\nabla_{\rho}\mathcal{H}^{\nu\rho}.
\end{align}
This can be clearly separated into three parts: the sum of the relativistic Euler equation for each fluid, the flux tube self-tension force where the electromagnetic field contribution is subtracted from the vorticity tensor, and the Lorentz force. The last of these has the standard relativistic form in a magnetizable medium and reduces to $(\nabla\times\mathbf{H}_T)\times\mathbf{B}/4\pi$ nonrelativistically.

As is suggested by our recovering the same Maxwell equations as~\citep{Carter1998,Glampedakis2011,Gusakov2016a}, the meaning of $\mathbf{H}$ is somewhat subjective -- we have discretion to choose between the field which obeys the Maxwell equation Eq.~(\ref{eq:LondonFieldMaxwellEquation}), or (up to proportionality constants) the free energy per length associated with adding a flux tube. The first option is more appropriate to electrodynamics problems e.g. ``find $\mathbf{H}$ given $\mathbf{j}_e$.'' However, in this case the solution may also involve surface current densities, as in London's original derivation of his eponymous field. 
The second option, the $\mathbf{H}_T$ defined in Eq.~(\ref{eq:HfieldThermodynamicDef}), is more appropriate to (magneto)hydrodynamics because the Lorentz force contains $(\nabla\times\mathbf{H}_T)\times\mathbf{B}/4\pi$. It also corresponds more closely to condensed matter literature where the systems are typically nondynamical, but it is also consistent with the stress tensor in~\citet{Easson1977}, which does not refer to a system in thermodynamic equilibrium and which leads to a force equation~\citep{Akgun2008}. Note that this ``thermodynamic'' field $\mathbf{H}_T$ has a curl that may be largely unrelated to the current density. Given $\mathbf{H}_T$ as a function of density and $\mathbf{B}$, we can compute equilibrium models (e.g. axisymmetric) and their perturbations (i.e. using Faraday's Law to compute changes in $\mathbf{B}$ and conservation laws to compute changes in density for given displacement field). The third option, $\mathbf{H}=\mathbf{B}$~\citep{Mendell1991b}, only works if the curl of the magnetic field due to the flux tubes/magnetized vortex lines is zero so that it has no effect on the current.

\section{Conclusion}

This article has extended the elegant convective variational principle first developed by Carter to a finite temperature, fully general relativistic multifluid system including neutron superfluidity and proton superconductivity that is appropriate for use in studying the fluid dynamics of neutron star cores. The hydrodynamics includes the proton flux tubes and magnetized neutron vortex lines, with mutual friction and vortex pinning incorporated covariantly. Viscosity and heat conduction are also included in the equations of the motion to further extend the scope of the hydrodynamics. This is the first work to incorporate all of these contributions to a relativistic, variational procedure-based hydrodynamics simultaneously, though we note that there are other, equivalent relativistic formulations based on the Landau--Khalatnikov hydrodynamics~\citep{Gusakov2016,Gusakov2016a}. Our formulation has the practical advantage of using the distinct fluid species as degrees of freedom, including distinct currents for normal fluid and superfluid baryons which were neglected in the zero temperature calculation of~\citep{Carter1998}. One advantage of this choice is that it allows sources of buoyancy among the different fluids to emerge naturally.

The averaging procedure used to determine the form of the macroscopic action from the mesoscopic theory allowed us to find an approximate effective macroscopic theory, but not an exact term-by-term match. In particular, we were forced to ignore certain terms in the averaged mesoscopic stress-energy tensor, and to drop subdominant terms in partial derivatives of the electromagnetism-vorticity master function $\Lambda_{\text{EM+V}}$, to obtain a consistent macroscopic effective theory. In principle, one could use the averaged mesoscopic theory to perform calculations instead of the effective macroscopic theory based on it. Like previous attempts at obtaining vortex energy contributions starting from a mesoscopic theory, we made use of the rotation energy cancellation lemma to eliminate cross terms between the large-scale (but measurably small) London field and the magnetic field of the flux tubes and magnetized vortex lines. We verified this lemma's applicability to the energy density, but found that the averaged mesoscopic stress-energy tensor as a whole does not satisfy the lemma.

Based on the effective macroscopic theory found by averaging the mesoscopic theory, we have clarified the interpretation of the magnetic field in a type-II superconducting neutron star core. 
Using the thermodynamic definition of the $H$-field provided by Eq.~(\ref{eq:HfieldThermodynamicDef}), our result matches that found in the condensed matter literature; that is, in the low flux tube density limit, $H$ is the first critical field $H_{c1}$ for proton superconductivity. The sourced Maxwell equations found using the effective macroscopic theory only involve the London field, which is why it has previously, and alternatively, been identified with the magnetic $H$-field in the MHD of~\citep{Carter1998,Glampedakis2011,Gusakov2016a}. We emphasize that the MHD based on both of these options is equivalent-- the difference is a matter of how terms are grouped together in the equations of motion. We are able to combine the charged fluid equations of motions into a single equation and show that the Lorentz force is the relativistic analog of $(\nabla\times\mathbf{H})\times\mathbf{B}/4\pi$, whereas in previous versions of relativistic MHD the same forces would be distributed among different terms where such an identification would be obscured.

\begin{acknowledgments}
The authors would like to thank Armen Sedrakian, Erbil G\"{u}gercino\u{g}lu, Vasiliy Dommes, Mikhail Gusakov and the referee for many useful comments. P.B.R. was supported by Cornell's Boochever Fellowship for Spring 2020.
\end{acknowledgments}

\appendix

\section{Alternative form of stress-energy tensor}
\label{app:SETensorExpansion}

Starting from Eq.~(\ref{eq:StressEnergyTensorwithViscosity}), insert the forms of $n_x^{\mu}$, $\mu^x_{\nu}$, $s^{\mu}$ and $\Theta_{\nu}$ as given in Section~\ref{sec:Conduction} and $\kappa^{\Sigma}_{\mu\nu}$ as given by Eq.~(\ref{eq:KappaDefinition}) to obtain
\begin{widetext}
\begin{align}
T^{\mu\nu}={}&(n_n\mu_n+n_p\mu_p+n_e\mu_e+n_m\mu_m+n^*_{\overline{n}}\mu^*_{\overline{n}}+n^*_{\overline{p}}\mu^*_{\overline{p}}+s^*T^*)u^{\mu}u^{\nu} +\Psi g^{\mu\nu}\nonumber +2u^{(\mu}q^{\nu)}+2n^*_{\overline{n}}\mu^*_{\overline{n}}u^{(\mu}v_{\overline{n}}^{\nu)}+2n^*_{\overline{p}}\mu^*_{\overline{p}}u^{(\mu}v_{\overline{p}}^{\nu)}
\\
{}& +2\mathcal{A}^{\overline{n}\overline{p}}n^*_{\overline{n}}n^*_{\overline{p}}v_{\overline{n}}^{(\mu}v_{\overline{p}}^{\nu)} +\frac{\mathcal{B}^s}{T^*}q^{\mu}q^{\nu}+\mathcal{B}^{\overline{n}}(n^*_{\overline{n}})^2v_{\overline{n}}^{\mu}v_{\overline{n}}^{\nu}+\mathcal{B}^{\overline{p}}(n^*_{\overline{p}})^2v_{\overline{p}}^{\mu}v_{\overline{p}}^{\nu}+\frac{1}{4\pi}\mathcal{K}^{\mu\rho}F^{\nu}_{\ \rho}+\sum_{\overline{x}}\lambda^{\mu\rho}_{\overline{x}}w^{\nu}_{\overline{x}\rho}+\sum_{\Sigma}\tau^{\mu\nu}_{\Sigma} \nonumber
\\
={}&(\Psi-\Lambda)u^{\mu}u^{\nu} + \Psi g^{\mu\nu} + \left(\frac{\mathcal{B}^s}{T^*}q^{\mu}q^{\nu}+2u^{(\mu}q^{\nu)}\right)+\frac{1}{4\pi}\mathcal{K}^{\mu\rho}F^{\nu}_{\ \rho}+\sum_{\overline{x}=\overline{n},\overline{p}}\lambda^{\mu\rho}_{\overline{x}}w^{\nu}_{\overline{x}\rho}+\sum_{\Sigma}\tau^{\mu\nu}_{\Sigma} \nonumber
\\
{}& +\left( \left[\mathcal{B}^{\overline{n}}n^*_{\overline{n}}v_{\overline{n}}^{\mu}+\mathcal{A}^{\overline{n}\overline{p}}n^*_{\overline{p}}v_{\overline{p}}^{\mu}\right]n^*_{\overline{n}}v_{\overline{n}}^{\nu}+2\mu^*_{\overline{n}}u^{(\mu}n^*_{\overline{n}}v_{\overline{n}}^{\nu)}\right)+\left( \left[\mathcal{B}^{\overline{p}}n^*_{\overline{p}}v_{\overline{p}}^{\mu}+\mathcal{A}^{\overline{n}\overline{p}}n^*_{\overline{n}}v_{\overline{n}}^{\mu}\right]n^*_{\overline{p}}v_{\overline{p}}^{\nu}+2\mu^*_{\overline{p}}u^{(\mu}n^*_{\overline{p}}v_{\overline{p}}^{\nu)}\right) \nonumber
\\
{}&+\left(\frac{\mathcal{B}^s}{T^*}q^2+\mathcal{B}^{\overline{n}}(n^*_{\overline{n}})^2v^2_{\overline{n}}+\mathcal{B}^{\overline{p}}(n^*_{\overline{p}})^2v^2_{\overline{p}}+2\mathcal{A}^{\overline{n}\overline{p}}n^*_{\overline{n}}n^*_{\overline{p}}v^{\rho}_{\overline{n}}v^{\overline{p}}_{\rho}\right) u^{\mu}u^{\nu}.
\label{eq:StressEnergyTensorwithViscosityExpansion}
\end{align}
\end{widetext}
The first term proportional to $u^{\mu}u^{\nu}$ is found by rewriting $\Psi$, given by Eq.~(\ref{eq:GeneralizedPressurewithViscosity}), as
\begin{align}
\Psi={}&\Lambda+n_n\mu_n+n_p\mu_p+n_e\mu_e+n_m\mu_m+n_{\overline{n}}\mu_{\overline{n}}+n_{\overline{p}}\mu_{\overline{p}} \nonumber
\\
{}&+s^*T^*-\frac{\mathcal{B}^s}{T^*}q^2-\frac{1}{2}\sum_{\Sigma}\tau^{\rho}_{\Sigma\rho}.
\label{eq:GeneralizedPressurewithViscosityExpansion}
\end{align}
The similarity of this form of $T^{\mu\nu}$ to that of a single perfect fluid~\citep{Weinberg1972} is now evident; this form is effectively the same as that for such a fluid, plus electromagnetism, vorticity and viscosity, with differences depending on the relative motion of heat and the superfluids separated out.

\section{Full details of mesoscopic stress-energy tensor and averaging procedure}
\label{app:MesoscopicAveraging}

We continue from the main text immediately following the introduction of the mesoscopic Lagrangian and stress-energy tensor Eq.~(\ref{eq:MesoscopicSETensor}). On the mesoscopic scale, currents around vortex lines/flux tubes are represented within the currents $\tilde{n}_{\overline{x}}^{\mu}$, not by using the vorticity tensors $w^{\mu\nu}_{\overline{x}}$ as is the case in the macroscopic dynamics. We incorporate these purely ``mesoscale'' currents $\delta v^{\mu}_{\overline{x}}$ by defining $\tilde{n}_{\overline{x}}^{\mu}$ as
\begin{equation}
\tilde{n}^{\mu}_{\overline{x}}\equiv \tilde{n}_{\overline{x}}\gamma(\delta v^2_{\overline{x}})(u^{\mu}_{\overline{x}}+\delta v^{\mu}_{\overline{x}}), \quad \overline{x}=\overline{n},\overline{p},
\label{eq:MesoscopicRelativeCurrents}
\end{equation}
where $u^{\mu}_{\overline{x}}$ is defined as in Eq.~(\ref{eq:SuperfluidCurrent}). $\tilde{n}^{\mu}_{\overline{x}}$ satisfies the normalization condition $-\tilde{n}^{\mu}_{\overline{x}}\tilde{n}_{\mu}^{\overline{x}}=\tilde{n}_{\overline{x}}^2$, since $u^{\mu}_{\overline{x}}\delta v_{\mu}^{\overline{x}}=0$ as a result of the approximation that the vortex lines/flux tubes move with their respective superfluid. We enforce that the $\delta v^{\mu}_{\overline{x}}$ average to zero over scales larger than the typical cross-section of a vortex line/flux tube, and that any large-scale average part of a relative velocity between the normal fluid and superfluids is included in $u^{\mu}_{\overline{x}}$. $\tilde{n}_{\overline{x}}$ is the number density of species $\overline{x}$ measured in the frame comoving with the total current of that species, and it is related to the number density $n_{\overline{x}}$ in the frame of the bulk flow (the frame of $u^{\mu}_{\overline{x}}$) by
\begin{equation}
n_{\overline{x}}=-u^{\mu}_{\overline{x}}\tilde{n}_{\mu}^{\overline{x}}=\tilde{n}_{\overline{x}}\gamma(\delta v^2_{\overline{x}}).
\label{eq:BulkFrameNumberDensity}
\end{equation}
Note that the for the normal fluid species, $\tilde{n}_{x}^{\mu}=n_{x}^{\mu}$.

We first expand out the terms in Eq.~(\ref{eq:MesoscopicSETensor}), removing any dependence on the vortex line/flux tube mesoscale currents from the master function $\tilde{\Lambda}$ and replace it with $\tilde{\Lambda}_0$, which represents only the internal energy of the fluid and the kinetic energy of macroscopic currents. Following~\cite{Prix2000}, we write
\begin{align}
\tilde{\Lambda}(\tilde{n}_x^2,\tilde{\alpha}^2_{xy})={}&\tilde{\Lambda}_0(n^2_{x},\alpha^2_{xy}) \nonumber
\\
{}&+\sum_{x,y}\left(\frac{\partial\tilde{\Lambda}_0}{\partial n_x^2}\mathfrak{d} (n_x^2)+\frac{\partial\tilde{\Lambda}_0}{\partial \alpha_{xy}^2}\mathfrak{d} (\alpha_{xy}^2)\right)
\label{eq:MesoscopicMasterFunctionExpansion}
\end{align}
where the ``$0$'' subscript denotes the master function with the $\delta v^{\mu}_{\overline{x}}$ removed, and where
\begin{align}
\mathfrak{d}(n_x^2)={}&\tilde{n}_x^2-n^2_x=\begin{cases}-n_{\overline{x}}^2\delta v^2_{\overline{x}}, & x=\overline{n},\overline{p}, 
\nonumber
\\
0, & \text{otherwise},
\end{cases}
\\
\mathfrak{d}(\alpha_{xy}^2)={}&\tilde{\alpha}_{xy}^2-\alpha^2_{xy}=-\tilde{n}^{\sigma}_x\tilde{n}^y_{\sigma}+n^{\sigma}_xn^y_{\sigma}
\nonumber
\\
={}&\begin{cases} n_pn_{\overline{n}}v^{\sigma}_{\overline{n}}\delta v_{\sigma}^{\overline{n}}, & x=\overline{n}, y=p, 
\\
n_nn_{\overline{n}}v^{\sigma}_{\overline{n}}\delta v_{\sigma}^{\overline{n}}, & x=\overline{n}, y=n, 
\\
n_nn_{\overline{p}}v^{\sigma}_{\overline{p}}\delta v_{\sigma}^{\overline{p}}, & x=\overline{p}, y=n, 
\\
n_pn_{\overline{p}}v^{\sigma}_{\overline{p}}\delta v_{\sigma}^{\overline{p}}, & x=\overline{p}, y=p, 
\\
n_{\overline{n}}n_{\overline{p}}\left[\gamma(v^2_{\overline{n}})(v^{\sigma}_{\overline{p}}-v^{\sigma}_{\overline{n}})\delta v^{\overline{p}}_{\sigma}\right.
\nonumber
\\
\qquad\quad+\gamma(v^2_{\overline{p}})(v^{\sigma}_{\overline{n}}-v^{\sigma}_{\overline{p}})\delta v^{\overline{n}}_{\sigma}
\nonumber
\\
\qquad\quad\left.-\delta v_{\sigma}^{\overline{n}}\delta v^{\sigma}_{\overline{p}}\right], & x=\overline{n},y=\overline{p}, 
\\
0, & \text{otherwise},
\end{cases}
\end{align}
where $u^{\mu}\delta v_{\mu}^{\overline{x}}=-v^{\mu}_{\overline{x}}\delta v_{\mu}^{\overline{x}}$ was used. For the normal fluids, $\tilde{n}^{\mu}_x$ simply equals $n^{\mu}_x$, since the normal fluid currents are unchanged by the inclusion of the mesoscale currents. We have kept only terms that are order $\delta v^2_{\overline{x}}$ in the mesoscopic-scale velocities. 

The partial derivatives of $\tilde{\Lambda}_0$ in Eq.~(\ref{eq:MesoscopicMasterFunctionExpansion}) are identified with the entrainment coefficients as defined in the macroscopic theory (Eq.~(\ref{eq:ChemPotCoeffDefs})). We use the physical arguments presented at the end of Section~\ref{sec:RelationtoPhysParams} to reduce the number of entrainment coefficients i.e. $\mathcal{A}^{np}=\mathcal{A}^{n\overline{p}}=\mathcal{A}^{p\overline{n}}=\mathcal{A}^{\overline{np}}$. Hence $\tilde{\Lambda}(\tilde{n}_x^2,\tilde{\alpha}^2_{xy})$ becomes
\begin{align}
\tilde{\Lambda}(\tilde{n}_x^2,\tilde{\alpha}^2_{xy})={}&\tilde{\Lambda}_0 + \frac{1}{2}\mathcal{B}^{\overline{n}}n^{2}_{\overline{n}}\delta v^2_{\overline{n}}+\frac{1}{2}\mathcal{B}^{\overline{p}}n^{2}_{\overline{p}}\delta v^2_{\overline{p}} \nonumber
\\
{}&-\mathcal{A}^{\overline{np}}\left[n_pn_{\overline{n}}v^{\sigma}_{\overline{n}}\delta v_{\sigma}^{\overline{n}}+n_nn_{\overline{p}}v^{\sigma}_{\overline{p}}\delta v_{\sigma}^{\overline{p}}\right.
\nonumber
\\
{}&\qquad\quad+n_{\overline{n}}n_{\overline{p}}\left(\gamma(v^2_{\overline{n}})(v^{\sigma}_{\overline{p}}-v^{\sigma}_{\overline{n}})\delta v^{\overline{p}}_{\sigma}\right.
\nonumber
\\
{}&\qquad\quad\left.\left.+\gamma(v^2_{\overline{p}})(v^{\sigma}_{\overline{n}}-v^{\sigma}_{\overline{p}})\delta v^{\overline{n}}_{\sigma}-\delta v_{\sigma}^{\overline{n}}\delta v^{\sigma}_{\overline{p}}\right)\right]
\nonumber
\\
{}&-\mathcal{A}^{n\overline{n}}n_nn_{\overline{n}}v^{\sigma}_{\overline{n}}\delta_{\sigma}^{\overline{n}}-\mathcal{A}^{p\overline{p}}n_pn_{\overline{p}}v^{\sigma}_{\overline{p}}\delta_{\sigma}^{\overline{p}}.
\end{align}

It is also convenient to define a mesoscale superfluid neutron canonical momentum covector
\begin{equation}
\delta\pi^{\overline{n}}_{\mu}\equiv\mathcal{B}^{\overline{n}}n_{\overline{n}}\delta v^{\overline{n}}_{\mu}+\mathcal{A}^{\overline{np}}n_{\overline{p}}\delta v^{\overline{p}}_{\mu},
\label{eq:DeltaPiN}
\end{equation}
which is simply the part of $\tilde{\mu}^{\overline{n}}_{\mu}$ that depends on the mesoscale velocities $\delta v^{\overline{x}}_{\mu}$. Note that, because the neutron superfluid is not coupled to the electromagnetic field, we could also have called $\delta\pi^{\overline{n}}_{\mu}$ simply $\delta\mu^{\overline{n}}_{\mu}$. The definition of $\delta\pi^{\overline{n}}_{\mu}$ will simply some terms of the stress-energy tensor immensely by canceling terms which couple $\delta v^{\overline{n}}_{\mu}$ and $\delta v^{\overline{p}}_{\mu}$.

Combining these definitions and results with Eq.~(\ref{eq:GusakovEntrainmentParameters}) and Eq.~(\ref{eq:NFNeutronMuVec2}--\ref{eq:ChemPotDefT}), the mesoscopic stress-energy tensor resulting from  Eq.~(\ref{eq:MesoscopicMasterFunctionExpansion}) is
\begin{align}
\tilde{T}^{\mu\nu}={}&\sum_xn_x^{\mu}\mu^{\nu}_x
+\frac{1}{\mathcal{B}^{\overline{n}}}\delta \pi^{\mu}_{\overline{n}}\delta \pi^{\nu}_{\overline{n}}
+\frac{1}{Y_{pp}}n^2_{\overline{p}}\delta v^{\mu}_{\overline{p}}\delta v^{\nu}_{\overline{p}}
\nonumber
\\
{}&+\frac{2}{\mathcal{B}^{\overline{n}}}\mu^{(\mu}_{\overline{n}}\delta\pi^{\nu)}_{\overline{n}}-2en_{\overline{p}}A^{(\mu}_{\text{L}}\delta v^{\nu)}_{\overline{p}}
\nonumber
\\
{}&+\Bigg[\tilde{\Lambda}_0+\sum_xn_x\mu_x-\frac{1}{2\mathcal{B}^{\overline{n}}}\delta\pi^2_{\overline{n}}-\frac{1}{2Y_{pp}}n^2_{\overline{p}}\delta v^2_{\overline{p}}
\nonumber
\\
{}&\qquad-\frac{1}{\mathcal{B}^{\overline{n}}}\mu^{\sigma}_{\overline{n}}\delta\pi_{\sigma}^{\overline{n}}+en_{\overline{p}}A^{\sigma}_{\text{L}}\delta v_{\sigma}^{\overline{p}}\Bigg]g^{\mu\nu}
\nonumber
\\
{}&+\frac{1}{4\pi}\left(\tilde{F}^{\mu\rho}\tilde{F}^{\nu}_{\ \rho}-\frac{1}{4}\tilde{F}^{\sigma\rho}\tilde{F}_{\sigma\rho}g^{\mu\nu}\right).
\label{eq:MesoscopicSETensor2}
\end{align}
where $n_x^{\mu}$ and $\mu^{\nu}_x$ are the number current and conjugate momenta as defined in terms of the macroscopic currents in Section~\ref{sec:RelationtoPhysParams}. We have used Eq.~(\ref{eq:NFNeutronMuVec2}--\ref{eq:ChemPotDefT}) plus the definition of the London four-potential
\begin{equation}
A_{\mu}^{\text{L}}\equiv-\frac{1}{e}\left[\left(\mu^*_{\overline{p}}+\frac{Y_{np}}{Y_{pp}}\mu^*_{\overline{n}}\right)u_{\mu}+\frac{n^*_{\overline{p}}}{Y_{pp}}v_{\mu}^{\overline{p}}\right],
\label{eq:LondonFourPotential}
\end{equation}
in writing $\tilde{T}^{\mu\nu}$ this way. The non-electromagnetic part of this stress-energy tensor has been separated into three parts: those which only depend on large-scale flows, those which depend on mixed large-scale--mesoscale flows, and those which only depend on mesoscale flows. We now want to make this separation for the electromagnetic part of $\tilde{T}^{\mu\nu}$, and also to determine the form of the mesoscale superconducting proton velocity $\delta v^{\mu}_{\overline{p}}$ and the mesoscale superfluid neutron canonical momentum $\delta\pi^{\overline{n}}_{\mu}$.

\subsection{Mesoscopic treatment of vortex lines, flux tubes and magnetic fields}

We first calculate $\delta\pi^{\overline{n}}_{\mu}$. In its rest frame, the canonical 3-momenta for a single quantized vortex line will have the following form~\cite{Glampedakis2011}
\begin{equation}
\delta\boldsymbol{\pi}_{\overline{x}}=\frac{\hbar}{2\varpi}\hat{\mathbf{e}}_{\varphi},
\label{eq:VortexLinePi}
\end{equation}
where $\varpi$ and $\varphi$ are the cylindrical radius and azimuthal angle in a coordinate system in which the vortex line lies along the $z$-axis. For a single vortex line labeled $a$, this is expressed in covariant form as
\begin{equation}
\delta\pi_{\mu}^{\overline{x},a}=\frac{\hbar}{2\varpi^2_{\overline{x},a}}\varepsilon_{\mu\nu\alpha\beta}u_{L,\overline{x}}^{\nu}\hat{t}^{\alpha}_{\overline{x},a}\varpi^{\beta}_{\overline{x},a},
\label{eq:VortexLinePiCovariant}
\end{equation}
where $\varpi^{\beta}_{\overline{x},a}$ points radially outward from the vortex line/flux tube and $\hat{t}^{\alpha}_{\overline{x},a}$ is the unit tangent vector to the vortex line/flux tube. This form is also consistent with Eq.~(\ref{eq:QuantizationofW}--\ref{eq:VorticityVector2}) and the definition of $w^{\overline{x}}_{\mu\nu}$, since for a single vortex line/flux tube in its rest frame we have~\cite{Andersson2011}
\begin{equation}
\varepsilon^{\sigma\alpha\mu\nu}u^{L,\overline{x}}_{\sigma}\nabla_{\mu}\delta\pi_{\nu}^{\overline{x},a}=\frac{h}{2}\delta^2(\varpi_{\overline{x},a})\hat{t}^{\alpha}_{\overline{x},a},
\label{eq:VortexLineCurl}
\end{equation}
where $\delta^2(\varpi_a)$ is a two-dimensional delta function at the position of the vortex line labeled ``$a$''. So whenever $\delta\pi^{\overline{n}}_{\mu}$ appears in Eq.~(\ref{eq:MesoscopicSETensor2}), it is replaced with a sum of  Eq.~(\ref{eq:VortexLinePiCovariant}) over the line labels $a$ and with $u^{L,\overline{x}}_{\sigma}=u^{\overline{x}}_{\sigma}$ since we work in the approximation that the vortex lines move with the superfluid.

We next find the form of the electromagnetic tensor and $\delta v^{\mu}_{\overline{p}}$. We use the splitting Eq.~(\ref{eq:FmunuMesoscopicDivision}) for $\tilde{F}^{\mu\nu}$. Along with the split of $\tilde{F}^{\mu\nu}$ in Eq.~(\ref{eq:FmunuMesoscopicDivision}), we split the Maxwell equation Eq.~(\ref{eq:MesoscopicMaxwellEquation}) into large-scale and mesoscale parts, with the mesoscale proton current $\propto\delta v^{\mu}_{\overline{p}}$ sourcing the mesoscale fields and the other parts of the current sourcing the large-scale field (the London field). Using Eq.~(\ref{eq:NormalFluidCurrent},\ref{eq:SuperfluidCurrent},\ref{eq:MesoscopicRelativeCurrents}) and assuming local charge neutrality
\begin{equation}
\nabla_{\nu}\tilde{F}^{\mu\nu}_{\text{L}}=4\pi e n^*_{\overline{p}}v^{\mu}_{\overline{p}}, \quad \nabla_{\nu}(\tilde{F}^{\mu\nu}_{\overline{p}}+\tilde{F}^{\mu\nu}_{\overline{n}})=4\pi e n_{\overline{p}}\delta v^{\mu}_{\overline{p}}.
\label{eq:MesoscopicMaxwellEquationSplit}
\end{equation}
We can also split the four-potential $\tilde{A}_{\mu}$ into large-scale and mesoscale contributions $A^{\text{L}}_{\mu}$ and $\delta A_{\mu}=\delta A^{\overline{p}}_{\mu}+\delta A^{\overline{n}}_{\mu}$ respectively where $\tilde{F}_{\mu\nu}^{\overline{x}}=2\nabla_{[\mu}\delta A^{\overline{x}}_{\nu]}$. Then, defining a mesoscale canonical momentum covector for the protons analogously to Eq.~(\ref{eq:DeltaPiN})
\begin{equation}
\delta\pi^{\overline{p}}_{\mu}\equiv\mathcal{B}^{\overline{p}}n_{\overline{p}}\delta v^{\overline{p}}_{\mu}+\mathcal{A}^{\overline{np}}n_{\overline{n}}\delta v^{\overline{n}}_{\mu}+e\delta A_{\mu},
\end{equation}
and combining it with Eq.~(\ref{eq:DeltaPiN}) to eliminate $\delta v_{\mu}^{\overline{n}}$, we obtain
\begin{align}
\frac{1}{Y_{pp}}n_{\overline{p}}\delta v^{\overline{p}}_{\mu}+e\delta A_{\mu}=\delta\pi^{\overline{p}}_{\mu}+\frac{Y_{np}}{Y_{pp}}\delta\pi^{\overline{n}}_{\mu},
\label{eq:LondonEquationDerivation}
\end{align}
or, eliminating $\delta v^{\overline{p}}_{\mu}$ with Eq.~(\ref{eq:MesoscopicMaxwellEquationSplit}) and using $\Phi_{\overline{p}}=h/(2e)$ and $\Phi_{\overline{n}}=Y_{np}\Phi_{\overline{p}}/Y_{pp}$ and $\Lambda_*=(4\pi e^2Y_{pp})^{-1/2}$,
\begin{align}
\sum_{\overline{x}}\left(\nabla_{\nu}\tilde{F}^{\mu\nu}_{\overline{x}}-\frac{1}{\Lambda^2_*}\delta A_{\overline{x}}^{\mu}\right)=\frac{2}{h\Lambda^2_*}\sum_{\overline{x}}\Phi_{\overline{x}}\delta\pi^{\mu}_{\overline{x}}.
\label{eq:LondonEquationDerivation2}
\end{align}

Eq.~(\ref{eq:LondonEquationDerivation}) is used to obtain the London equation for proton flux tubes or magnetized neutron vortex lines. We will assume that the magnetic fields due to the flux tubes or magnetized vortex lines will have negligible overlap, and so we can fix $\overline{x}$ to be either $\overline{p}$ or $\overline{n}$ and drop the other contribution to Eq.~(\ref{eq:LondonEquationDerivation2}). As a consequence of Eq.~(\ref{eq:NoElectricField}) with $u^{L,\overline{x}}_{\mu}=u^{\overline{x}}_{\mu}$, we have
\begin{equation}
\tilde{F}^{\mu\nu}_{\overline{x}}=-\varepsilon^{\mu\nu\alpha\beta}u^{\overline{x}}_{\alpha}\delta B^{\overline{x}}_{\beta},
\label{eq:FmunufromB}
\end{equation}
where $\delta B_{\mu}^{\overline{x}}$ is the magnetic field due to the flux tubes/vortex lines measured in their rest frame. Then contracting with $\varepsilon^{\rho\sigma\eta\mu}u_{\sigma}^{\overline{x}}\nabla_{\eta}$, ignoring spatial curvature (which is a very good approximation for microscopic structures like vortex lines and flux tubes), assuming the flux tubes/vortex lines move with their respective superfluid and using Eq.~(\ref{eq:VortexLineCurl}), we obtain
\begin{align}
\nabla^2\delta B^{\mu}_{\overline{x}}-\frac{\delta B^{\mu}_{\overline{x}}}{\Lambda_*^2}=-\frac{\Phi_{\overline{x}}}{\Lambda_*^2}\sum_a\hat{t}^{\mu}_{\overline{x},a}\delta^2(\varpi_{\overline{x},a}).
\label{eq:LondonEquation}
\end{align}
This is the London equation in the comoving frame, whose solutions are the magnetic fields for flux tubes/magnetized vortex lines. $\nabla^2$ is the usual flat space Laplacian, and $\Lambda_*$ is the London length. The right-hand side of the equation is a sum over flux tubes/magnetized vortex lines labeled by index $a$ and represented as two-dimensional delta functions. The solutions in the comoving frame for single flux tubes/magnetized vortex lines take the familiar form~\cite{Prix2000}
\begin{equation}
\delta B_{\overline{x},a}^{\mu}=\hat{t}_{\overline{x},a}^{\mu}\frac{\Phi_{\overline{x}}K_0(\varpi_{\overline{x},a}/\Lambda_*)}{2\pi\Lambda_*^2 x^{\overline{x}}_0K_1(x^{\overline{x}}_0)}\equiv \delta B_{\overline{x},a}(\varpi_{\overline{x},a})\hat{t}_{\overline{x},a}^{\mu}, 
\label{eq:IndividualFTVLBField}
\end{equation}
where $K_n(x)$ is the modified Bessel function of the second kind of order $n$ and  $x^{\overline{x}}_0\equiv\xi_{x}/\Lambda_*$. Flux in the core of the flux tubes/vortex lines, included in e.g.~\cite{Prix2000}, is ignored here.

Using Eq.~(\ref{eq:FmunufromB},\ref{eq:IndividualFTVLBField}), the mesoscale electromagnetic field tensors are
\begin{equation}
\tilde{F}_{\mu\nu}^{\overline{x}}=-\sum_a\varepsilon_{\mu\nu\alpha\beta}u^{\alpha}_{\overline{x}}\delta B^{\beta}_{\overline{x},a}.
\label{eq:FmunuMesoscopicSum}
\end{equation}
Hence we can replace $\delta v^{\mu}_{\overline{p}}$ in Eq.~(\ref{eq:MesoscopicSETensor2}) using the mesoscale Maxwell equation Eq.~(\ref{eq:MesoscopicMaxwellEquationSplit}) with the gradient of Eq.~(\ref{eq:FmunuMesoscopicSum}). 

\subsection{Averaging the mesoscopic stress-energy tensor}
\label{app:AveragingSETensor}							 

We now average the mesoscopic stress-energy tensor, Eq.~(\ref{eq:MesoscopicSETensor2}). As noted before, the non-electromagnetic part of this equation consists of purely large-scale flow terms, purely mesoscale flow terms, and mixed terms. Though $\delta \pi^{\overline{n}}_{\mu}$ does not average to zero, we will absorb any effect of the large scale--small scale superfluid neutron momentum term $\propto\mu_{\overline{n}}^{(\mu}\delta\pi^{\nu)}_{\overline{n}}$ or $\mu^{\sigma}_{\overline{n}}\delta\pi_{\sigma}^{\overline{n}}g^{\mu\nu}$ in $\tilde{T}^{\mu\nu}$ into the purely small-scale superfluid neutron momentum terms using a cutoff length. We thus treat $\langle\delta \pi^{\overline{n}}_{\mu}\rangle=0$ and so both of the aforementioned terms vanish upon averaging. The purely large-scale flow terms $\sum_xn^{\mu}_x\mu^{\nu}_x+(\tilde{\Lambda}_0+\sum_xn_x\mu_x)g^{\mu\nu}$, do not change upon averaging, and have exact matches in the macroscopic stress-energy tensor as we will demonstrate in the next section. We label the remaining terms $\Delta\tilde{T}^{\mu\nu}$:
\begin{align}
\Delta\tilde{T}^{\mu\nu}={}&\frac{1}{\mathcal{B}^{\overline{n}}}\left[\delta\pi^{\mu}_{\overline{n}}\delta\pi^{\nu}_{\overline{n}}-\frac{1}{2}\delta\pi^2_{\overline{n}}g^{\mu\nu}\right]
\nonumber
\\
{}&+\frac{n^2_{\overline{p}}}{Y_{pp}}\left[\delta v^{\mu}_{\overline{p}}\delta v^{\nu}_{\overline{p}}-\frac{1}{2}\delta v^2_{\overline{p}}g^{\mu\nu}\right]
\nonumber
\\
{}&-2eA_{\text{L}}^{(\mu}\delta v^{\nu)}_{\overline{p}}+eA^{\text{L}}_{\sigma}\delta v^{\sigma}_{\overline{p}}g^{\mu\nu}
\nonumber
\\
{}&+\frac{1}{4\pi}\Bigg[\tilde{F}^{\mu\rho}_{\overline{p}}\tilde{F}^{\nu}_{\overline{p}\rho}+\tilde{F}^{\mu\rho}_{\overline{n}}\tilde{F}^{\nu}_{\overline{n}\rho}+\tilde{F}^{\mu\rho}_{\text{L}}\tilde{F}^{\nu}_{\text{L}\rho}
\nonumber
\\
{}&\qquad+2\tilde{F}^{(\mu}_{\ \overline{p}\rho}\tilde{F}^{\nu)\rho}_{\text{L}}+2\tilde{F}^{(\mu}_{\overline{n}\ \rho}\tilde{F}^{\nu)\rho}_{\text{L}}+2\tilde{F}^{(\mu}_{\overline{p}\ \rho}\tilde{F}^{\nu)\rho}_{\overline{n}}\Bigg]
\nonumber
\\
{}&-\frac{g^{\mu\nu}}{16\pi}\Bigg[\tilde{F}^{\sigma\rho}_{\overline{p}}\tilde{F}_{\sigma\rho}^{\overline{p}}+\tilde{F}^{\sigma\rho}_{\overline{n}}\tilde{F}_{\sigma\rho}^{\overline{n}}+\tilde{F}^{\sigma\rho}_{\text{L}}\tilde{F}_{\sigma\rho}^{\text{L}}
\nonumber
\\
{}&\qquad+2\tilde{F}^{\sigma\rho}_{\overline{p}}\tilde{F}_{\sigma\rho}^{\text{L}}+2\tilde{F}^{\sigma\rho}_{\overline{n}}\tilde{F}_{\sigma\rho}^{\text{L}}+2\tilde{F}^{\sigma\rho}_{\overline{p}}\tilde{F}_{\sigma\rho}^{\overline{n}}\Bigg].
\label{eq:MesoscopicSETensor3}
\end{align}

We now integrate Eq.~(\ref{eq:MesoscopicSETensor3}) over a surface of size $\sim \ell^2$, then replace the quantities in the mesoscopic stress-energy tensor with averaged quantities. First we consider
\begin{equation}
\tilde{T}^{\mu\nu}_{\pi}\equiv\frac{1}{\mathcal{B}^{\overline{n}}}\left[\delta\pi^{\nu}_{\overline{n}}\delta\pi^{\nu}_{\overline{n}}-\frac{1}{2}\delta\pi^2_{\overline{n}}g^{\mu\nu}\right].
\end{equation}
$\delta\pi_{\overline{n}}^{\mu}$ is replaced by a sum over individual vortex lines. For this purpose, we rewrite Eq.~(\ref{eq:VortexLinePi}) as 
\begin{equation}
\delta\pi_{\overline{n},a}^{\mu}=\frac{\hbar}{2\varpi_{\overline{n},a}}(-\sin\varphi_{\overline{n},a}\hat{\zeta}_{\overline{x},a}^{\mu}+\cos\varphi_{\overline{n},a}\hat{\eta}_{\overline{x},a}^{\mu})
\end{equation} 
where $\varphi_{\overline{n},a}$ is an azimuthal angle measured around a vortex line labeled $a$ and $\hat{\zeta}_{\overline{n},a}^{\mu}$ and $\hat{\zeta}_{\overline{n},a}^{\mu}$ are mutually orthogonal unit vectors which are also orthogonal to both $u^{\mu}_{\overline{n}}$ and $\hat{t}^{\mu}_{\overline{n},a}$.  When integrating over a surface area $\sim\ell^2$ in the plane perpendicular to the average vortex line tangent vector $\hat{t}^{\mu}_{\overline{x}}=\langle\hat{t}^{\mu}_{\overline{x},a}\rangle$, the sum over different vortex lines is replaced with a multiplication by the areal density of vortex lines $\mathcal{N}_{\overline{n}}$. We also replace the other vectors with their average values over the area of integration $\hat{\zeta}^{\mu}_{\overline{n}}=\langle\hat{\zeta}^{\mu}_{\overline{n},a}\rangle$, $\hat{\eta}^{\mu}_{\overline{n}}=\langle\hat{\eta}^{\mu}_{\overline{n},a}\rangle$. This means we only need to consider the integral for a single vortex line, integrating radially from the coherence length (since we're ignoring the core) to a cutoff radius $r^{\text{cut}}_n$:
\begin{align}
\langle\tilde{T}^{\mu\nu}_{\pi}\rangle={}&\mathcal{N}_{\overline{n}}\frac{\hbar^2}{4\mathcal{B}^{\overline{n}}}\int^{2\pi}_0d\varphi_{\overline{n}}\int^{r_n^{\text{cut}}}_{\xi_n} \frac{d\varpi_{\overline{n}}}{\varpi_{\overline{n}}} \nonumber 
\\
{}&\quad\times\Bigg(\sin^2\varphi_{\overline{n}}\hat{\zeta}^{\mu}_{\overline{n}}\hat{\zeta}^{\nu}_{\overline{n}}+\cos^2\varphi_{\overline{n}}\hat{\eta}_{\overline{n}}^{\mu}\hat{\eta}_{\overline{n}}^{\nu} \nonumber 
\\
{}&\quad\qquad-2\sin\varphi_{\overline{n}}\cos\varphi_{\overline{n}}\hat{\zeta}_{\overline{n}}^{(\mu}\hat{\eta}_{\overline{n}}^{\nu)}-\frac{1}{2}g^{\mu\nu}\Bigg) \nonumber
\\
={}&\mathcal{N}_{\overline{n}}\frac{\pi\hbar^2}{8\mathcal{B}^{\overline{n}}}\ln\left(\frac{r^{\text{cut}}_n}{\xi_n}\right)\left(\hat{\zeta}_{\overline{n}}^{\mu}\hat{\zeta}_{\overline{n}}^{\nu}+\hat{\eta}_{\overline{n}}^{\mu}\hat{\eta}_{\overline{n}}^{\nu}-g^{\mu\nu}\right).
\label{eq:VLPiIntegral}
\end{align}
The neutron vortex line cutoff radius $r^{\text{cut}}_n$ accounts for the long-range nature of the vortex lines and incorporates the effect of interactions between them. $\langle\tilde{T}^{\mu\nu}_{\pi}\rangle$ thus absorbs the $\propto\delta\pi_{\overline{n}}\mu_{\overline{n}}$ terms in Eq.~(\ref{eq:MesoscopicSETensor2}) that we argued average to zero earlier. Based on~\citet{Tkachenko1966} and~\citet{Sonin2016}, we expect $r^{\text{cut}}_n\approx 0.0712(\xi_n\mathcal{N}_{\overline{n}})^{-1}$. Additionally we have
\begin{equation}
g^{\mu\nu}=-u_{\overline{x}}^{\mu}u_{\overline{x}}^{\nu}+\hat{\zeta}_{\overline{x}}^{\mu}\hat{\zeta}_{\overline{x}}^{\nu}+\hat{\eta}_{\overline{x}}^{\mu}\hat{\eta}_{\overline{x}}^{\nu}+\hat{t}_{\overline{x}}^{\mu}\hat{t}_{\overline{x}}^{\nu},
\label{eq:MetricUnitVectors}
\end{equation}
so we can write
\begin{align}
\langle\tilde{T}^{\mu\nu}_{\pi}\rangle=\mathcal{N}_{\overline{n}}\frac{\pi\hbar^2}{8\mathcal{B}^{\overline{n}}}\ln\left(\frac{0.0712}{\mathcal{N}_{\overline{n}}\xi^2_n}\right)\left(u^{\mu}_{\overline{n}}u^{\nu}_{\overline{n}}-\hat{t}^{\mu}_{\overline{n}}\hat{t}^{\nu}_{\overline{n}}\right),
\label{eq:AMTSFNeutronMomentumEnergy}
\end{align}
Eq.~(\ref{eq:AMTSFNeutronMomentumEnergy}) has the general form of the stress-energy tensor for a single string along $\hat{t}^{\mu}_{\overline{n}}$~\cite{Vilenkin1994}.

Next consider
\begin{align}
\tilde{T}^{\mu\nu}_{v}\equiv\frac{n^{2}_{\overline{p}}}{Y_{pp}}\Bigg[{}&\delta v^{\mu}_{\overline{p}}\delta v^{\nu}_{\overline{p}}-\frac{1}{2}\delta v^2_{\overline{p}}g^{\mu\nu}\Bigg].
\end{align}
At this point, we neglect the interactions between different flux tubes/vortex lines and consider only their self-energy contributions. This allows us to simplify in our averaging integral and again integrate only over a surface locally perpendicular to the flux lines/vortex tubes, then multiply by the relevant areal number density $\mathcal{N}_{\overline{x}}$. Using Eq.~(\ref{eq:MesoscopicMaxwellEquationSplit},\ref{eq:FmunufromB}) and ignoring spatial curvature, we find an integral very similar to Eq.~(\ref{eq:VLPiIntegral})
\begin{align}
\langle \tilde{T}^{\mu\nu}_{v}\rangle={}&\sum_{\overline{x}}\frac{\mathcal{N}_{\overline{x}}\Phi_{\overline{x}}^2}{16\pi^3\Lambda^4_*}\int^{2\pi}_{0}d\varphi_{\overline{x}}\int^{r^{\text{cut}}_x}_{\xi_x}\frac{d\varpi_{\overline{x}}\varpi_{\overline{x}}K_1^2(\varpi_{\overline{x}}/\Lambda_*)}{[x_0^{\overline{x}}K_1(x_0^{\overline{x}})]^2} \nonumber
\\
{}&\quad\times\Bigg(\sin^2\varphi_{\overline{x}}\hat{\zeta}_{\overline{x}}^{\mu}\hat{\zeta}_{\overline{x}}^{\nu}+\cos^2\varphi_{\overline{x}}\hat{\eta}_{\overline{x}}^{\mu}\hat{\eta}_{\overline{x}}^{\nu} \nonumber 
\\
{}&\quad\qquad-2\sin\varphi_{\overline{x}}\cos\varphi_{\overline{x}}\hat{\zeta}_{\overline{x}}^{(\mu}\hat{\eta}_{\overline{x}}^{\nu)}-\frac{1}{2}g^{\mu\nu}\Bigg) \nonumber
\\
={}&\sum_{\overline{x}}\frac{\mathcal{N}_{\overline{x}}\Phi^2_{\overline{x}}}{16\pi^2\Lambda^2_*}\ln\left(\frac{0.681\Lambda_*}{\xi_x}\right) \nonumber
\\
{}&\qquad\times\left(\hat{\zeta}_{\overline{x}}^{\mu}\hat{\zeta}_{\overline{x}}^{\nu}+\hat{\eta}_{\overline{x}}^{\mu}\hat{\eta}_{\overline{x}}^{\nu}-g^{\mu\nu}\right).
\label{eq:FTVIntegral}
\end{align}
where we use the definition of the London length and an identical coordinate system as was used to compute $\langle T^{\hat{\mu}\hat{\nu}}_{\pi}\rangle$. We take the large cutoff radius $r^{\text{cut}}_x\rightarrow\infty$ limit and use the approximation
\begin{equation}
\frac{K_0(x_0^{\overline{x}})K_2(x_0^{\overline{x}})}{K^2_1(x_0^{\overline{x}})}-1\approx 2\left[\ln\left(\frac{1}{x_0^{\overline{x}}}\right)-0.384\right], \quad x_0^{\overline{x}}\ll 1.
\end{equation} 

Now we consider the electromagnetic field tensor terms in the mesoscopic stress-energy tensor. As we have previously discussed, the overlap between the magnetic fields due to different flux tubes and magnetized vortex lines is negligibly small, and hence we neglect the $\tilde{F}_{\overline{p}}\tilde{F}_{\overline{n}}$ terms in Eq.~(\ref{eq:MesoscopicSETensor3}). The London field is the same before and after averaging, and so the $\tilde{F}^2_{\text{L}}$ terms are unchanged by averaging other than removing the tilde. We combine the London field-vortex line/flux tube field cross terms in Eq.~(\ref{eq:MesoscopicSETensor3}) with the terms depending on the London four-potential to give
\begin{align}
\tilde{T}^{\mu\nu}_{B,\text{cross}}=\sum_{\overline{x}}\Bigg({}&-\frac{1}{2\pi}A_{\text{L}}^{(\mu}\nabla_{\rho}\tilde{F}^{\nu)\rho}_{\overline{x}}+\frac{1}{2\pi}\tilde{F}^{(\mu}_{\overline{x}\ \rho}\tilde{F}^{\nu)\rho}_{\text{L}}
\nonumber
\\
{}&+\frac{1}{4\pi}A^{\text{L}}_{\sigma}\nabla_{\rho}\tilde{F}^{\sigma\rho}_{\overline{x}}g^{\mu\nu}-\frac{1}{8\pi}\tilde{F}^{\sigma\rho}_{\overline{x}}\tilde{F}_{\sigma\rho}^{\text{L}}g^{\mu\nu}\Bigg),
\label{eq:TmunuMixed}
\end{align}
where we use the second equation in Eq.~(\ref{eq:MesoscopicMaxwellEquationSplit}) to replace $\delta v^{\mu}_{\overline{p}}$. In taking the average of this term, we approximate that the relative velocities $v^{\mu}_{\overline{x}}$ are small so $u^{\mu}_{\overline{x}}\approx u^{\mu}$, and we work in the common rest frame of the fluids. We also assume that the lines can be regarded as straight to lowest order.

First consider the first term on the right-hand side of Eq.~(\ref{eq:TmunuMixed}). Our treatment is similar to that of~\citet{Baym1983} for the terms in the energy per unit length coupling large-scale rotation and the flow around superfluid vortex lines. From Eq.~(\ref{eq:LondonFourPotential}), we approximate $A_{\text{L}}^{\mu}$ to be
\begin{equation}
A_{\text{L}}^{\mu}\approx K(n_x)u^{\mu},
\end{equation}
where $K(n_x)$ is a function of the number densities and can be approximated as constant over length scales $\ell$. To lowest order $u^{\mu}$ is pure rotation, so
\begin{equation}
u^{\mu}\approx-\delta^{\mu}_0+\delta^{\mu}_i\epsilon^i_{km}\Omega^kr^m,
\end{equation}
for radial position and rotational velocity three-vectors $r^m$ and $\Omega^k$. Ignoring electric fields associated with the vortex lines/flux tubes, the only nonzero components of the first two terms of Eq.~(\ref{eq:TmunuMixed}) will be the spatial components $\mu,\nu=i,j$, since $\delta^{\mu}_0\nabla_{\rho}\tilde{F}^{\nu\rho}_{\overline{x}}$ will average to zero. Expanding $r^m=r_0^m+\varpi_{\overline{x}}(\hat{\zeta}^m_{\overline{x}}\cos\varphi_{\overline{x}}+\hat{\eta}^m_{\overline{x}}\sin\varphi_{\overline{x}})$ near a vortex line/flux tube of species $\overline{x}$, where $\hat{\zeta}^i$ and $\hat{\zeta}^i$ are the three-vector versions of the spacelike four-vectors $\hat{\zeta}^{\mu_{\overline{x}}}$ and $\hat{\zeta}^{\mu}_{\overline{x}}$, we again perform the averaging integral for a single line and then multiply by the number density $\mathcal{N}_{\overline{x}}$:
\begin{align}
\bigg\langle-{}&\frac{1}{4\pi}A_{\text{L}}^{\mu}\nabla_{\rho}\tilde{F}^{\nu\rho}_{\overline{x}}\bigg\rangle \approx -\frac{1}{4\pi}\left\langle u^{i}\nabla_{\rho}\tilde{F}^{j\rho}_{\overline{x}}\right\rangle\delta^{\mu}_i\delta^{\nu}_j
\nonumber
\\
={}&-\frac{\mathcal{N}_{\overline{x}}\Phi_{\overline{x}}K(n_x)}{8\pi^2\Lambda^3_*}
\nonumber
\\
{}&\times\delta^{\mu}_i\delta^{\nu}_j\epsilon^i_{\ km}\Omega^k\int^{2\pi}_0 d\varphi_{\overline{x}}    (\hat{\zeta}^m_{\overline{x}}\cos\varphi_{\overline{x}}+\hat{\eta}^m_{\overline{x}}\sin\varphi_{\overline{x}})   
\nonumber
\\
{}&\times(\sin\varphi_{\overline{x}}\hat{\zeta}^j_{\overline{x}}-\cos\varphi_{\overline{x}}\hat{\eta}^j_{\overline{x}})\int^{r_x^{\text{cut}}}_{\xi_x} d\varpi_{\overline{x}}\varpi_{\overline{x}}^2\frac{K_1(\varpi_{\overline{x}}/\Lambda_*)}{x_0^{\overline{x}}K_1(x_0^{\overline{x}})}
\nonumber
\\
\approx{}&-\frac{\mathcal{N}_{\overline{x}}\Phi_{\overline{x}}K(n_x)}{4\pi}\left(\hat{t}^i_{\overline{x}}\Omega^j-\delta^{ij}\Omega^k\hat{t}^{\overline{x}}_k\right)\delta^{\mu}_i\delta^{\nu}_j,
\end{align}
where we use $\hat{t}^i_{\overline{x}}=\epsilon^i_{\ km}\hat{\zeta}^k_{\overline{x}}\hat{\eta}^m_{\overline{x}}$ and again approximate $x^{\overline{x}}_0\ll 1$, $r^{\text{cut}}_x\rightarrow\infty$. So
\begin{align}
\left\langle -\frac{1}{2\pi}A_{\text{L}}^{(\mu}\nabla_{\rho}\tilde{F}^{\nu)\rho}_{\overline{x}}\right\rangle\approx{}&-\frac{\mathcal{N}_{\overline{x}}\Phi_{\overline{x}}K(n_x)}{4\pi}\delta^{\mu}_i\delta^{\nu}_j
\nonumber
\\
{}&\times\left((\hat{t}^i_{\overline{x}}\Omega^j+\hat{t}^j_{\overline{x}}\Omega^i)-2\delta^{ij}\Omega^k\hat{t}^{\overline{x}}_k\right)
\label{eq:FirstTmunuCrossTerm}
\end{align}
Ignoring electric fields, only the spatial components $\mu,\nu=i,j$ of the second term in Eq.~(\ref{eq:TmunuMixed}) survive, so using $F^{\mu\nu}_{\text{L}}=\tilde{F}^{\mu\nu}_{\text{L}}\approx 2K(n_x)\epsilon^{ij}_{\ k}\Omega^k\delta^{\mu}_i\delta^{\nu}_j$ and $\tilde{F}^{\mu\nu}_{\overline{x}}\approx \sum_a\epsilon^{ij}_{\ k}\delta B^k_{\overline{x},a}\delta^{\mu}_i\delta^{\nu}_j$, we find
\begin{align}
\left\langle\frac{1}{4\pi}\tilde{F}^{\mu}_{\overline{x}\rho}\tilde{F}^{\nu\rho}_{\text{L}}\right\rangle \approx{}& \frac{\mathcal{N}_{\overline{x}}\Phi_{\overline{x}}K(n_x)}{4\pi^2\Lambda^2_*}(\hat{t}^i_{\overline{x}}\Omega^j-\delta^{ij}\Omega^k\hat{t}^{\overline{x}}_k)\int^{2\pi}_0 d\varphi_{\overline{x}}
\nonumber
\\
{}& \times\int^{r_x^{\text{cut}}}_{\xi_x} d\varpi_{\overline{x}}\varpi_{\overline{x}}\frac{K_0(\varpi_{\overline{x}}/\Lambda_*)}{x_0^{\overline{x}}K_1(x_0^{\overline{x}})}   \delta^{\mu}_i\delta^{\nu}_j
\nonumber
\\
\approx{}&\frac{\mathcal{N}_{\overline{x}}\Phi_{\overline{x}}K(n_x)}{2\pi}(\hat{t}^i_{\overline{x}}\Omega^j-\delta^{ij}\Omega^k\hat{t}^{\overline{x}}_k)\delta^{\mu}_i\delta^{\nu}_j,
\end{align}
so
\begin{equation}
\left\langle\frac{1}{2\pi}\tilde{F}^{(\mu}_{\overline{x}\rho}\tilde{F}^{\nu)\rho}_{\text{L}}\right\rangle\approx\left\langle\frac{1}{\pi}A_{\text{L}}^{(\mu}\nabla_{\rho}\tilde{F}^{\nu)\rho}_{\overline{x}}\right\rangle.
\label{eq:NonCancellationofCrossTerms}
\end{equation}
The first two terms in Eq.~(\ref{eq:TmunuMixed}) thus partially cancel each other upon averaging under these approximations. The remaining terms in Eq.~(\ref{eq:TmunuMixed}), those proportional to $g^{\mu\nu}$, entirely cancel each other, which is demonstrated by taking the trace of Eq.~(\ref{eq:NonCancellationofCrossTerms}). That these remaining two terms cancel is consistent with the rotation energy cancellation lemma applied in~\cite{Carter2000,Prix2000,Glampedakis2011}, since nonrelativistically the energy density is simply the coefficient of the metric in the stress-energy tensor. However, we do not find that there is a tensorial version of the rotation energy cancellation lemma that eliminates all such cross terms from $\langle\tilde{T}^{\mu\nu}\rangle$. Additionally, higher order (in velocity over $c$) corrections to the energy density will appear due to contractions between the four-velocity of the reference frame in which the energy density is measured and the non-canceled part of the first two terms in Eq.~(\ref{eq:TmunuMixed}).

The final terms to average in $\Delta \tilde{T}^{\mu\nu}$ are
\begin{equation}
\tilde{T}^{\hat{\mu}\hat{\nu}}_{B,\overline{x}}=\frac{1}{4\pi}\left[\tilde{F}^{\mu\rho}_{\overline{x}}\tilde{F}^{\nu}_{\overline{x}\rho}-\frac{1}{4}\tilde{F}^{\sigma\rho}_{\overline{x}}\tilde{F}^{\overline{x}}_{\sigma\rho}g^{\mu\nu}\right].
\end{equation}
Using Eq.~(\ref{eq:FmunufromB}) and again ignoring interactions between different flux tubes and vortex lines, we can compute the average of $\tilde{T}^{\hat{\mu}\hat{\nu}}_{B}$ analogously to Eq.~(\ref{eq:VLPiIntegral},\ref{eq:FTVIntegral}), giving
\begin{align}
\langle\tilde{T}^{\mu\nu}_{B,\overline{x}}\rangle={}&\sum_{\overline{x}}\frac{\mathcal{N}_{\overline{x}}\Phi^2_{\overline{x}}}{16\pi^3\Lambda^4_*}\int^{2\pi}_0d\varphi_{\overline{x}}\int^{r^{\text{cut}}_x}_{\xi_x}\frac{d\varpi_{\overline{x}}\varpi_{\overline{x}}K^2_0(\varpi_{\overline{x}}/\Lambda_*)}{[x_0^{\overline{x}}K_1(x_0^{\overline{x}})]^2} \nonumber
\\
{}&\qquad\times\left(u^{\mu}_{\overline{x}}u^{\nu}_{\overline{x}}+\frac{1}{2}g^{\mu\nu}-\hat{t}^{\hat{\mu}}_{\overline{x}}\hat{t}^{\nu}_{\overline{x}}\right) \nonumber
\\
={}&\sum_{\overline{x}}\frac{\mathcal{N}_{\overline{x}}\Phi_{\overline{x}}^2}{16\pi^2\Lambda^2_*}\left(u^{\mu}_{\overline{x}}u^{\nu}_{\overline{x}}+\frac{1}{2}g^{\mu\nu}-\hat{t}^{\mu}_{\overline{x}}\hat{t}^{\nu}_{\overline{x}}\right),
\label{eq:FxFxIntegral}
\end{align}
where we used the approximation $K^2_0(x_0^{\overline{x}})/K^2_1(x_0^{\overline{x}})\approx 0$ for $x_0^{\overline{x}}\ll 1$.

Using Eq.~(\ref{eq:VorticityVector1},\ref{eq:VorticityVector2},\ref{eq:MetricUnitVectors}) we can rewrite the following contractions of tensors as
\begin{align}
w^{\rho\sigma}_{\overline{x}}w^{\overline{x}}_{\rho\sigma}={}& 2W_{\overline{x}}^{\sigma}W^{\overline{x}}_{\sigma}=2(\mathcal{N}_{\overline{x}}\Phi_{\overline{p}}e)^2=4X_{\overline{x}},
\\
w^{\mu\rho}_{\overline{x}}w^{\nu}_{\overline{x}\rho}={}&(g^{\mu\nu}+u^{\mu}_{\overline{x}}u^{\nu}_{\overline{x}})W^{\sigma}_{\overline{x}}W_{\sigma}^{\overline{x}}-W^{\mu}_{\overline{x}}W^{\nu}_{\overline{x}} \nonumber
\\
={}&(\mathcal{N}_{\overline{x}}\Phi_{\overline{p}}e)^2\left[g^{\mu\nu}+u^{\mu}_{\overline{x}}u^{\nu}_{\overline{x}}-\hat{t}^{\mu}_{\overline{x}}\hat{t}^{\nu}_{\overline{x}}\right]
\nonumber
\\
={}&(\mathcal{N}_{\overline{x}}\Phi_{\overline{p}}e)^2\left[\hat{\zeta}_{\overline{x}}^{\mu}\hat{\zeta}_{\overline{x}}^{\nu}+\hat{\eta}_{\overline{x}}^{\mu}\hat{\eta}_{\overline{x}}^{\nu}\right].
\label{eq:WContractionUnitVectors}
\end{align}
Replacing terms in Eq.~(\ref{eq:MesoscopicSETensor3}) with their respective averages and using Eq.~(\ref{eq:WContractionUnitVectors}), we finally obtain
\begin{align}
\langle\tilde{T}^{\mu\nu}\rangle={}&\sum_x n^{\mu}_x\mu^{\nu}_x
+\left(\tilde{\Lambda}_0+\sum_xn_x\mu_x\right)g^{\mu\nu}
\nonumber
\\
{}&+\sum_{\overline{x}}\left\langle\frac{1}{4\pi}\tilde{F}^{(\mu}_{\overline{x}\rho}\tilde{F}^{\nu)\rho}_{\text{L}}\right\rangle
\nonumber 
\\
{}&+\frac{1}{4\pi}\left(F^{\mu\rho}_{\text{L}}F^{\nu}_{\text{L}\rho}-\frac{1}{4}F^{\sigma\rho}_{\text{L}}F^{\text{L}}_{\sigma\rho}g^{\mu\nu}\right)
\nonumber
\\
{}&+\sum_{\overline{x}}\frac{\mathcal{E}_{v,\overline{x}}}{\mathcal{N}_{\overline{x}}(\Phi_{\overline{p}}e)^2}\left(w^{\mu\rho}_{\overline{x}}w^{\nu}_{\overline{x}\rho}-\frac{1}{2}w^{\sigma\rho}_{\overline{x}}w^{\overline{x}}_{\sigma\rho}g^{\mu\nu}\right)
 \nonumber
\\
{}&+\sum_{\overline{x}}\frac{1}{32\pi^2\mathcal{N}_{\overline{x}}\Lambda^2_*e^2}w^{\mu\rho}_{\overline{x}}w^{\nu}_{\overline{x}\rho},
\label{eq:MesoscopicSETensor4}
\end{align}
which is Eq.~(\ref{eq:MesoscopicSETensor4Main}) in the main text, and where the $\mathcal{E}_{v,\overline{x}}$ are defined in Eq.~(\ref{eq:VLEnergyP}--\ref{eq:VLEnergyN}).

\subsection{Matching to the macroscopic stress-energy tensor}
\label{app:MatchingSETensors}

We now match the macroscopic stress-energy tensor as found in Eq.~(\ref{eq:StressEnergyTensor}) with the averaged mesoscopic stress-energy tensor Eq.~(\ref{eq:MesoscopicSETensor4}). To begin, expanding out the $\sum_x n^{\rho}_x\mu^x_{\rho}$ terms using the definitions of the currents and conjugate momenta from Section~\ref{sec:RelationtoPhysParams}. This gives
\begin{align}
T^{\mu\nu}={}&\sum_xn_x^{\mu}\mu_x^{\nu}+\frac{1}{4\pi}\mathcal{K}^{\mu\rho}F^{\nu}_{\ \rho}+\sum_{\overline{x}}\lambda^{\mu\rho}_{\overline{x}} w^{\nu}_{{\overline{x}}\rho}
\nonumber
\\
{}&+\hspace{-0.5mm}\left(\hspace{-0.5mm}\Lambda+\hspace{-0.5mm}\sum_xn_x\mu_x\hspace{-0.5mm}\right)\hspace{-0.4mm}g^{\mu\nu}.
\label{eq:MacroSETensor}
\end{align}
The macroscopic current terms in Eq.~(\ref{eq:MacroSETensor}) and Eq.~(\ref{eq:MesoscopicSETensor4}) match, so we now focus on matching the remaining terms.

We postulate the Lorentz-invariant form of the macroscopic master function $\Lambda$ as in Eq.~(\ref{eq:MacroscopicMasterFunctionSplit}) and identify $\Lambda_0=\tilde{\Lambda}_0$. Using Eq.~(\ref{eq:MacroscopicMasterFunctionSplit}--\ref{eq:LambdaNDefinition}) in Eq.~(\ref{eq:MacroSETensor}) and then substituting $F^{\mu\nu}$ using Eq.~(\ref{eq:FmunuMesoscopicAverage}) gives
\begin{widetext}
\begin{align}
\Delta T^{\mu\nu}={}&\Lambda_{\text{EM+V}}g^{\mu\nu}-\frac{\partial\Lambda_{\text{EM+V}}}{\partial X_F}F^{\mu\rho}_{\text{L}}F^{\nu}_{\text{L}\rho}-\left[\frac{1}{e^2}\frac{\partial\Lambda_{\text{EM+V}}}{\partial X_F}+\frac{2}{e}\frac{\partial\Lambda_{\text{EM+V}}}{\partial Y_{\overline{p}}}+\frac{\partial\Lambda_{\text{EM+V}}}{\partial X_{\overline{p}}}\right]w^{\mu\rho}_{\overline{p}}w^{\nu}_{\overline{p}\rho} \nonumber
\\
{}&-\left[\frac{Y_{np}^2}{e^2Y_{pp}^2}\frac{\partial\Lambda_{\text{EM+V}}}{\partial X_F}+\frac{2Y_{np}}{eY_{pp}}\frac{\partial\Lambda_{\text{EM+V}}}{\partial Y_{\overline{n}}}+\frac{\partial\Lambda_{\text{EM+V}}}{\partial X_{\overline{n}}}\right]w^{\mu\rho}_{\overline{n}}w^{\nu}_{\overline{n}\rho} \nonumber
\\
{}&+2\left[\frac{1}{e}\frac{\partial\Lambda_{\text{EM+V}}}{\partial X_F}+\frac{\partial\Lambda_{\text{EM+V}}}{\partial Y_{\overline{p}}}\right]F^{\rho(\mu}_{\text{L}}w^{\nu)}_{\overline{p}\rho}+2\left[\frac{Y_{np}}{eY_{pp}}\frac{\partial\Lambda_{\text{EM+V}}}{\partial X_F}+\frac{\partial\Lambda_{\text{EM+V}}}{\partial Y_{\overline{n}}}\right]F^{\rho(\mu}_{\text{L}}w^{\nu)}_{\overline{n}\rho}
\nonumber
\\
{}&+2\left[\frac{Y_{np}}{e^2Y_{pp}}\frac{\partial\Lambda_{\text{EM+V}}}{\partial X_F}+\frac{Y_{np}}{eY_{pp}}\frac{\partial\Lambda_{\text{EM+V}}}{\partial Y_{\overline{p}}}+\frac{1}{e}\frac{\partial\Lambda_{\text{EM+V}}}{\partial Y_{\overline{n}}}+\frac{\partial\Lambda_{\text{EM+V}}}{\partial Z}\right]w^{\rho(\mu}_{\overline{p}}w^{\nu)}_{\overline{n}\rho},
\label{eq:MacroSETensor2}
\end{align}
\end{widetext}
where $\Delta T^{\mu\nu}$ is defined to only include those terms in the macroscopic stress-energy tensor which do not have an exact matching term in the mesoscopic stress-energy tensor, but including all of the electromagnetic terms.

We can now match terms by comparing Eq.~(\ref{eq:MacroSETensor2}) to $\langle\Delta \tilde{T}^{\mu\nu}\rangle$ (the last four lines of Eq.~(\ref{eq:MesoscopicSETensor4})). We first note that since the $\langle\tilde{F}^{(\mu}_{\overline{x}\rho}\tilde{F}^{\nu)\rho}_{\text{L}}\rangle/4\pi$ terms in  Eq.~(\ref{eq:MesoscopicSETensor4}) do not have a corresponding term proportional to the metric (such a term having been eliminated by the rotation energy cancellation lemma), there is no way to incorporate such a term into the macroscopic effective theory. If we try to include this term in the theory, say by matching to the third line of Eq.~(\ref{eq:MacroSETensor2}) using $\langle\tilde{F}^{(\mu}_{\overline{x}\rho}\tilde{F}^{\nu)\rho}_{\text{L}}\rangle=\Phi_{\overline{x}}/(e\Phi_{\overline{p}})w^{(\mu}_{\overline{x}\rho}\tilde{F}^{\nu)\rho}_{\text{L}}$, we will find that the partial derivatives of $\Lambda_{\text{EM+V}}$ will be inconsistent with the definition of $\Lambda_{\text{EM+V}}$ found by matching the terms proportional to $g^{\mu\nu}$. We thus exclude these terms from the matching procedure and from the resulting macroscopic effective theory. For this reason, a reader might choose to use the averaged mesoscopic theory rather than the effective theory for dynamical calculations, though the terms missing from the macroscopic theory are relatively unimportant for dynamics.

Matching the London magnetic field squared terms requires
\begin{equation}
\frac{\partial\Lambda_{\text{EM+V}}}{\partial X_F}=-\frac{1}{4\pi}.
\label{eq:LondonFieldSquaredMatch}
\end{equation}
Matching to the London field--flux tube/vortex line field cross terms, which are all zero in the averaged mesoscopic theory after dropping the $\langle\tilde{F}^{(\mu}_{\overline{x}\rho}\tilde{F}^{\nu)\rho}_{\text{L}}\rangle/4\pi$ term, and using Eq.~(\ref{eq:LondonFieldSquaredMatch}), we require
\begin{align}
\frac{\partial\Lambda_{\text{EM+V}}}{\partial Y_{\overline{p}}}=\frac{1}{4\pi e}, \qquad\frac{\partial\Lambda_{\text{EM+V}}}{\partial Y_{\overline{n}}}=\frac{Y_{np}}{4\pi eY_{pp}}.
\label{eq:CrossTermNMatch}
\end{align}
The flux tube/vortex line cross term in Eq.~(\ref{eq:MacroSETensor2}) are zero as a result of our ignoring their interactions in the averaged mesoscopic theory. In accordance with Eq.~(\ref{eq:LondonFieldSquaredMatch}--\ref{eq:CrossTermNMatch}), this requires
\begin{equation}
\frac{\partial\Lambda_{\text{EM+V}}}{\partial Z}=-\frac{Y_{np}}{4\pi e^2Y_{pp}}.
\label{eq:PartialZMatch}
\end{equation}
Matching to terms proportional to $w^{\mu\rho}_{\overline{x}}w^{\nu}_{\overline{x}\rho}$ gives
\begin{align}
\frac{\partial\Lambda_{\text{EM+V}}}{\partial X_{\overline{p}}}\hspace{-0.4mm}={}&\hspace{-0.8mm}-\hspace{-0.4mm}\frac{1}{4\pi e^2}\hspace{-0.4mm}-\hspace{-0.4mm}\frac{1}{\mathcal{N}_{\overline{p}}(\Phi_{\overline{p}}e)^2}\left(\hspace{-0.4mm}\mathcal{E}_{v,\overline{p}}+\hspace{-0.5mm}\frac{\Phi^2_{\overline{p}}}{32\pi^2\Lambda_*^2}\right),
\label{eq:PartialXpMatch}
\\
\frac{\partial\Lambda_{\text{EM+V}}}{\partial X_{\overline{n}}}\hspace{-0.4mm}={}&\hspace{-0.8mm}-\hspace{-0.4mm}\frac{Y_{np}^2}{Y_{pp}^2}\hspace{-0.3mm}\left[\hspace{-0.4mm}\frac{1}{4\pi e^2}\hspace{-0.4mm}+\hspace{-0.4mm}\frac{1}{\mathcal{N}_{\overline{n}}(\Phi_{\overline{n}}e)^2}\left(\hspace{-0.4mm}\mathcal{E}_{v,\overline{n}}+\hspace{-0.5mm}\frac{\Phi^2_{\overline{n}}}{32\pi^2\Lambda_*^2}\right)\hspace{-0.4mm}\right]\hspace{-0.8mm}.
\label{eq:PartialXnMatch}
\end{align}

Matching terms proportional to $g^{\mu\nu}$ gives the same $\Lambda_{\text{EM+V}}$ as Eq.~(\ref{eq:LambdaEMV1}). Rewriting this in terms of the scalars $X_F,X_{\overline{x}},Y_{\overline{x}},Z$ and taking the partial derivatives of $\Lambda_{\text{EM+V}}$ with respect to each scalar, we obtain the same results as in Eq.~(\ref{eq:LondonFieldSquaredMatch}--\ref{eq:PartialZMatch}). However, we do not completely recover Eq.~(\ref{eq:PartialXpMatch}--\ref{eq:PartialXnMatch}) and miss additional vortex line/flux tube magnetic field energy contributions $\propto\Phi_{\overline{x}}^2/(32\pi^2\Lambda_*^2)$ (in fact, one-half the magnetic field energy per unit length). In the strong type-II superconductivity limit, the missing terms would be irrelevant and so both ways to find $\Lambda_{\text{EM+V}}$ would be consistent. We drop them regardless of the physical limit, which is equivalent to dropping the last line in Eq.~(\ref{eq:MesoscopicSETensor4}). We also gain an extra term in Eq.~(\ref{eq:PartialXnMatch}) because of the $\mathcal{N}_{\overline{x}}$-dependence of the vortex line energy cutoff radius in $\mathcal{E}_{v,\overline{n}}$. This contribution is argued to be small in Eq.~(\ref{eq:EvnDerivativeArgument}).

That we cannot obtain a completely consistent macroscopic master function and stress-energy tensor from averaging the mesoscopic theory is not entirely surprising, as we had no reason to believe this was possible before we began. We can at least have an approximate effective macroscopic theory by using the $\Lambda_{\text{EM+V}}$ found by matching terms proportional to $g^{\mu\nu}$ between the averaged mesoscopic and macroscopic theories and then ignoring terms in the stress-energy tensor inconsistent with this-- fortunately there are only three such terms, and in the strong type-II superconductivity limit and for $d_n\gg\xi_n$ only the $\langle\tilde{F}^{(\mu}_{\overline{x}\rho}\tilde{F}^{\nu)\rho}_{\text{L}}\rangle/4\pi$ term is not negligible.

\bibliographystyle{apsrev4-2}
\bibliography{library,textbooks}

\end{document}